\documentclass[oneside,english]{elsart}

\usepackage{times}
\usepackage[T1]{fontenc}
\usepackage[latin1]{inputenc}
\usepackage{wrapfig}
\usepackage{subfigure}
\usepackage{color}
\usepackage{graphicx}
\usepackage{float}

\usepackage{amsmath,amsfonts,amssymb}

\DeclareMathAlphabet{\bi}{OT1}{cmr}{bx}{it}

\newcommand{\wh}[1]{ { \widehat{#1} } }
\newcommand{\wwh}[1]{ {\wh{\wh{#1}}} }
\newcommand{\wwhc}[1]{ {\wwh{\,#1\,}} }

\newcommand{\RE}{{\text{Re}}}
\newcommand{\IM}{{\text{Im}}}

\newcommand{\nph}{{m+\frac{1}{2}}}
\newcommand{\nmh}{{m-\frac{1}{2}}}

\newcommand{\nout}{ {n_0^{\text{ext}} } } 
\newcommand{\Einc}{ { E_{\text{inc}}^{0} } } 
\newcommand{\lri}{\nu}  
\newcommand{\hk}{{\tilde{h}}}
\newcommand{\epscrit}{\epsilon_{\rm c}}  

\newcommand{\ddEp}{ {E''_{m+}} }  
\newcommand{\ddEm}{ {E''_{(m+1)-}} } 

\newcommand{\DS}{{ \bvec{F}(\bvec{E}) }}
\newcommand{\DSm}{{ F_m ( \bvec{E} ) }}

\newcommand{\eqdef}{ = }

\newcommand{\oh}[1]{ { \mathcal{O}\left(h^{#1}\right) } }
	
\newcommand{\odesq} {\mathcal{O}\left(|\delta E|^2\right)}

\newcommand{\bvec}[1] {{\bf {#1}}}
\newcommand{\hbvec}[1] { {\wh{\bvec{#1}} }}

\newcommand{\lf}{{\rm left}}
\newcommand{\rt}{{\rm right}}

\makeatletter

\usepackage{cite}
\usepackage{babel}
\makeatother
\begin{document}
\begin{frontmatter}

\title{High-order numerical method for the nonlinear Helmholtz equation with material 
discontinuities in one space dimension}

\author[Tel-Aviv]{G. Baruch}
\ead{guybar@tau.ac.il}
\ead[url]{http://www.tau.ac.il/$\sim$guybar}
\author[Tel-Aviv]{G. Fibich}
\ead{fibich@tau.ac.il}
\ead[url]{http://www.math.tau.ac.il/$\sim$fibich}
\author[Raleigh]{S. Tsynkov\thanksref{cor}}
\ead{tsynkov@math.ncsu.edu}
\ead[url]{http://www.math.ncsu.edu/$\sim$stsynkov}
\address[Tel-Aviv]{Department of Applied Mathematics,
School of Mathematical Sciences, Tel Aviv University, Ramat Aviv, Tel Aviv 69978, Israel}
\address[Raleigh]{Department of Mathematics,
North Carolina State University, Box 8205, Raleigh, NC 27695, USA}
\thanks[cor]{Corresponding author. Phone: +1-919-515-1877, Facsimile: +1-919-513-7336.
The research of this author was supported by the US NSF, Grant \#~DMS-0509695, and by the US Air Force,
Grants \#~FA9550-04-1-0118 and \#~FA9550-07-1-0170.}

\begin{abstract}
The nonlinear Helmholtz equation~(NLH) models the propagation of electromagnetic waves in Kerr media, and describes  a range of important phenomena in nonlinear optics and in other areas. 
In our previous work, we developed a fourth order method for its numerical  solution that involved an iterative solver  based on freezing the nonlinearity.
The method enabled a direct simulation of nonlinear self-focusing in the nonparaxial regime, and a quantitative prediction of backscattering. 
However, our simulations showed that there is a threshold value for the magnitude of the nonlinearity, above which the iterations diverge.

In this study, we numerically solve the one-dimensional NLH using a Newton-type nonlinear solver.
Because the Kerr nonlinearity contains absolute values of the field, the NLH has to be recast as a system of two real equations in order to apply Newton's method.
Our numerical simulations show that Newton's method converges rapidly and, in contradistinction with the iterations based on freezing the nonlinearity, enables computations for very high levels of nonlinearity. 

In addition, we introduce a novel compact finite-volume fourth order discretization for the NLH with material discontinuities.
Our computations corroborate the design fourth order convergence of the method.

The one-dimensional results of the current paper create a foundation for the 
analysis of multi-dimensional problems in the future.

\end{abstract}
\end{frontmatter}

\section{\label{sec:intro}Introduction}

\subsection{Background}
\label{sec:backgr}

The nonlinear Helmholtz equation~(NLH)
\begin{equation}
    \Delta E(\bvec{x}) + \frac{\omega_0^2}{c^2}n^2 E = 0,\qquad 
	n^2(\bvec{x},|E|) = n_0^2(\bvec{x}) + 2n_0(\bvec{x})n_2(\bvec{x})|E|^{2}, 
    \label{eq:NLH}
\end{equation}
governs the propagation of linearly-polarized, time-harmonic electromagnetic waves in Kerr-type dielectrics.
Here, $\bvec{x} = [x_1,\ldots,x_D]$ are the spatial coordinates,
$E=E(\bvec{x})$ denotes the scalar electric field,
$\omega_0$ is the laser frequency, $c$ is the speed of light in vacuum,
$\Delta=\partial_{x_1}^2+\ldots+\partial_{x_D}^2$ is the $D$-dimensional Laplacian, 
$n_0$ is the linear index of refraction,
and $n_2$ is the Kerr coefficient.
In this study, we consider the case of an inhomogeneous medium in which both $n_0$ and $n_2$ can vary in space.
We assume that the medium is lossless, i.e., that $n_0$ and $n_2$ are real.
Furthermore, we consider only the case in which 
the electric field $E$ and the material coefficients $n_0$ and $n_2$ 
vary in one spatial direction that we identify with the direction of propagation and denote by~$z$. 
Hence, equation~(\ref{eq:NLH}) reduces to the one-dimensional cubic NLH:
\begin{equation}
    \frac{d^2E(z)}{dz^2}
    + \frac{\omega_0^2}{c^2}\left(n_0^2(z) + 2n_0(z)n_2(z)\left|E\right|^2\right) E = 0.
    \label{eq:1DNLH-2}
\end{equation}
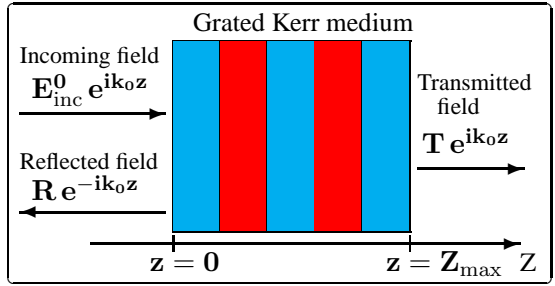
\begin{wrapfigure}{r}{0.5\textwidth}
	\begin{center}

\setlength{\unitlength}{1302sp}%
\begingroup\makeatletter\ifx\SetFigFont\undefined%
\gdef\SetFigFont#1#2#3#4#5{%
  \reset@font\fontsize{#1}{#2pt}%
  \fontfamily{#3}\fontseries{#4}\fontshape{#5}%
  \selectfont}%
\fi\endgroup%
\begin{picture}(10374,5349)(439,-5248)
\thicklines
{\color[rgb]{0,0,0}\put(3451,-3886){\vector(-1, 0){2775}}
}%
{\color[rgb]{0,0,0}\put(8251,-3061){\vector( 1, 0){2025}}
}%
{\color[rgb]{0,0,0}\put(676,-2011){\vector( 1, 0){2775}}
}%
\thinlines
{\color[rgb]{0,0,0}\put(556,-5131){\oval(210,210)[bl]}
\put(556,-16){\oval(210,210)[tl]}
\put(10696,-5131){\oval(210,210)[br]}
\put(10696,-16){\oval(210,210)[tr]}
\put(556,-5236){\line( 1, 0){10140}}
\put(556, 89){\line( 1, 0){10140}}
\put(451,-5131){\line( 0, 1){5115}}
\put(10801,-5131){\line( 0, 1){5115}}
}%
{\color[rgb]{0,0,0}\put(3600,-4261){\framebox(900,3630){\textcolor{cyan}{\rule{0.25in}{1in}}}}
}%
{\color[rgb]{0,0,0}\put(4500,-4261){\framebox(900,3630){\textcolor{red}{\rule{0.25in}{1in}}}}
}%
{\color[rgb]{0,0,0}\put(5400,-4261){\framebox(900,3630){\textcolor{cyan}{\rule{0.25in}{1in}}}}
}%
{\color[rgb]{0,0,0}\put(6300,-4261){\framebox(900,3630){\textcolor{red}{\rule{0.25in}{1in}}}}
}%
{\color[rgb]{0,0,0}\put(7200,-4261){\framebox(900,3630){\textcolor{cyan}{\rule{0.25in}{1in}}}}
}%
{\color[rgb]{0,0,0}\put(3601,-4336){\line( 0,-1){300}}
\put(3601,-4636){\line( 0, 1){300}}
}%
\thicklines
{\color[rgb]{0,0,0}\put(2026,-4486){\vector( 1, 0){8175}}
}%
\thinlines
{\color[rgb]{0,0,0}\put(8101,-4336){\line( 0,-1){300}}
\put(8101,-4636){\line( 0, 1){300}}
}%
\put(3976,-436){\makebox(0,0)[lb]{\smash{{\SetFigFont{10}{12.0}{\rmdefault}{\mddefault}{\updefault}{\color[rgb]{0,0,0}Grated Kerr medium}%
}}}}
\put(10201,-5011){\makebox(0,0)[lb]{\smash{{\SetFigFont{10}{12.0}{\rmdefault}{\mddefault}{\updefault}{\color[rgb]{0,0,0}Z}%
}}}}
\put(3151,-5011){\makebox(0,0)[lb]{\smash{{\SetFigFont{11}{13.2}{\familydefault}{\mddefault}{\updefault}{\color[rgb]{0,0,0}$\bf z=0$}%
}}}}
\put(8326,-2761){\makebox(0,0)[lb]{\smash{{\SetFigFont{11}{13.2}{\familydefault}{\mddefault}{\updefault}{\color[rgb]{0,0,0}$\bf T \, e^{ik_0z}$}%
}}}}
\put(676,-1036){\makebox(0,0)[lb]{\smash{{\SetFigFont{9}{10.8}{\rmdefault}{\mddefault}{\updefault}{\color[rgb]{0,0,0}Incoming field}%
}}}}
\put(676,-3061){\makebox(0,0)[lb]{\smash{{\SetFigFont{9}{10.8}{\rmdefault}{\mddefault}{\updefault}{\color[rgb]{0,0,0}Reflected field}%
}}}}
\put(901,-3661){\makebox(0,0)[lb]{\smash{{\SetFigFont{11}{13.2}{\familydefault}{\mddefault}{\updefault}{\color[rgb]{0,0,0}$\bf R\,e^{-ik_0z}$}%
}}}}
\put(901,-1711){\makebox(0,0)[lb]{\smash{{\SetFigFont{11}{13.2}{\familydefault}{\mddefault}{\updefault}{\color[rgb]{0,0,0}$\bf E_{\rm inc}^0 \, e^{ik_0z}$}%
}}}}
\put(8251,-1561){\makebox(0,0)[lb]{\smash{{\SetFigFont{9}{10.8}{\rmdefault}{\mddefault}{\updefault}{\color[rgb]{0,0,0}Transmitted }%
}}}}
\put(8626,-2011){\makebox(0,0)[lb]{\smash{{\SetFigFont{9}{10.8}{\rmdefault}{\mddefault}{\updefault}{\color[rgb]{0,0,0} field}%
}}}}
\put(7651,-5011){\makebox(0,0)[lb]{\smash{{\SetFigFont{11}{13.2}{\familydefault}{\mddefault}{\updefault}{\color[rgb]{0,0,0}$\bf z=Z_{\max}$}%
}}}}
\end{picture}%
		\caption{ A grated Fabry-Perot etalon.}
		\label{fig:physical_setup}
	\end{center}
\end{wrapfigure}
The ordinary differential equation~(\ref{eq:1DNLH-2}) arises, for example, when modeling nonlinear optical devices, such as the Fabry-Perot~etalon~\cite{born-wolf}, see Figure~\ref{fig:physical_setup}.
This device consists of a layer or slab of Kerr medium located between $\left.0\leq z\leq Z_{\max}\right.$.
The Kerr slab is surrounded by a linear homogeneous medium, so that $\left.n_0\equiv \nout \right.$ and $\left.n_2\equiv0\right.$ for $\left.z<0\right.$ and for $\left.z>Z_{\max}\right.$.
We consider the case when an incoming plane wave $\left.E=\Einc e^{ik_0z}\right.$ impinges normally on the slab at the interface $z=0$. Here, $
	\left.k_0=\frac{\omega_0}{c}\nout\right.
$ is the linear wavenumber in the surrounding linear medium.
Let us define \[
	\lri(z)=(n_0(z)/\nout)^2, \qquad
	\epsilon(z)=2n_2(z)n_0(z)/(\nout)^2.
\]
Then, equation~(\ref{eq:1DNLH-2}) transforms into
\begin{equation}
    \frac{d^2E(z)}{dz^2}
    + k_0^2 \left(\lri(z) + \epsilon(z)\left|E\right|^2\right) E = 0,
    \label{eq:1DNLH-1}
\end{equation}
where $\lri\equiv 1$ and $\epsilon\equiv 0$ for $z<0$ and for $z>Z_{\max}$.

We assume that the Kerr material is either homogeneous, i.e.,
\begin{equation}
	\label{eq:homogeneous-material}
	\lri(z)\equiv\lri^\text{int},	\qquad
	\epsilon(z)\equiv\epsilon^\text{int},	\qquad 
	0\leq z \leq Z_{\max},
\end{equation}
or layered (piecewise-constant). 
The latter case corresponds to a one-dimensional grating (see Figure~\ref{fig:physical_setup}), 
where for some given partition:
\begin{subequations}
	\label{eqs:grated_material}
	\begin{equation}
    	0 = \tilde{z}_1 < \dots < \tilde{z}_l < \dots < \tilde{z}_L = Z_{\max},
	\end{equation}
	we have:
	\begin{equation}
		\lri(z)\equiv\tilde{\lri}_l,\quad
    	\epsilon(z)\equiv\tilde{\epsilon}_l, \quad \text{for} \quad
    	z\in\left(\tilde{z}_l, \tilde{z}_{l+1}\right).
	\end{equation}
\end{subequations}
At the interfaces $\tilde{z}_l$, the boundary conditions for Maxwell's equations 
imply continuity of the field~$E(z)$ and its first derivative~$\frac{dE}{dz}$ 
(see Appendix~\ref{app:E_z-continuity}).
Note that, the material coefficients $\lri(z)$ and $\epsilon(z)$ are, generally speaking, discontinuous at the Kerr medium boundaries $z=0$ and $z=Z_{\max}$.

When equation~(\ref{eq:1DNLH-1}) is considered on the interval $0\leq z\leq Z_{\max}$, it needs to be supplemented by boundary conditions at $z=0$ and $z=Z_{\max}$. 
Outside of this interval, the field propagates linearly with $\lri\equiv1$ and $\epsilon\equiv0$.
\begin{subequations}
	Therefore, for $\left.z\leq0\right.$, the total field is composed of a given incoming wave and the unknown reflected wave
	\begin{equation}
		E(z)=\Einc e^{ik_0z}+Re^{-ik_0z}. \label{eq:inc_ref_field}
	\end{equation}
	For $\left.z\geq Z_{\max}\right.$, the electric field is given by the transmitted wave
	\begin{equation}
		E(z)=T e^{ik_0z}. \label{eq:transmitted_field}
	\end{equation}
\end{subequations}
The transmitted and reflected waves shall be interpreted as outgoing with respect to the domain of interest $[0,Z_{\max}]$. 
Note that the left-traveling wave $Re^{-ik_0z}$ contains the field reflected from the interface $z=0$  (i.e., the reflection per se), as well as the field generated by nonlinear backscattering inside the interval $[0,Z_{\max}]$.

The transmitted field (\ref{eq:transmitted_field}) satisfies a Sommerfeld-type homogeneous differential relation at $z=Z_{\max}+$:
\begin{equation*}
	\left.
		\left( \frac{d}{dz}-ik_0 \right) E 
	\right|_{z=Z_{\max}+} 
	= \left.
		\left( \frac{d}{dz}-ik_0 \right)  T e^{ik_0z}  
	\right|_{z=Z_{\max}+} 
	= 0.
\end{equation*}
Hence, continuity of $E$ and $\frac{dE}{dz}$  at $z=Z_{\max}$ yields the following boundary condition:
\begin{subequations}
    \label{eq:CM_TWBCs_1}
	\begin{equation}
    \label{eq:CM_TWBCs_1_r}
		\left.
			\left( \frac{d}{dz}-ik_0 \right) E 
		\right|_{z=Z_{\max}} 
		= 0.
	\end{equation}
Similarly,  at $z=0-$ we can write, see (\ref{eq:inc_ref_field}):
	\begin{equation*}
		\left.
			\left( \frac{d}{dz}+ik_0 \right) E 
		\right|_{z=0-} 
		= \left.
			\left( \frac{d}{dz}+ik_0 \right) 
			\left( \Einc e^{ik_0z}+Re^{-ik_0z} \right) 
		\right|_{z=0-} 
		= 2ik_0 \Einc.
	\end{equation*} 
Hence, the continuity of $E$ and $\frac{dE}{dz}$ at $z=0$ leads to the boundary condition:
	\begin{equation}
    \label{eq:CM_TWBCs_1_l}
		\left.
			\left( \frac{d}{dz}+ik_0 \right) E 
		\right|_{z=0} 
		= 2ik_0 \Einc.
	\end{equation} 
\end{subequations}
The boundary conditions~(\ref{eq:CM_TWBCs_1_r}) and~(\ref{eq:CM_TWBCs_1_l}) enable the propagation of outgoing waves from inside the interval $[0,Z_{\max}]$  toward its exterior.
In addition, the boundary condition~(\ref{eq:CM_TWBCs_1_l}) prescribes the given incoming wave $\Einc e^{ik_0z}$ at the left boundary $z=0$, and is therefore referred to as the {\em two-way} boundary condition.

The problem~(\ref{eq:1DNLH-1}),~(\ref{eq:CM_TWBCs_1}) can be rescaled as 
follows: \[
	\tilde{E} = E/\Einc, \qquad \tilde{\epsilon} = \epsilon|\Einc|^2.
\]
Hence, we can assume hereafter with no loss of generality that
\begin{equation} 
	\Einc=1. \label{eq:Einc_unity}
\end{equation}
Under this rescaling, a variation in  $\epsilon$ represents a variation in the input beam power~$|\Einc|^2$.

\begin{figure}[t]
        \noindent
        \begin{centering}
                \includegraphics[clip,scale=0.25]{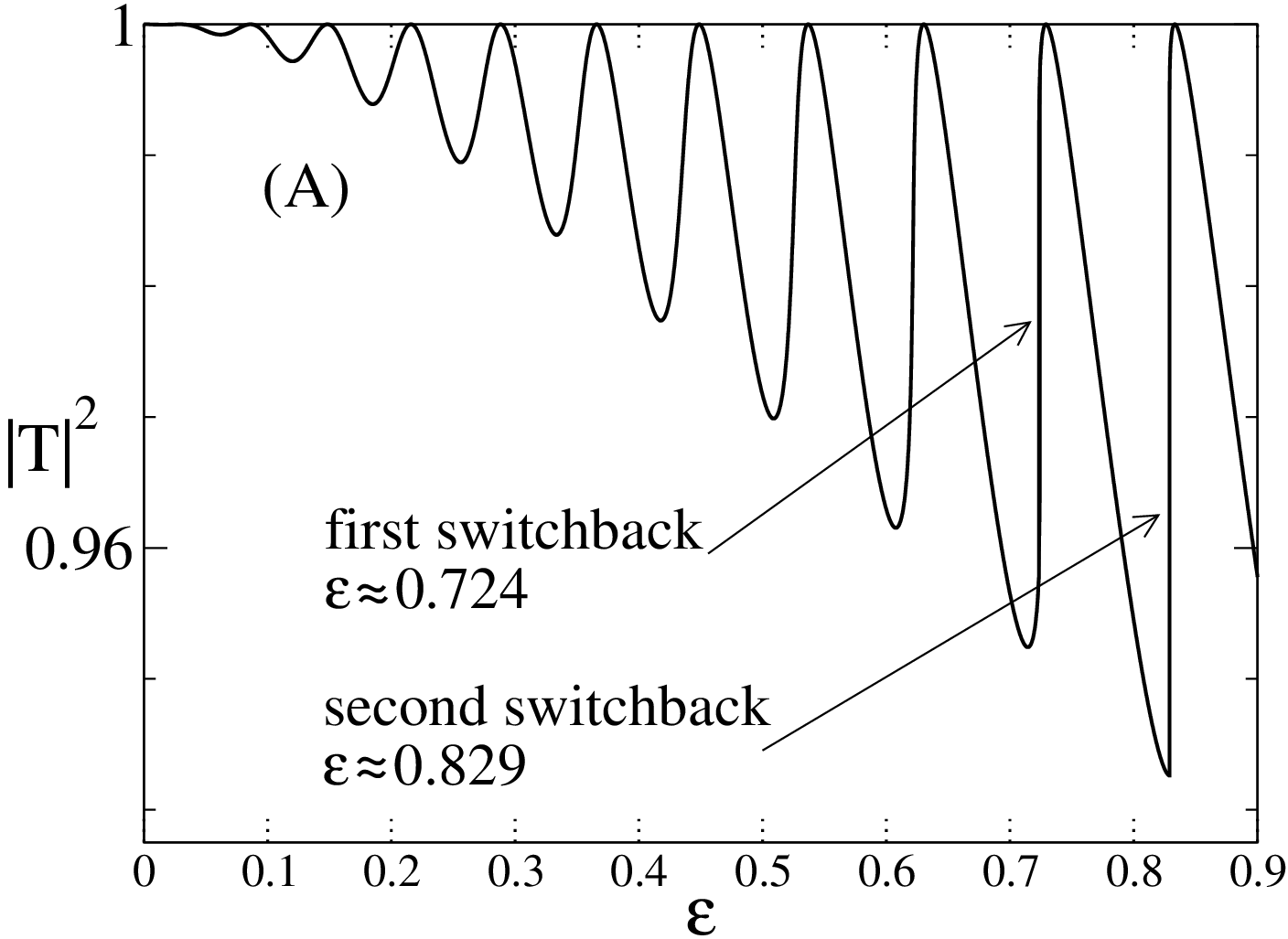}
                \includegraphics[clip,scale=0.25]{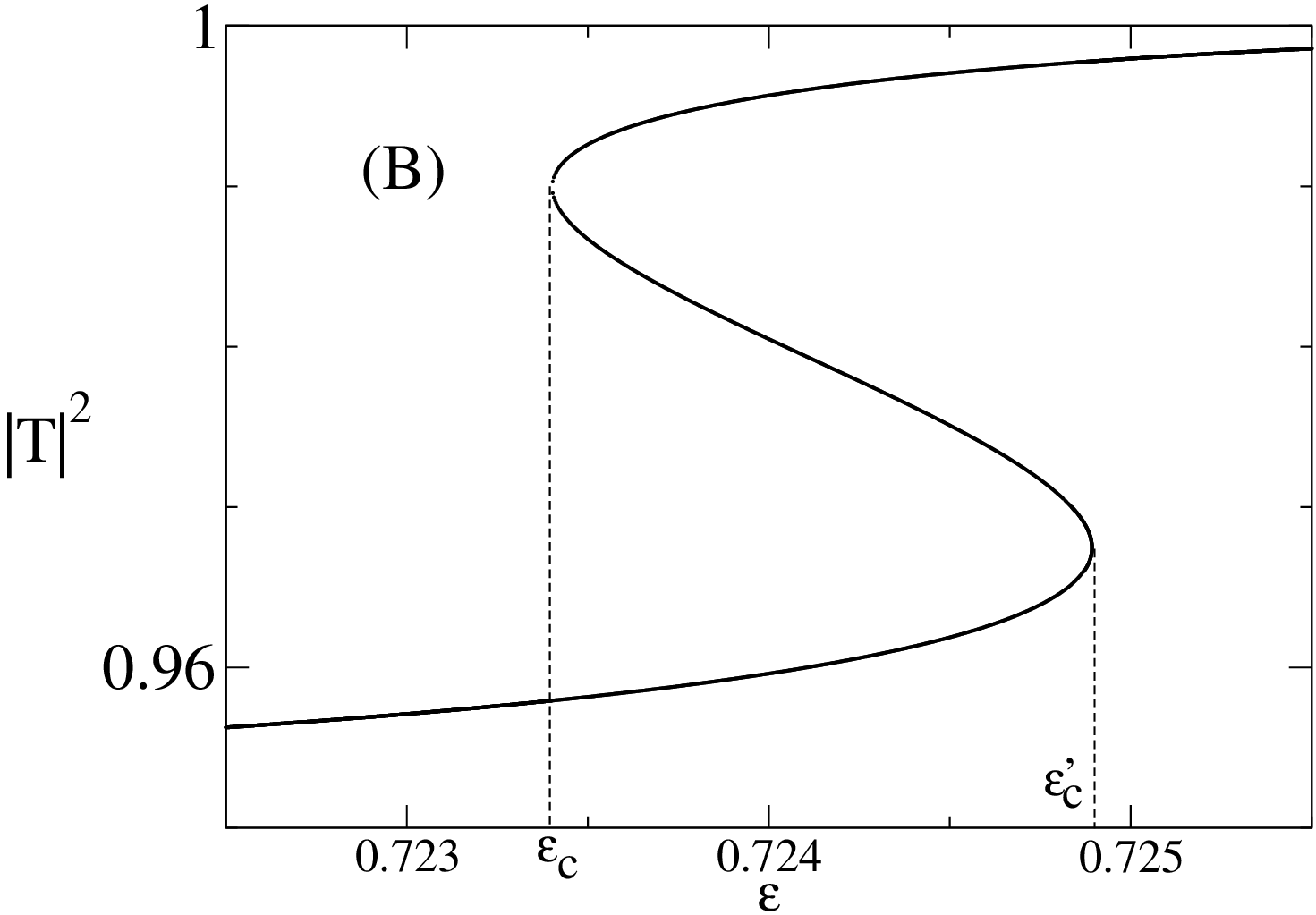}
                \caption{
                        \label{fig:transmittance}
                        (A) The transmittance $\left|T\right|^{2}$ as a function
                        of $\epsilon$ for the solution of the one-dimensional NLH~(\ref{eq:1DNLH-1})
                        with $\lri\equiv1$, $k_0=8$ and $Z_{\max}=10$.
                        (B) Zoom-in on the first region of switchback-type 
						nonuniqueness for $
                                0.7234 \approx \epsilon_{c} \leq \epsilon
                                \leq \epsilon_{c}' \approx 0.7249
                        $.
                }
        \end{centering}
\end{figure}

Closed form solutions for equation~(\ref{eq:1DNLH-1}) in a homogeneous medium (\ref{eq:homogeneous-material}) were first obtained by Wilhelm \cite{Wilhelm} for a real-valued field, and by Marburger and Felber \cite{Marburger-Felber:75} for a complex-valued field.
These solutions were later used by Chen and Mills~\cite{chen-mills-PRB-87} to solve equation~(\ref{eq:1DNLH-1}) with the boundary conditions~(\ref{eq:CM_TWBCs_1}), as follows.
Since the NLH~(\ref{eq:1DNLH-1}) is a second order ODE, the boundary condition~(\ref{eq:CM_TWBCs_1_r}) at $z=Z_{\max}$, together with a choice of the transmitted field amplitude $T$, 
constitute an initial value problem at $z=Z_{\max}$ that has a unique solution~$E=E(z;\,T,\epsilon)$. \footnote{ 
	Note that as $|E(Z_{\max})|=|T|$, a choice of $T$ is equivalent to a choice of~$E$ at~$\left.z=Z_{\max}\right.$.
}
For an arbitrary value of $T$, the solution ${E(z;T)}$ does not, generally speaking, satisfy the boundary condition~(\ref{eq:CM_TWBCs_1_l}) at $z=0$.
One can therefore use a shooting approach to find the value(s) of $T=T(\epsilon)$ for which the solutions of the initial value problem also satisfy~(\ref{eq:CM_TWBCs_1_l}) [and hence the full problem (\ref{eq:1DNLH-1}),~(\ref{eq:CM_TWBCs_1}),~(\ref{eq:Einc_unity})].
When the nonlinearity $\epsilon$ is small, the function $T=T(\epsilon)$ is single-valued, see Figure~\ref{fig:transmittance}(A).
When the nonlinearity exceeds a certain threshold $\epsilon>\epsilon_{c}$, the function $T=T(\epsilon)$ becomes multi-valued, which implies nonuniqueness of the solution. 
The nonuniqueness occurs at certain intervals of $\epsilon$ and  is of a switchback type, see Figure~\ref{fig:transmittance}(B). 
In the physics literature, this behavior is often referred to as bistability.

In a subsequent paper~\cite{chen-mills-PRB-87-ML}, Chen and Mills extended their approach to the case of piecewise-constant material coefficients (\ref{eqs:grated_material}), which corresponds to the formulation that we analyze numerically in this paper, see Section~\ref{sub:formulation}.
Knapp, Papanicolaou and White~\cite{Knaap-Papanicolau-89} considered the case of a large homogeneous slab and a weak nonlinearity. 
They showed that the threshold for nonuniqueness $\epscrit$ scales as $Z_{\max}^{-3}$. 
They also treated random media, which we do not consider here.

In addition to analytical studies, equation~(\ref{eq:1DNLH-1}) was also studied numerically
using a shooting approach~\cite{baghdasaryan-99,midrio-01,kwan-lu-04,petracek-06}
which is conceptually similar to the one of 
Chen and Mills \cite{chen-mills-PRB-87,chen-mills-PRB-87-ML}.
Unlike~\cite{chen-mills-PRB-87,chen-mills-PRB-87-ML},
however, in these studies, for each value of~$T$ at $Z_{\rm max}$    
the Cauchy problem is solved numerically, rather than analytically. 
The advantage of this approach  over~\cite{chen-mills-PRB-87,chen-mills-PRB-87-ML}
is that it can be applied to media
with a smooth variation of material 
properties~\cite{baghdasaryan-99,midrio-01,kwan-lu-04,petracek-06}
and to lossy materials~\cite{baghdasaryan-99}, 
as opposed to only piecewise-constant media in~\cite{chen-mills-PRB-87,chen-mills-PRB-87-ML}.
The main shortcoming of the shooting approach, however, 
is that it cannot be generalized to
multidimensional problems.

The NLH can also be solved numerically as a full boundary value problem.
In our previous work~\cite{FT:01,FT:04,BFT:05}, we solved the multidimensional NLH~(\ref{eq:NLH}) for the homogeneous Kerr medium with $\lri\equiv1$ and  $\epsilon\equiv const$.\footnote{Note 
that the $\lri\equiv1$ corresponds to the case for which the linear index of refraction $n_0(\bvec{x})$ is the same both inside and outside the Kerr medium.
}
To do that, we developed and implemented nonlocal two-way 
boundary conditions similar to~(\ref{eq:CM_TWBCs_1});
they provided a key element of the numerical methodology.
In~\cite{Suryanto:2002,Suryanto:2003}, Suryanto \etal\ used a finite element scheme for solving the one-dimensional NLH~(\ref{eq:1DNLH-1}) subject to the two-way boundary conditions.
The finite element approximation constructed in~\cite{Suryanto:2002,Suryanto:2003} allowed for material discontinuities at the grid nodes. 
This approximation was of a mixed order; the linear terms of~(\ref{eq:1DNLH-1})  were approximated with 
fourth order accuracy, while the nonlinearity was approximated with second order accuracy.

Let us emphasize that at the points where the material coefficients $\lri$ and/or $\epsilon$ 
are discontinuous, the second derivative of the solution $E(z)$ is
discontinuous. 
The presence of discontinuities in the solution must be properly accounted for when building a numerical approximation of equation~(\ref{eq:1DNLH-1}). 
In particular, a naive high-order approximation may lose its accuracy as the grid is refined.
In this context, we note that {\em the coefficient $\epsilon$ is always discontinuous at least at $z=0$ and $z=Z_{\max}$}.
Such a discontinuity cannot be addressed by a scheme that assumes smoothness across the boundary, such as the standard (five-point) fourth order central-difference scheme used in our previous work~\cite{FT:01,FT:04,BFT:05}.
Indeed, we have observed in~\cite{BFT:05} a deterioration of the fourth order accuracy at fine grid resolutions.

In the current paper, we present a novel fourth order numerical scheme for the NLH~(\ref{eq:1DNLH-1}) based on a {\em compact approximation of finite volume type}.
The use of integration over the grid cells allows us to correctly account for the discontinuities in $\lri(z)$ and $\epsilon(z)$ both at the outer boundaries and inside the Kerr medium.
The fourth order accuracy is attained on a compact three node stencil by using the differential equation~(\ref{eq:1DNLH-1}) to eliminate the leading terms of the truncation error. 
A similar equation-based approach was used by Singer and Turkel in~\cite{singer-turkel-98} to obtain a compact high order approximation for the linear Helmholtz equation.
As we shall see, however, construction of  a compact approximation for finite volumes, and especially in the nonlinear case is considerably more complex.
In particular, 
we need to use Birkhoff-Hermite interpolation to approximate the field between the grid nodes with fourth order accuracy.
To the best of our knowledge, this is the first time ever that a genuine fourth order scheme is built for the NLH with discontinuous coefficients.

While we analyze the formal accuracy of our schemes, a theoretical error estimate is beyond the scope of this paper, because the problem is nonlinear.
Instead, we evaluate the numerical error experimentally, and demonstrate that the schemes possess the anticipated rate of convergence.
Moreover, in Appendix~\ref{app:linear_interface} we provide a convergence proof for a linear problem with a material discontinuity, in which the material coefficient $\lri$ is in the form of a step function.
In this case, we can obtain closed form solutions for both the continuous equation and its discrete counterpart, and use them to establish the error estimates.
Note that this simple setup captures the key features of our treatment of material discontinuities by finite volumes, and illustrates that the scheme indeed has the design rate of grid convergence.

The second key improvement offered by the current paper is in the methodology used to solve the nonlinear equations on the grid.
Previously~\cite{FT:01,FT:04,BFT:05}, we solved the NLH by simple iterations based on freezing the nonlinearity; a similar approach was also employed by Suryanto \etal\ in~\cite{Suryanto:2002,Suryanto:2003}. 
While this approach has allowed us to obtain a number of interesting solutions to the NLH with a weak nonlinearity, for somewhat stronger nonlinearities the iterations would cease to converge \cite{FT:01,FT:04,BFT:05, Suryanto:2002,Suryanto:2003}.
In order to overcome this limitation, in this paper we solve the NLH~(\ref{eq:1DNLH-1}) using Newton's iterations. 
Applying Newton's method to the NLH is not straightforward though, since the nonlinearity in~(\ref{eq:1DNLH-1}) is nondifferentiable in the sense of Frech\'et.
We recall that the solutions of the NLH~(\ref{eq:1DNLH-1}) must be complex valued, otherwise it is impossible to adequately describe traveling waves in the time-harmonic context.\footnote{This 
is reflected by the fact that the boundary conditions (\ref{eq:CM_TWBCs_1}) are complex.}
Hence, to obtain a proper Newton's linearization we recast the complex equation~(\ref{eq:1DNLH-1}) as a system of two real equations.
In the literature, Newton's method has been applied to similar problems. For example, in
the work of G\'omez-Garde\~nes, et~al.~\cite{gomez-04}, the authors solve the steady-state
nonlinear Schr\"odinger
equation on a lattice by Newton's method (see also \cite{aubry-98,aubry-97,aubry-96,aubry-94}).
Our particular
implementation of Newton's method for the NLH
leads to a block tridiagonal structure of the Jacobians, which enables an efficient
inversion.
We also note that the application of Newton's method to a higher order discretization of the the NLH with material discontinuities brings along additional complications (Section~\ref{sec:Newtons}).

Our computations show that the use of Newton's iterations leads to a very considerable improvement in performance over the previous "frozen-nonlinearity" iterative methods~\cite{FT:01,FT:04,BFT:05,Suryanto:2002,Suryanto:2003}, as it enables robust numerical solution of the NLH for strong nonlinearities.
In fact, solutions can be computed for nonlinearities far above the threshold of nonuniqueness, and even for  the nonlinearities that lead to material breakdown in an actual physical setting.
Note that in the latter case, the Kerr model itself becomes inapplicable.

The paper is organized as follows: 
In Section~\ref{sub:formulation} we present a summary of the mathematical formulation.
In Section~\ref{sec:discrete-approx}, we describe our discrete approximation.
We begin with the finite volume formulation (Section~\ref{sub:FV_formulation}), then introduce two second order approximations (Section~\ref{sub:SO_approx} and Section~\ref{sub:alt_SO_approx}) and the fourth order approximation (Section~\ref{sub:FO_approx}), and finally construct the boundary conditions in the discrete setting (Section~\ref{sub:BCs}).
In Section~\ref{sec:Newtons}, we build a Newton's solver for the  Frech\'et nondifferentiable NLH.
To clarify the presentation, we first illustrate the approach for a single variable (Section~\ref{sub:newtons-1D}), then generalize to multivariable nondifferentiable functions (Section~\ref{sub:newtons-MD}), apply the method to the three discrete approximations of the NLH (Section~\ref{sub:stencil-variation}), and finally discuss the choice of the initial guess (Section~\ref{sub:choice}).
A summary of the numerical method is given in Section~\ref{sec:method-summary}.
Numerical computations are performed in Section~\ref{sec:results}, examining the convergence of the iterations and the computational error of the methods (Section~\ref{sub:Newtons-convergence} and Section~\ref{sub:accuracy}, respectively).
We conclude with a discussion in Section~\ref{sec:discussion}.

\subsection{\label{sub:formulation}Summary of the formulation}

In the current paper, we will be solving the one-dimensional NLH [cf.\ (\ref{eq:1DNLH-1})]:
\begin{subequations}
	\label{eqs:formulation}
	\begin{equation}
	    \frac{d^2E(z)}{dz^2}
	    + k_0^2\left(\lri(z) + \epsilon(z)\left|E\right|^2\right) E = 0,
		\qquad 0 < z < Z_{\max},
	    \label{eq:1DNLH}
	\end{equation}
	subject to the boundary conditions [cf.\ (\ref{eq:CM_TWBCs_1_r}), (\ref{eq:CM_TWBCs_1_l})]: 	
	\begin{equation}
	    \label{eq:CM_TWBCs}
		\left.
			\left( \frac{d}{dz}+ik_0 \right) E 
		\right|_{z=0} 
			= 2ik_0,
		\qquad
		\left.
			\left( \frac{d}{dz}-ik_0 \right) E 
		\right|_{z=Z_{\max}} = 0.
	\end{equation} 
	In formulae~(\ref{eq:1DNLH}),~(\ref{eq:CM_TWBCs}), we assume the
	scaling $\Einc=1$, see~(\ref{eq:Einc_unity}).
	The medium on the interval $[0,Z_{\max}]$ can have piecewise-constant material coefficients:
	\begin{equation}
	\label{eq:intervals}
	    \lri(z)\equiv\tilde{\lri}_l, \quad
	    \epsilon(z)\equiv\tilde{\epsilon}_l, 
		\quad \text{for} \quad
	    z \in \left( \tilde{z}_l,\tilde{z}_{l+1} \right).
	\end{equation}
	For simplicity only, we assume a uniform partition into $L-1$ 
	homogeneous slabs of equal width
	$
		\Delta z = \frac{Z_{\max}}{L-1}
	$: 
	\begin{equation}
		\label{eq:partition}
		\tilde{z}_l=(l-1)\Delta z, \quad 
		l=1,\dots,L.
	\end{equation}
\end{subequations}
	
The homogeneous case (\ref{eq:homogeneous-material}) corresponds to the case $L=2$.
At the interfaces $\tilde{z}_l$, the solution $E(z)$ and its first derivative $\frac{dE}{dz}$ are continuous, but the second derivative $\frac{d^2E}{dz^2}$ is discontinuous. 
Away from the interfaces, i.e., inside every interval~(\ref{eq:intervals}), the material coefficients $\lri(z)$ and $\epsilon(z)$ are constant, and the NLH~(\ref{eq:1DNLH}) implies that the field $E(z)$ is infinitely differentiable.

\section{\label{sec:discrete-approx}Discrete approximation}

In this section, we present our discretization of problem~(\ref{eqs:formulation}).
First, we introduce an integral formulation of the NLH~(\ref{eq:1DNLH}) (see
Section~\ref{sub:FV_formulation})
and discretize it on the grid (Section~\ref{sub:SO_approx}, \ref{sub:alt_SO_approx},
and \ref{sub:FO_approx}).
Then, we implement the boundary conditions~(\ref{eq:CM_TWBCs}) in a fully discrete framework (Section~\ref{sub:BCs}).

\subsection{\label{sub:FV_formulation} Integral formulation}

Let $a,\,b\in[0,Z_{\max}]$, $a<b$, and let us integrate equation~(\ref{eq:1DNLH})
between the points $a$ and $b$ with respect to $z$. 
Since $\frac{dE}{dz}$ is continuous everywhere, we obtain:
\begin{equation}
\label{eq:claw}
\frac{dE(b)}{dz}- \frac{dE(a)}{dz} + k_0^2 \int_a^b \left(
            \lri(z)+\epsilon(z)\left|E\right|^2
        \right) E \, dz =0.
\end{equation}
Equation~(\ref{eq:claw}) can be interpreted as the integral
conservation law that corresponds to the NLH~(\ref{eq:1DNLH}). 
It is easy to see that for sufficiently smooth solutions the
two formulations are equivalent. 
Indeed, if we require that the integral relation~(\ref{eq:claw}) hold {\em for any pair of points} $a$ and $b$, then at every point $z_0$ where $\frac{d^2E}{dz^2}$ exists the NLH~(\ref{eq:1DNLH}) can be reconstructed from the conservation law~(\ref{eq:claw}) by a straightforward passage to the limit: $a\to z_0-0$, $b\to z_0+0$.
However, the integral formulation~(\ref{eq:claw}) makes sense even when the  differential equation per se loses its validity because of insufficient regularity of the solution, i.e., when the material coefficients undergo jump discontinuities and the second derivative $\frac{d^2E}{dz^2}$ becomes discontinuous.

Let us introduce a uniform grid of $M$ nodes on the interval
$0\leq z\leq Z_{\max}$:
\begin{subequations}
	\begin{equation}
		\label{eq:grid}
    	z_m=(m-1)h,\quad \text{ where } \quad 
    	h=\frac{Z_{\max}}{M-1},\quad 
    	m=1,\dots,M.
	\end{equation}
	We choose $h$ so that $\Delta z$ of~(\ref{eq:partition}) 
	is an integer multiple of $h$.
	This choice guarantees that material discontinuities will only 
	be located at the grid nodes,
	i.e., that both $\lri(z)$ and $\epsilon(z)$ will be constant within 
	each grid cell:
	\begin{equation}
		\label{eq:piecewise-const-coefs}
		\lri(z) \equiv \lri_m,\quad  
		\epsilon(z) \equiv \epsilon_m,\qquad  
		z\in\left(z_m,z_{m+1}\right).
	\end{equation}
\end{subequations}

To approximate the NLH on the grid~(\ref{eq:grid}), we apply
the integral relation~(\ref{eq:claw}) between the midpoints of every 
two neighboring cells, i.e., for $
	[a,b]=[z_\nmh,z_\nph]$, $m=1,2,\ldots,M$.
Then, using formula~(\ref{eq:piecewise-const-coefs}), we arrive at
\begin{equation}
    \label{eq:VI-integrals}
\begin{aligned}
      \left.
        \frac{dE}{dz}
    \right|_{z_\nmh}^{z_\nph}             
        + & \> k_0^2\lri_{m-1}                   
            \int_{z_\nmh}^{z_m} E\,dz
        +   k_0^2\epsilon_{m-1}                
            \int_{z_\nmh}^{z_m}
                \left|E\right|^2 E\,dz\\
        + & \> k_0^2\lri_m                       
            \int_{z_m}^{z_\nph}
                E\,dz
        +   k_0^2\epsilon_m                    
            \int_{z_m}^{z_\nph}
                \left|E\right|^2 E\,dz = 0.
\end{aligned}
\end{equation}
Equation~(\ref{eq:VI-integrals}) relates integrals of the unknown 
continuous function~$E(z)$ with its derivatives at~$z=z_{m\pm\frac{1}{2}}$.
We will approximate the individual terms in (\ref{eq:VI-integrals})
using the nodal values $E(z_m)\equiv E_m$, $m=1,\ldots,M$, of the field.
The resulting scheme will be equivalent to compact finite differences on the regions 
of smoothness of the solution, where it could also be obtained without
using the integral formulation (see Section~\ref{sub:disc}
for further discussion of an approach alternative 
to the use of integral formulation).
Otherwise, i.e., near the discontinuities, 
the scheme will approximate the integral relation~(\ref{eq:claw}), and hence~(\ref{eq:VI-integrals}), 
rather than the differential equation~(\ref{eq:1DNLH}). 

Recall that the material coefficients $\lri(z)$ and $\epsilon(z)$ 
are constant in between the grid nodes and consequently,
$E(z)$ is infinitely differentiable within each grid cell.
Hence, all the integrands in~(\ref{eq:VI-integrals}) can be approximated
with fourth order accuracy using cubic polynomials. 
Together with a fourth order approximation of the derivatives, this yields a fourth order compact scheme for the NLH~(\ref{eq:1DNLH}), see Section~\ref{sub:FO_approx}.
An even simpler piecewise linear approximation of $E(z)$ yields a second order compact scheme, 
and we will describe its two different versions, 
in Sections~\ref{sub:SO_approx} and \ref{sub:alt_SO_approx}.
In addition to providing a reference point for comparison, the second order schemes allow us to introduce the general framework and notations exploited later for building the more complex fourth order method.

\subsection{\label{sub:SO_approx}Second order approximation}

We approximate the first term on the left-hand side of~(\ref{eq:VI-integrals})
using central differences:
\begin{equation}
\label{eq:SO_undivided}
\left.
        \frac{dE}{dz}
    \right|_{z_\nmh}^{z_\nph} = 
            \frac{ E_{m+1}-E_m }{ h }
            -
            \frac{ E_m - E_{m-1} }{ h }
        + \oh{2} .
\end{equation}
Without assuming any additional regularity of $E(z)$ beyond the continuity 
of its first derivative, we merely have the difference of two fluxes approximated with second order accuracy.\footnote{The 
flux difference on the right-hand side of~(\ref{eq:SO_undivided}) is exactly the same as we would have obtained if we approximated the second derivative $\frac{d^2E}{dz^2}$ by the standard piecewise linear Galerkin finite elements, see, e.g.,~\cite{ciarlet-02}; having a continuous first derivative of $E(z)$ is sufficient for building this approximation.
}
If, however, the material coefficients are continuous at~$z_m$, i.e., 
if $\lri_m=\lri_{m-1}$ and $\epsilon_m=\epsilon_{m-1}$, then
$\frac{d^2E}{dz^2}$ and higher derivatives exist and are continuous as well.
In this case, if we divide the  undivided second difference on the right-hand side 
of~(\ref{eq:SO_undivided})  by $h$, then a straightforward Taylor-based argument will 
yield a second order central-difference approximation of $\frac{d^2E}{dz^2}$:
\begin{equation}
\label{eq:central}
	 \frac{ E_{m+1}-2E_m + E_{m-1}}{ h^2 } 
	  = 
	\left.
        \frac{d^2E}{dz^2}
    \right|_{z_m}+\oh{2}.
\end{equation}

To approximate the third integral on the left-hand side of~(\ref{eq:VI-integrals}), we linearly interpolate $E(z)$ on the interval $[z_m,z_\nph]$:
\begin{equation}
    \label{eq:SO_linint_E}
    E(z) \equiv
        E \left( z_m+h\zeta \right)
    =
        \left( 1-\zeta \right) E_m
        + \zeta E_{m+1} + \oh{2},
    \quad 
        \zeta \in
            \left[ 0, \frac{1}{2} \right].
\end{equation}
Then, substituting expression~(\ref{eq:SO_linint_E}) into the third integral
of~(\ref{eq:VI-integrals}), we have:
\begin{equation}
\label{eq:SO_lin_int}
\begin{aligned}
     \int_{z_m}^{z_\nph}E\,dz 
		=  &\>
        h\int_0^{1/2}
            \left[
                (1-\zeta)E_m
                + \zeta E_{m+1}
            \right]
            \,d\zeta + {\mathcal O}(h^3)\\
        = &\>
            \frac{3h}{8} E_m
            +\frac{h}{8}E_{m+1} + {\mathcal O}(h^3).
\end{aligned}
\end{equation}
Likewise, we can linearly interpolate the cubic term 
$\left|E\right|^2E$ on $[z_m,z_\nph]$ to obtain:
\begin{equation*}
    \int_{z_m}^{z_\nph}
        \left|E\right|^2
        E\,dz
    = 
        \frac{3h}{8} \left|E_m\right|^2 E_m
        +\frac{h}{8} \left|E_{m+1}\right|^2 E_{m+1}
		+ {\mathcal O}(h^3).
\end{equation*}
The expressions for the subinterval $[ z_\nmh, z_m ]$ are derived similarly, we merely  replace 
$
    \lri_m,\>
    \epsilon_m$, and
    $E_{m+1}
$ with $
    \lri_{m-1},\>
    \epsilon_{m-1}$, and
    $E_{m-1}$, respectively.
Finally, by assembling all the terms  we arrive at the following second order approximation of the integral relation (\ref{eq:VI-integrals}) for $m=1,2,\ldots,M$:
\begin{subequations}
\begin{equation}
 \label{eq:SO_stencil}
\begin{aligned}
    ~ h\DSm \stackrel{\rm def}{=}
         & \frac{E_{m+1} - E_m}{h} -\frac{E_m- E_{m-1}}{h}\\
        & \> + hk_0^2 \lri_{m-1} 
		\frac{
            E_{m-1} + 3 E_m
        }{
            8
        } 
        + hk_0^2 \lri_m 
		\frac{
			3 E_m + E_{m+1}
        }{
            8
        } \\
       & \> +
        hk_0^2 \epsilon_{m-1} 
		\frac{
            \left| E_{m-1} \right|^2 E_{m-1}
            + 3 \left|E_m\right|^2E_m
        }{
            8
        }\\
		& \> + hk_0^2 \epsilon_m 
		\frac{
            3 \left|E_m\right|^2E_m
            + \left| E_{m+1} \right|^2 E_{m+1}
        }{
            8
        }=0. 
\end{aligned}
\end{equation}
The vector $\bvec{E}=\left[ E_1,\dots,E_M\right]^T$ was used as an argument of $\DSm$ in formula~(\ref{eq:SO_stencil}), because $F_m$ operates on $E_{m-1}$, $E_m$, and $E_{m+1}$.
Hence, for the interface nodes $m=1$ and $m=M$, the system of equations~(\ref{eq:SO_stencil}) requires the addition of the ghost nodes $m=0$ and $m=M+1$, respectively.
The value of the field at the ghost nodes will be determined by the boundary conditions, see Section~\ref{sub:BCs}.
Note also that the notation $E_m$ needs to be interpreted differently in different expressions. 
Namely, in~(\ref{eq:SO_undivided}),~(\ref{eq:SO_linint_E}),~(\ref{eq:SO_lin_int}) and similar formulae that introduce approximation of the individual terms in~(\ref{eq:VI-integrals}), $E_m$ denotes the value of the exact continuous solution of~(\ref{eqs:formulation}) on the grid~(\ref{eq:grid}). 
In formula~(\ref{eq:SO_stencil}), however, $E_m$ denotes the approximate discrete solution, which we calculate numerically.

If the material coefficients $\lri$ and $\epsilon$ are continuous at $z_m$, i.e., if $\left.\lri_{m-1}=\lri_m\right.$ and $\left.\epsilon_{m-1}=\epsilon_m\right.$, then $\frac{d^2E}{dz^2}$ exists at this point along with higher order derivatives.
In that case, equation (\ref{eq:SO_stencil}) reduces to 
%
\begin{equation}
    \label{eq:SO_stencil2}
    \begin{aligned}
    \DSm  = & \>
		\frac{E_{m+1} - 2E_m+E_{m-1}}{h^2}
        + k_0^2\lri_m\frac{
            E_{m-1}  + 6 E_m + E_{m+1}
        }{
            8
        } \\
   & \> + k_0^2\epsilon_m \frac{
            \left| E_{m-1} \right|^2 E_{m-1}
            + 6 \left|E_m\right|^2E_m
            + \left| E_{m+1} \right|^2 E_{m+1}
        }{
            8
        }= 0. 
        \end{aligned}
\end{equation}
\end{subequations}
In scheme (\ref{eq:SO_stencil2}), the second derivative $\frac{d^2E}{dz^2}$ 
is approximated by the conventional second order central differences (\ref{eq:central}), but the 
non-differentiated terms are evaluated as weighted sums over three neighboring nodes
rather than pointwise.

\subsection{\label{sub:alt_SO_approx}Alternative second order approximation}

Instead of interpolating the cubic term $\left|E\right|^2E$ as in Section~\ref{sub:SO_approx}, one can substitute the linear interpolation~(\ref{eq:SO_linint_E}) into the corresponding integrals of~(\ref{eq:VI-integrals}). 
This approach is slightly more cumbersome. 
As we will see in Section~\ref{sub:FO_approx}, however, it will enable the construction of the fourth order compact discretization.

It is convenient to adopt a tensor notation. First, we recast formula~(\ref{eq:SO_linint_E}) as
\begin{equation*}
    E\left( z_m + h\zeta \right)
    =
        \sum_{i=0}^1 F_i(\zeta) E_{m+i} + \oh{2},
    \quad  \text{where} \quad
        F_0 = 1 - \zeta ,
    \quad 
        F_1 = \zeta.
\end{equation*}
This representation, when substituted into the linear integral term of~(\ref{eq:VI-integrals}), provides an equivalent alternative form of equation~(\ref{eq:SO_lin_int}):
\begin{equation*}
     \int_{z_m}^{z_\nph}E\,dz
    = 
        h\sum_{i=0}^1
            \underbrace{ \left(
                    \int_0^{\frac{1}{2}}
                        F_i
                    \,d\zeta
            \right) }_{f_i}
            E_{m+i} +\oh{3}
        =
        h\sum_{i=0}^{1}
            f_i E_{m+i} + \oh{3},
\end{equation*}
while its substitution into the cubic term $|E|^2E=E^*E^2$ yields:
\begin{eqnarray*}
    \lefteqn{  
		\int_{z_m}^{z_\nph} 
			\left| E \right|^2 E\,dz
	} \\
    & & \quad = 
     h\int_{0}^{1/2}
        \left(
            \sum_{i=0}^1    F_i(\zeta) E_{m+i}^*
        \right)
        \left(
            \sum_{j=0}^1    F_j(\zeta) E_{m+j}
        \right)
		\left(
            \sum_{k=0}^1    F_k(\zeta) E_{m+k}
        \right)
    d\zeta +\oh{3} \\
    & & \quad =
    h\sum_{i,j,k=0}^1
        \underbrace{ \left(
            \int_0^{\frac{1}{2}}
                F_i F_j F_k
            \,d\zeta
        \right) }_{ g_{ijk} }
        E_{m+i}^*
        E_{m+j}
        E_{m+k} +\oh{3} \\
   & &\quad = h\sum_{i,j,k=0}^1
        g_{ijk}
            E_{m+i}^*
            E_{m+j}
            E_{m+k} +\oh{3}.
\end{eqnarray*}
The constants $f_i$ and $g_{ijk}$ in the previous formulae are defined as
\begin{equation*}
    f_i = \int_0^{\frac{1}{2}}
        F_i
    \,d\zeta,
    \qquad 
    g_{ijk} = \int_0^{\frac{1}{2}}
        F_i F_j F_k
    \,d\zeta,\qquad i,\,j,\,k=0,1.
\end{equation*}
Evaluation of these integrals yields:
\begin{align*}
	~& f_0  = \frac{3}{8}, 
	\quad  
    f_1 = \frac{1}{8}, \\
    ~ & g_{000} = \frac{15}{64},
    \quad 
    g_{001} = \frac{11}{192},
    \quad
    g_{011} = \frac{5}{192},
    \quad
    g_{111} = \frac{1}{64}.
\end{align*}
Note that the tensor elements $g_{ijk}$ are symmetric with respect to any permutation of the indices $i$, $j$, and $k$, e.g., $g_{011}=g_{101}=g_{110}$.

Altogether, the integrals over~$[z_m,z_{m+\frac{1}{2}}]$
in~(\ref{eq:VI-integrals}) are approximated as
\begin{align*}
     & \int_{z_m}^{z_\nph}
        \left(
            \lri_m E +
            \epsilon_m \left| E \right|^2 E
        \right)
    dz
     = \\ & \qquad  
    h\lri_m \sum_{i=0}^1
        f_i E_{m+i}
    + h\epsilon_m \sum_{i,j,k=0}^1
        g_{ijk}
            E_{m+i}^*
            E_{m+j}
            E_{m+k} + {\mathcal O}(h^3),
\end{align*}
and the integrals over $[z_{m-\frac{1}{2}},z_m]$ are approximated
the same way.
Hence, the alternative second order discretization of the integral relation
(\ref{eq:VI-integrals})
can be written as
\begin{subequations}
\begin{equation}
    \label{eq:alt_SO_stencil}
\begin{aligned}
     h\DSm  \stackrel{\rm def}{=} & \>
        \frac{
            E_{m+1} - E_m}{h} - \frac{E_m-E_{m-1}}{h} \\
        & + \> hk_0^2\lri_{m-1} \sum_{i=0}^1
            f_i E_{m-i}
        + hk_0^2\epsilon_{m-1} \sum_{i,j,k=0}^1
            g_{ijk}
                E_{m-i}^*
                E_{m-j}
                E_{m-k}  \\
         &+\>  hk_0^2\lri_m \sum_{i=0}^1
            f_i E_{m+i}
        + hk_0^2\epsilon_m \sum_{i,j,k=0}^1
            g_{ijk}
                E_{m+i}^*
                E_{m+j}
                E_{m+k} = 0,
\end{aligned}
\end{equation}
where $m=1,\dots,M$.
Similarly to~(\ref{eq:SO_stencil}), $E_m$ in formula~(\ref{eq:alt_SO_stencil}) should be interpreted as the approximate solution on the grid~(\ref{eq:grid}), and its values 
at the ghost nodes $m=0$ and $m=M+1$ are determined in Section~\ref{sub:BCs}.
Again, if $\lri$ and  $\epsilon$ are continuous and $E$ is smooth at $z_m$, then scheme~(\ref{eq:alt_SO_stencil}) reduces to a central-difference second order scheme for the NLH~(\ref{eq:1DNLH}):
\begin{equation}
\label{eq:alt_SO_stencil2}
\begin{aligned}
     \DSm  = & 
        \frac{
            E_{m+1} - 2E_m+E_{m-1}}{h^2} 
        +  k_0^2\lri_m  
			\sum_{i=0}^1 f_i ( E_{m-i} + E_{m+i} ) \\
        &\> + k_0^2\epsilon_m 
		\sum_{i,j,k=0}^1 g_{ijk} 
		\left(
                E_{m-i}^*	E_{m-j}	E_{m-k} 
                + E_{m+i}^*	E_{m+j}	E_{m+k} 
		\right) = 0. 
\end{aligned}
\end{equation}
\end{subequations}
Note that the linear terms in~(\ref{eq:alt_SO_stencil}) and~(\ref{eq:alt_SO_stencil2}) are identical to those in~(\ref{eq:SO_stencil}) and~(\ref{eq:SO_stencil2}), respectively,
they are merely expressed in a different form.

\subsection{\label{sub:FO_approx}Equation-based fourth order approximation}

In this section, we build a compact fourth order discretization for the integral
relation~(\ref{eq:VI-integrals}).
The general idea of all compact schemes is to use the original differential equation to obtain the higher order derivatives that could help cancel the leading terms of the truncation error and thus improve the order of accuracy. 
This idea has been implemented, e.g., by Singer and Turkel in~\cite{singer-turkel-98} for a finite-difference approximation of the linear Helmholtz equation.
Hereafter, we adopt some elements of their {\em equation-based} approach.
As we shall see though, some additional complications arise when this approach is applied to the approximation of the integral relation~(\ref{eq:VI-integrals}), which, in particular, involves nonlinearity.

The differential equation~(\ref{eq:1DNLH}) inside the grid cells can be used to evaluate the one-sided second derivatives
at the grid nodes as follows: 
\begin{subequations}
    \label{eq:1S_2nd_derivatives_1DNLH}
    \begin{align}
        \ddEp \stackrel{\rm def}{=}
            \left.
                \frac{d^2E}{dz^2}
            \right|_{ z=z_{m}+ }
        = &\>
            -k_0^2
            \left(
                \lri_m
                + \epsilon_m \left|E_m\right|^2
            \right)
            E_m,\\
        \ddEm \stackrel{\rm def}{=}
            \left.
                \frac{d^2E}{dz^2}
            \right|_{ z=z_{m+1}-}
        = & \>
            -k_0^2
            \left(
                \lri_m
                + \epsilon_m \left|E_{m+1} \right|^2
            \right)
            E_{m+1}.
    \end{align}
\end{subequations}
Subsequently, formulae~(\ref{eq:1S_2nd_derivatives_1DNLH}) will be used to approximate each of the five terms on the left-hand side of~(\ref{eq:VI-integrals}) with fourth order accuracy.

To approximate the fluxes
$E'_{m\pm\frac{1}{2}}$ in~(\ref{eq:VI-integrals}), 
we first use the Taylor expansion:
\begin{equation*}
    E'_\nph  = 
        \frac{ E_{m+1}-E_m }{ h }
        - \frac{h^2}{24} E_\nph^{(3)}
		+ \oh{4}.
\end{equation*}
Then, we approximate the third derivative $E_\nph^{(3)}$ with second order
accuracy and use~(\ref{eq:1S_2nd_derivatives_1DNLH}), which yields:
\begin{align*}
    E_\nph^{(3)}  = &\>
        \frac{ \ddEm-\ddEp }{h}
        + \oh{2} \\
     = & \>
        \frac{
            - k_0^2 \left(
                \lri_m
                + \epsilon_m \left|E_{m+1}\right|^2
            \right) E_{m+1}
            + k_0^2 \left(
                \lri_m
                + \epsilon_m \left|E_m\right|^2
            \right) E_m
        }{
            h
        } + \oh{2} .
\end{align*}
Finally, we introduce the dimensionless grid size \[\hk=k_0h\] and obtain:
\begin{align*}
    E'_\nph  =  &\>
        \frac{1}{h}
            \left(
                1+\frac{\hk^2}{24}
                \left(
                    \lri_m
                    + \epsilon_m \left|E_{m+1}\right|^2
                \right)
            \right) E_{m+1} \\
            & - \> 
			\frac{1}{h}\left(
                1+\frac{\hk^2}{24}
                \left(
                    \lri_m
                    + \epsilon_m \left|E_m\right|^2
                \right)
            \right) E_m
         + \oh{4}.
\end{align*}
We repeat the calculation for $E'_\nmh$. Altogether, the flux difference, i.e., the first term in~(\ref{eq:VI-integrals}), is approximated as
\begin{align*}
    \left.
        \frac{dE}{dz} 
    \right|_{z_\nmh}^{z_\nph}  
		= &\> 
		\frac{ E_{m+1}-E_m}{h}\left(1+\lri_m\frac{\hk^2}{24}\right)
        -\frac{E_{m}-E_{m-1}}{h}\left(1+\lri_{m-1}\frac{\hk^2}{24}\right)  \\
        & + \> 
        	\epsilon_m\frac{\hk^2}{24} 
				\frac{
					\left|E_{m+1}\right|^2 E_{m+1}- 
					\left|E_m\right|^2 E_m
				}{h} \\ 
		& - \> 
			\epsilon_{m-1}\frac{\hk^2}{24} 
				\frac{
					\left|E_{m}\right|^2 E_{m}-
                	\left|E_{m-1}\right|^2 E_{m-1}
				}{h} 
        + \oh{4}.
\end{align*}

Next, we approximate the four integral terms in~(\ref{eq:VI-integrals}).
To do that, we build 
fourth order polynomial approximations of the integrands.
The following lemma is instrumental for this purpose.

\begin{lem}
	\label{lem:cubic-interp}
	Let $E\in C^4([z_m,z_{m+1}])$. 
	Let the values $E_m=E(z_m)$ and~$E_{m+1}=E(z_{m+1})$ be known along with the values of the one-sided second derivatives~$\ddEp$ and~$\ddEm$.
	Then, the function $E(z)$ is approximated with fourth order accuracy:
	\begin{subequations}
		\label{eq:cubic_interp_E}
		\begin{equation}
		\label{eq:cubic_interp_E_a}
		    E( z_m + \zeta h ) = P_3(\zeta) + \oh{4},
			\qquad 
			z \in \left[z_m,z_{m+1}\right],
		\end{equation}
		by the Hermite-Birkhoff cubic polynomial:
		\begin{equation}
		\label{eq:cubic_interp_E_b}
			\begin{aligned}
		    	P_3(\zeta) = &\>
		    	\left(
		        	E_m -
		        	\frac{h^2}{6} \ddEp
			    \right) \left(
		    	    1-\zeta
		    	\right)
		    	+ \frac{h^2}{6} \ddEp
		        	\left( 1-\zeta \right)^3 \\
		        	& \> + \left(
		            	E_{m+1} -
		            	\frac{h^2}{6} \ddEm
		        	\right) \zeta
		        	+ \frac{h^2}{6} \ddEm \zeta^3.
				\end{aligned}
		\end{equation}
	\end{subequations}	
Moreover, given  $E_m$, $E_{m+1}$, $\ddEp$, and $\ddEm$, the polynomial~(\ref{eq:cubic_interp_E_b}) is unique.
\end{lem}
\begin{pf} See Appendix~\ref{app:interp_accuracy}. \end{pf}

Note that, in general, for the construction of~$P_3$ on a given individual interval $[z_m,z_{m+1}]$, it is unimportant that the derivatives in formula (\ref{eq:cubic_interp_E_b}) are one-sided. 
We only use one-sided derivatives in order to be able to use the result in the context of discrete approximation on the entire grid, when the material coefficients and hence second derivatives of the solution can undergo jumps at the grid nodes.
We also note that the cubic polynomials built in accordance with Lemma~\ref{lem:cubic-interp} are not 
equivalent to the standard cubic splines, see Section~\ref{sub:disc} for more detail.

Substituting expressions~(\ref{eq:1S_2nd_derivatives_1DNLH}) into formula~(\ref{eq:cubic_interp_E_b}), we obtain a fourth order approximation of $E(z)$ on $\left[z_m,z_{m+1}\right]$:
\begin{align*}
    E\left(z_m+\zeta h\right) 
	 = &\>
        \left(
            1+\frac{\hk^2}{6}
            \left(
                \lri_m + \epsilon_m \left|E_m\right|^2
            \right)
        \right) E_m
        \left( 1-\zeta \right)\\
         &\> - \frac{\hk^2}{6}
        \left(
            \lri_m + \epsilon_m \left|E_m\right|^2
        \right) E_m
        \left(1-\zeta\right)^{3}\\
         &\> + \left(
            1+\frac{\hk^2}{6}
            \left(
                \lri_m + \epsilon_m \left|E_{m+1}\right|^2
            \right)
        \right) E_{m+1} \zeta\\
         &\> - \frac{\hk^2}{6}
        \left(
            \lri_m + \epsilon_m \left|E_{m+1}\right|^2
        \right) E_{m+1} \zeta^3 
		+\oh{4}.
\end{align*}
For convenience, let us  rewrite the previous expression as
\begin{equation}
    \label{eq:FO_compact_E_zeta}
    E(z_m+\zeta h) =
    \sum_{i=0}^{3}  F_i(\zeta; \hk,\lri_m) v_i^+ +\oh{4},
\end{equation}
where 
\begin{eqnarray*}
     F_0(\zeta;\hk,\lri) = 
        (1-\zeta) \left(
            1+\lri \frac{\hk^2}{6}
                \left(
                    1-\left(1-\zeta\right)^2
                \right)
        \right),
	& \qquad & 
	F_2(\zeta;\hk,\lri) = 
        \zeta \left(
            1+\lri \frac{\hk^2}{6}
            \left( 1-\zeta^2 \right)
        \right),\\
    F_1(\zeta;\hk,\lri) = 
        \frac{\hk^2}{6}
        \left( 1-\zeta \right)
        \left(
            1-\left(1-\zeta\right)^2
        \right),
	& & 
	F_3(\zeta;\hk,\lri) = 
        \frac{\hk^2}{6} \zeta
        \left( 1-\zeta^2 \right),
\end{eqnarray*}
and
\begin{equation*}
    v_0^+ = E_m, \quad
    v_1^+ = \epsilon_m \left|E_m\right|^2 E_m, \quad 
    v_2^+ = E_{m+1},\quad
    v_3^+ = \epsilon_m \left|E_{m+1}\right|^2 E_{m+1}.
\end{equation*}
Then,
substituting  expression~(\ref{eq:FO_compact_E_zeta}) for $E(z)$ 
into the last two integral terms
of~(\ref{eq:VI-integrals}) and evaluating the integrals with respect to
 $\zeta$, we have: 
%
{\allowdisplaybreaks
    \begin{align*}
         \int_{z_m}^{z_\nph} E\,dz  = &\>
        h\sum_{i=0}^3
            \underbrace{ \left(
                \int_0^{\frac{1}{2}} F_i(\zeta; \lri_m, \hk) \,d\zeta
            \right) }_{ f_i }
            v_i^+ + \oh{5}
        = h\sum_{i=0}^3 f_i v_i^+ +{\mathcal O}(h^5),\\
         \int_{z_m}^{z_\nph}
            \left|E\right|^2E\,dz = &\>
        h\sum_{i,j,k=0}^3
            \underbrace{ \left(
                \int_0^{\frac{1}{2}} F_i F_j F_k \,d\zeta
            \right) }_{ g_{ijk} }
            (v_i^+)^*v_j^+v_k^+ +\oh{5} \\
        = &\> h\sum_{i,j,k=0}^3
            g_{ijk} \cdot (v_i^+)^* v_j^+ v_k^+ +\oh{5}.
    \end{align*}
    }
The constants $f_i$ and $g_{ijk}$ in the previous formulae are defined as
\begin{equation}
\begin{gathered}
	\label{eq:FO_tensor_elements}
    f_i(\lri,\hk) =
        \int_0^{\frac{1}{2}} F_i(\zeta; \lri,\hk) \,d\zeta
    , \quad 
    g_{ijk}(\lri, \hk) =
        \int_0^{\frac{1}{2}} F_i F_j F_k \,d\zeta,\\
    i,\>j,\>k=0,\dots,3,
\end{gathered}
\end{equation}
and their values are given in Table~\ref{tab:compact-FO-coefs}.\footnote{A direct 
computation of all the tensor elements in~(\ref{eq:FO_tensor_elements}) could be quite tedious and prone to errors. 
	This task, however, can be efficiently automated, see Appendix~\ref{app:SE}.
} 
As in the case of the second order scheme (Section~\ref{sub:alt_SO_approx}), it is clear from the definition of the tensor elements $g_{ijk}$, formula~(\ref{eq:FO_tensor_elements}), that they are symmetric with respect to any permutation of the indices $\lbrace i,\,j,\,k \rbrace$.

Evaluation of the integrals of~(\ref{eq:VI-integrals})
for the interval $[z_\nmh,z_m]$ is nearly identical; it only
requires replacing
$(\lri_m, \epsilon_m)$ with $(\lri_{m-1}, \epsilon_{m-1})$
and $v_i^+$ with $v_i^-$, where 
\begin{equation*}
    v_0^- = E_m , \quad
    v_1^- = \epsilon_{m-1} \left|E_m\right|^2 E_m,
    \quad 
    v_2^- = E_{m-1}, \quad
    v_3^- = \epsilon_{m-1} \left|E_{m-1}\right|^2 E_{m-1}.
\end{equation*}

\newcommand{\MCaaaaaa}{
	 {\frac{15}{64}} 
	+
	 {\frac{9}{16}}\,\lri\,{(\frac{\hk}{4})}^{2} 
	+
	 {\frac{21}{32}}\,{\lri}^{2}{(\frac{\hk}{4})}^{4} 
	+
	 \frac{3}{10}\,{\lri}^{3}{(\frac{\hk}{4})}^{6} 
}
\newcommand{\MCaaaaac}{
	 \frac{3}{16}\,{(\frac{\hk}{4})}^{2} 
	+
	 {\frac{7}{16}}\,{(\frac{\hk}{4})}^{4}\lri 
	+
	 \frac{3}{10}\,{(\frac{\hk}{4})}^{6}{\lri}^{2} 
}
\newcommand{\MCaaaaba}{
	 {\frac{11}{192}} 
	+
	 {\frac{41}{144}}\,\lri\,{(\frac{\hk}{4})}^{2} 
	+
	 {\frac{1949}{4320}}\,{\lri}^{2}{(\frac{\hk}{4})}^{4} 
	+
	 {\frac{2791}{11340}}\,{\lri}^{3}{(\frac{\hk}{4})}^{6} 
}
\newcommand{\MCaaaabc}{
	 {\frac{11}{80}}\,{(\frac{\hk}{4})}^{2} 
	+
	 {\frac{577}{1680}}\,{(\frac{\hk}{4})}^{4}\lri 
	+
	 {\frac{2791}{11340}}\,{(\frac{\hk}{4})}^{6}{\lri}^{2} 
}
\newcommand{\MCaaacaa}{
	 \frac{3}{16}\,{(\frac{\hk}{4})}^{2} 
	+
	 {\frac{7}{16}}\,{(\frac{\hk}{4})}^{4}\lri 
	+
	 \frac{3}{10}\,{(\frac{\hk}{4})}^{6}{\lri}^{2} 
}
\newcommand{\MCaaacac}{
	 {\frac{7}{32}}\,{(\frac{\hk}{4})}^{4} 
	+
	 \frac{3}{10}\,{(\frac{\hk}{4})}^{6}\lri 
}
\newcommand{\MCaaacba}{
	 {\frac{53}{720}}\,{(\frac{\hk}{4})}^{2} 
	+
	 {\frac{845}{3024}}\,{(\frac{\hk}{4})}^{4}\lri 
	+
	 {\frac{2791}{11340}}\,{(\frac{\hk}{4})}^{6}{\lri}^{2} 
}
\newcommand{\MCaaacbc}{
	 {\frac{577}{3360}}\,{(\frac{\hk}{4})}^{4} 
	+
	 {\frac{2791}{11340}}\,{(\frac{\hk}{4})}^{6}\lri 
}
\newcommand{\MCaabaaa}{
	 {\frac{11}{192}} 
	+
	 {\frac{41}{144}}\,\lri\,{(\frac{\hk}{4})}^{2} 
	+
	 {\frac{1949}{4320}}\,{\lri}^{2}{(\frac{\hk}{4})}^{4} 
	+
	 {\frac{2791}{11340}}\,{\lri}^{3}{(\frac{\hk}{4})}^{6} 
}
\newcommand{\MCaabaac}{
	 {\frac{53}{720}}\,{(\frac{\hk}{4})}^{2} 
	+
	 {\frac{845}{3024}}\,{(\frac{\hk}{4})}^{4}\lri 
	+
	 {\frac{2791}{11340}}\,{(\frac{\hk}{4})}^{6}{\lri}^{2} 
}
\newcommand{\MCaababa}{
	 {\frac{5}{192}} 
	+
	 {\frac{23}{144}}\,\lri\,{(\frac{\hk}{4})}^{2} 
	+
	 {\frac{1379}{4320}}\,{\lri}^{2}{(\frac{\hk}{4})}^{4} 
	+
	 {\frac{2329}{11340}}\,{\lri}^{3}{(\frac{\hk}{4})}^{6} 
}
\newcommand{\MCaababc}{
	 {\frac{43}{720}}\,{(\frac{\hk}{4})}^{2} 
	+
	 {\frac{691}{3024}}\,{(\frac{\hk}{4})}^{4}\lri 
	+
	 {\frac{2329}{11340}}\,{(\frac{\hk}{4})}^{6}{\lri}^{2} 
}
\newcommand{\MCaabcaa}{
	 {\frac{11}{80}}\,{(\frac{\hk}{4})}^{2} 
	+
	 {\frac{577}{1680}}\,{(\frac{\hk}{4})}^{4}\lri 
	+
	 {\frac{2791}{11340}}\,{(\frac{\hk}{4})}^{6}{\lri}^{2} 
}
\newcommand{\MCaabcac}{
	 {\frac{577}{3360}}\,{(\frac{\hk}{4})}^{4} 
	+
	 {\frac{2791}{11340}}\,{(\frac{\hk}{4})}^{6}\lri 
}
\newcommand{\MCaabcba}{
	 {\frac{43}{720}}\,{(\frac{\hk}{4})}^{2} 
	+
	 {\frac{691}{3024}}\,{(\frac{\hk}{4})}^{4}\lri 
	+
	 {\frac{2329}{11340}}\,{(\frac{\hk}{4})}^{6}{\lri}^{2} 
}
\newcommand{\MCaabcbc}{
	 {\frac{463}{3360}}\,{(\frac{\hk}{4})}^{4} 
	+
	 {\frac{2329}{11340}}\,{(\frac{\hk}{4})}^{6}\lri 
}
\newcommand{\MCacaaaa}{
	 \frac{3}{16}\,{(\frac{\hk}{4})}^{2} 
	+
	 {\frac{7}{16}}\,{(\frac{\hk}{4})}^{4}\lri 
	+
	 \frac{3}{10}\,{(\frac{\hk}{4})}^{6}{\lri}^{2} 
}
\newcommand{\MCacaaac}{
	 {\frac{7}{32}}\,{(\frac{\hk}{4})}^{4} 
	+
	 \frac{3}{10}\,{(\frac{\hk}{4})}^{6}\lri 
}
\newcommand{\MCacaaba}{
	 {\frac{53}{720}}\,{(\frac{\hk}{4})}^{2} 
	+
	 {\frac{845}{3024}}\,{(\frac{\hk}{4})}^{4}\lri 
	+
	 {\frac{2791}{11340}}\,{(\frac{\hk}{4})}^{6}{\lri}^{2} 
}
\newcommand{\MCacaabc}{
	 {\frac{577}{3360}}\,{(\frac{\hk}{4})}^{4} 
	+
	 {\frac{2791}{11340}}\,{(\frac{\hk}{4})}^{6}\lri 
}
\newcommand{\MCacacaa}{
	 {\frac{7}{32}}\,{(\frac{\hk}{4})}^{4} 
	+
	 \frac{3}{10}\,{(\frac{\hk}{4})}^{6}\lri 
}
\newcommand{\MCacacac}{
	 \frac{3}{10}\,{(\frac{\hk}{4})}^{6} 
}
\newcommand{\MCacacba}{
	 {\frac{3257}{30240}}\,{(\frac{\hk}{4})}^{4} 
	+
	 {\frac{2791}{11340}}\,{(\frac{\hk}{4})}^{6}\lri 
}
\newcommand{\MCacacbc}{
	 {\frac{2791}{11340}}\,{(\frac{\hk}{4})}^{6} 
}
\newcommand{\MCacbaaa}{
	 {\frac{53}{720}}\,{(\frac{\hk}{4})}^{2} 
	+
	 {\frac{845}{3024}}\,{(\frac{\hk}{4})}^{4}\lri 
	+
	 {\frac{2791}{11340}}\,{(\frac{\hk}{4})}^{6}{\lri}^{2} 
}
\newcommand{\MCacbaac}{
	 {\frac{3257}{30240}}\,{(\frac{\hk}{4})}^{4} 
	+
	 {\frac{2791}{11340}}\,{(\frac{\hk}{4})}^{6}\lri 
}
\newcommand{\MCacbaba}{
	 {\frac{29}{720}}\,{(\frac{\hk}{4})}^{2} 
	+
	 {\frac{2743}{15120}}\,{(\frac{\hk}{4})}^{4}\lri 
	+
	 {\frac{2329}{11340}}\,{(\frac{\hk}{4})}^{6}{\lri}^{2} 
}
\newcommand{\MCacbabc}{
	 {\frac{2743}{30240}}\,{(\frac{\hk}{4})}^{4} 
	+
	 {\frac{2329}{11340}}\,{(\frac{\hk}{4})}^{6}\lri 
}
\newcommand{\MCacbcaa}{
	 {\frac{577}{3360}}\,{(\frac{\hk}{4})}^{4} 
	+
	 {\frac{2791}{11340}}\,{(\frac{\hk}{4})}^{6}\lri 
}
\newcommand{\MCacbcac}{
	 {\frac{2791}{11340}}\,{(\frac{\hk}{4})}^{6} 
}
\newcommand{\MCacbcba}{
	 {\frac{2743}{30240}}\,{(\frac{\hk}{4})}^{4} 
	+
	 {\frac{2329}{11340}}\,{(\frac{\hk}{4})}^{6}\lri 
}
\newcommand{\MCacbcbc}{
	 {\frac{2329}{11340}}\,{(\frac{\hk}{4})}^{6} 
}
\newcommand{\MCbaaaaa}{
	 {\frac{11}{192}} 
	+
	 {\frac{41}{144}}\,\lri\,{(\frac{\hk}{4})}^{2} 
	+
	 {\frac{1949}{4320}}\,{\lri}^{2}{(\frac{\hk}{4})}^{4} 
	+
	 {\frac{2791}{11340}}\,{\lri}^{3}{(\frac{\hk}{4})}^{6} 
}
\newcommand{\MCbaaaac}{
	 {\frac{53}{720}}\,{(\frac{\hk}{4})}^{2} 
	+
	 {\frac{845}{3024}}\,{(\frac{\hk}{4})}^{4}\lri 
	+
	 {\frac{2791}{11340}}\,{(\frac{\hk}{4})}^{6}{\lri}^{2} 
}
\newcommand{\MCbaaaba}{
	 {\frac{5}{192}} 
	+
	 {\frac{23}{144}}\,\lri\,{(\frac{\hk}{4})}^{2} 
	+
	 {\frac{1379}{4320}}\,{\lri}^{2}{(\frac{\hk}{4})}^{4} 
	+
	 {\frac{2329}{11340}}\,{\lri}^{3}{(\frac{\hk}{4})}^{6} 
}
\newcommand{\MCbaaabc}{
	 {\frac{43}{720}}\,{(\frac{\hk}{4})}^{2} 
	+
	 {\frac{691}{3024}}\,{(\frac{\hk}{4})}^{4}\lri 
	+
	 {\frac{2329}{11340}}\,{(\frac{\hk}{4})}^{6}{\lri}^{2} 
}
\newcommand{\MCbaacaa}{
	 {\frac{53}{720}}\,{(\frac{\hk}{4})}^{2} 
	+
	 {\frac{845}{3024}}\,{(\frac{\hk}{4})}^{4}\lri 
	+
	 {\frac{2791}{11340}}\,{(\frac{\hk}{4})}^{6}{\lri}^{2} 
}
\newcommand{\MCbaacac}{
	 {\frac{3257}{30240}}\,{(\frac{\hk}{4})}^{4} 
	+
	 {\frac{2791}{11340}}\,{(\frac{\hk}{4})}^{6}\lri 
}
\newcommand{\MCbaacba}{
	 {\frac{29}{720}}\,{(\frac{\hk}{4})}^{2} 
	+
	 {\frac{2743}{15120}}\,{(\frac{\hk}{4})}^{4}\lri 
	+
	 {\frac{2329}{11340}}\,{(\frac{\hk}{4})}^{6}{\lri}^{2} 
}
\newcommand{\MCbaacbc}{
	 {\frac{2743}{30240}}\,{(\frac{\hk}{4})}^{4} 
	+
	 {\frac{2329}{11340}}\,{(\frac{\hk}{4})}^{6}\lri 
}
\newcommand{\MCbabaaa}{
	 {\frac{5}{192}} 
	+
	 {\frac{23}{144}}\,\lri\,{(\frac{\hk}{4})}^{2} 
	+
	 {\frac{1379}{4320}}\,{\lri}^{2}{(\frac{\hk}{4})}^{4} 
	+
	 {\frac{2329}{11340}}\,{\lri}^{3}{(\frac{\hk}{4})}^{6} 
}
\newcommand{\MCbabaac}{
	 {\frac{29}{720}}\,{(\frac{\hk}{4})}^{2} 
	+
	 {\frac{2743}{15120}}\,{(\frac{\hk}{4})}^{4}\lri 
	+
	 {\frac{2329}{11340}}\,{(\frac{\hk}{4})}^{6}{\lri}^{2} 
}
\newcommand{\MCbababa}{
	 {\frac{1}{64}} 
	+
	 {\frac{5}{48}}\,\lri\,{(\frac{\hk}{4})}^{2} 
	+
	 {\frac{67}{288}}\,{\lri}^{2}{(\frac{\hk}{4})}^{4} 
	+
	 {\frac{47}{270}}\,{\lri}^{3}{(\frac{\hk}{4})}^{6} 
}
\newcommand{\MCbababc}{
	 {\frac{5}{144}}\,{(\frac{\hk}{4})}^{2} 
	+
	 {\frac{67}{432}}\,{(\frac{\hk}{4})}^{4}\lri 
	+
	 {\frac{47}{270}}\,{(\frac{\hk}{4})}^{6}{\lri}^{2} 
}
\newcommand{\MCbabcaa}{
	 {\frac{43}{720}}\,{(\frac{\hk}{4})}^{2} 
	+
	 {\frac{691}{3024}}\,{(\frac{\hk}{4})}^{4}\lri 
	+
	 {\frac{2329}{11340}}\,{(\frac{\hk}{4})}^{6}{\lri}^{2} 
}
\newcommand{\MCbabcac}{
	 {\frac{2743}{30240}}\,{(\frac{\hk}{4})}^{4} 
	+
	 {\frac{2329}{11340}}\,{(\frac{\hk}{4})}^{6}\lri 
}
\newcommand{\MCbabcba}{
	 {\frac{5}{144}}\,{(\frac{\hk}{4})}^{2} 
	+
	 {\frac{67}{432}}\,{(\frac{\hk}{4})}^{4}\lri 
	+
	 {\frac{47}{270}}\,{(\frac{\hk}{4})}^{6}{\lri}^{2} 
}
\newcommand{\MCbabcbc}{
	 {\frac{67}{864}}\,{(\frac{\hk}{4})}^{4} 
	+
	 {\frac{47}{270}}\,{(\frac{\hk}{4})}^{6}\lri 
}
\newcommand{\MCbcaaaa}{
	 {\frac{11}{80}}\,{(\frac{\hk}{4})}^{2} 
	+
	 {\frac{577}{1680}}\,{(\frac{\hk}{4})}^{4}\lri 
	+
	 {\frac{2791}{11340}}\,{(\frac{\hk}{4})}^{6}{\lri}^{2} 
}
\newcommand{\MCbcaaac}{
	 {\frac{577}{3360}}\,{(\frac{\hk}{4})}^{4} 
	+
	 {\frac{2791}{11340}}\,{(\frac{\hk}{4})}^{6}\lri 
}
\newcommand{\MCbcaaba}{
	 {\frac{43}{720}}\,{(\frac{\hk}{4})}^{2} 
	+
	 {\frac{691}{3024}}\,{(\frac{\hk}{4})}^{4}\lri 
	+
	 {\frac{2329}{11340}}\,{(\frac{\hk}{4})}^{6}{\lri}^{2} 
}
\newcommand{\MCbcaabc}{
	 {\frac{463}{3360}}\,{(\frac{\hk}{4})}^{4} 
	+
	 {\frac{2329}{11340}}\,{(\frac{\hk}{4})}^{6}\lri 
}
\newcommand{\MCbcacaa}{
	 {\frac{577}{3360}}\,{(\frac{\hk}{4})}^{4} 
	+
	 {\frac{2791}{11340}}\,{(\frac{\hk}{4})}^{6}\lri 
}
\newcommand{\MCbcacac}{
	 {\frac{2791}{11340}}\,{(\frac{\hk}{4})}^{6} 
}
\newcommand{\MCbcacba}{
	 {\frac{2743}{30240}}\,{(\frac{\hk}{4})}^{4} 
	+
	 {\frac{2329}{11340}}\,{(\frac{\hk}{4})}^{6}\lri 
}
\newcommand{\MCbcacbc}{
	 {\frac{2329}{11340}}\,{(\frac{\hk}{4})}^{6} 
}
\newcommand{\MCbcbaaa}{
	 {\frac{43}{720}}\,{(\frac{\hk}{4})}^{2} 
	+
	 {\frac{691}{3024}}\,{(\frac{\hk}{4})}^{4}\lri 
	+
	 {\frac{2329}{11340}}\,{(\frac{\hk}{4})}^{6}{\lri}^{2} 
}
\newcommand{\MCbcbaac}{
	 {\frac{2743}{30240}}\,{(\frac{\hk}{4})}^{4} 
	+
	 {\frac{2329}{11340}}\,{(\frac{\hk}{4})}^{6}\lri 
}
\newcommand{\MCbcbaba}{
	 {\frac{5}{144}}\,{(\frac{\hk}{4})}^{2} 
	+
	 {\frac{67}{432}}\,{(\frac{\hk}{4})}^{4}\lri 
	+
	 {\frac{47}{270}}\,{(\frac{\hk}{4})}^{6}{\lri}^{2} 
}
\newcommand{\MCbcbabc}{
	 {\frac{67}{864}}\,{(\frac{\hk}{4})}^{4} 
	+
	 {\frac{47}{270}}\,{(\frac{\hk}{4})}^{6}\lri 
}
\newcommand{\MCbcbcaa}{
	 {\frac{463}{3360}}\,{(\frac{\hk}{4})}^{4} 
	+
	 {\frac{2329}{11340}}\,{(\frac{\hk}{4})}^{6}\lri 
}
\newcommand{\MCbcbcac}{
	 {\frac{2329}{11340}}\,{(\frac{\hk}{4})}^{6} 
}
\newcommand{\MCbcbcba}{
	 {\frac{67}{864}}\,{(\frac{\hk}{4})}^{4} 
	+
	 {\frac{47}{270}}\,{(\frac{\hk}{4})}^{6}\lri 
}
\newcommand{\MCbcbcbc}{
	 {\frac{47}{270}}\,{(\frac{\hk}{4})}^{6} 
}
\newcommand{\MCCOEF}{1}
\begin{table}[ht]
\begin{centering}\begin{tabular}{|c|c|}
\hline 
coefficients&
explicit expression \tabularnewline
\hline
\hline 
$f_{0}\left(\lri\right)$&
$\frac{3}{8}\left(1+\lri\left(\frac{\hk}{4}\right)^{2}\right)$\tabularnewline
\hline 
$f_{1}\left(\lri\right)$&
$\frac{3}{8}\left(\frac{\hk}{4}\right)^{2}$\tabularnewline
\hline 
$f_{2}\left(\lri\right)$&
$\frac{1}{8}\left(1+\frac{7}{3}\lri\left(\frac{\hk}{4}\right)^{2}\right)$\tabularnewline
\hline 
$f_{3}\left(\lri\right)$&
$\frac{7}{24}\left(\frac{\hk}{4}\right)^{2}$\tabularnewline
\hline 
$g_{000}\left(\lri\right)$&
$\MCaaaaaa$\tabularnewline
\hline 
$g_{001}\left(\lri\right)$&
$\MCaaaaac$\tabularnewline
\hline 
$g_{011}\left(\lri\right)$&
$\MCaaacac$\tabularnewline
\hline 
$g_{111}\left(\lri\right)$&
$\MCacacac$\tabularnewline
\hline 
$g_{002}\left(\lri\right)$&
$\MCaaaaba$\tabularnewline
\hline 
$g_{012}\left(\lri\right)$&
$\MCaaacba$\tabularnewline
\hline 
$g_{003}\left(\lri\right)$&
$\MCaaaabc$\tabularnewline
\hline 
$g_{013}\left(\lri\right)$&
$\MCaaacbc$\tabularnewline
\hline 
$g_{112}\left(\lri\right)$&
$\MCacacba$\tabularnewline
\hline 
$g_{113}\left(\lri\right)$&
$\MCacacbc$\tabularnewline
\hline 
$g_{022}\left(\lri\right)$&
$\MCaababa$\tabularnewline
\hline 
$g_{122}\left(\lri\right)$&
$\MCacbaba$\tabularnewline
\hline 
$g_{023}\left(\lri\right)$&
$\MCaababc$\tabularnewline
\hline 
$g_{123}\left(\lri\right)$&
$\MCacbabc$\tabularnewline
\hline 
$g_{033}\left(\lri\right)$&
$\MCaabcbc$\tabularnewline
\hline 
$g_{133}\left(\lri\right)$&
$\MCacbcbc$\tabularnewline
\hline 
$g_{222}\left(\lri\right)$&
$\MCbababa$\tabularnewline
\hline 
$g_{223}\left(\lri\right)$&
$\MCbababc$\tabularnewline
\hline 
$g_{233}\left(\lri\right)$&
$\MCbabcbc$\tabularnewline
\hline 
$g_{333}\left(\lri\right)$&
$\MCbcbcbc$\tabularnewline
\hline
\end{tabular}\par\end{centering}

\caption{\label{tab:compact-FO-coefs} Coefficients (\ref{eq:FO_tensor_elements})
of the fourth-order compact approximation (\ref{eq:FO_stencil}).
Only 20 coefficients are given out of o total of 64, because $g_{ijk}$
are symmetric with respect to the permutations of indices, i.e. $g_{010}=g_{001}$,
$g_{310}=g_{013}$, etc. }
\end{table}

Finally, by combining the approximations  for all the individual terms in~(\ref{eq:VI-integrals}) we obtain the following fourth order scheme:
{\allowdisplaybreaks
\begin{subequations}
\begin{align}
    \nonumber
     h\DSm  \stackrel{\rm def}{=} &\>
	\frac{ E_{m+1}-E_m}{h}\left(1+\lri_m\frac{h^2k_0^2}{24}\right)
        -\frac{E_{m}-E_{m-1}}{h}\left(1+\lri_{m-1}\frac{h^2k_0^2}{24}\right)
        \\
    \nonumber
        &\> +
        \epsilon_m\frac{h^2k_0^2}{24} \frac{\left|E_{m+1}\right|^2 E_{m+1}-
                \left|E_m\right|^2 E_m}{h}\\ 
    \label{eq:FO_stencil}
        &\> - 
\epsilon_{m-1}\frac{h^2k_0^2}{24} \frac{\left|E_{m}\right|^2 E_{m}-
                \left|E_{m-1}\right|^2 E_{m-1}}{h}\\
    \nonumber
    & \> + hk_0^2\lri_{m-1} \sum_{i=0}^3
        f_i(\lri_{m-1}) v_i^-
    + hk_0^2\epsilon_{m-1} \sum_{i,j,k=0}^3
        g_{ijk}( \lri_{m-1} ) (v_i^-)^* v_j^- v_k^- \\
    \nonumber
     &\> + 
    hk_0^2\lri_m \sum_{i=0}^3
        f_i(\lri_m) v_i^+
    + hk_0^2\epsilon_m \sum_{i,j,k=0}^3
        g_{ijk}( \lri_m ) (v_i^+)^* v_j^+ v_k^+
		= 0,
\end{align}
where $m=1,\dots,M$.
As in the case of second order approximations, the value of the field on the ghost nodes $m=0$ and $m=M+1$ will be determined from the boundary conditions, see Section~\ref{sub:BCs}.

Similarly to the second order cases (Sections~\ref{sub:SO_approx} and \ref{sub:alt_SO_approx}), if $\lri$ and $\epsilon$ are continuous at a given node $z_m$, then $E$ is smooth at this location and the scheme~(\ref{eq:FO_stencil}) reduces to the following fourth order scheme for the differential equation~(\ref{eq:1DNLH}):

\begin{equation}	
\label{eq:FO_stencil2} 
\begin{aligned}
     \DSm  = &\>
		\frac{ E_{m+1}-2E_m+E_{m-1}}{h^2} 
		\left( 1 + \frac{h^2k_0^2}{24}\lri_m \right) \\
            & \> + \frac{k_0^2\epsilon_m}{24} 
			\left(
				|E_{m+1}|^2E_{m+1} - 2|E_m|^2E_m + |E_{m-1}|^2E_{m-1}
			\right)\\
     & \> + k_0^2 \lri_m 
		\sum_{i=0}^3 f_i(\lri_m) \left( v_i^- + v_i^+ \right)  \\
     & \> + k_0^2 \epsilon_m 
			\sum_{i,j,k=0}^3 g_{ijk}( \lri_m ) \left(
				(v_i^-)^* v_j^- v_k^-
				+ (v_i^+)^* v_j^+ v_k^+
			\right) 
	= 0. 
\end{aligned}
\end{equation}
\end{subequations}
}
Note that in the simplest case of a linear equation with constant coefficients, 
$\epsilon_m\equiv0$ and
$\nu_m\equiv\nu={\rm const}$, scheme (\ref{eq:FO_stencil2}) transforms into
\begin{equation}
\label{eq:FO_stencil2_cc} 
\begin{aligned}
		\frac{E_{m-1} -2E_m + E_{m+1}}{h^2}
		+ &\> k_0^2 \lri 
			\frac{E_{m-1} +4E_m + E_{m+1}}{6}\\
		+ &\> h^2 k_0^4 \lri^2
			\frac{7E_{m-1} +18E_m + 7E_{m+1}}{384}=0.
			\end{aligned}
\end{equation}
It can be verified that the scheme (\ref{eq:FO_stencil2_cc}) is equivalent (up to terms of order ${\cal O}(h^4)$ and higher) to the standard three-point fourth order compact approximation
\begin{equation}
\label{eq:FO_stencil2_std}
                \frac{E_{m-1} -2E_m + E_{m+1}}{h^2}
                +  k_0^2 \lri
                        \frac{E_{m-1} +10E_m + E_{m+1}}{12}=0. 
\end{equation}
of the linear constant coefficient Helmholtz equation \cite{singer-turkel-98}.

\subsection{\label{sub:BCs} Two-way boundary conditions }

We now derive the discrete version of the two-way boundary conditions~(\ref{eq:CM_TWBCs}) at the interface $z=0$ and $z=Z_{\max}$.
Recall that the two-way boundary condition~(\ref{eq:CM_TWBCs_1_l}) was constructed in Section~\ref{sec:intro} so as to facilitate the propagation of the outgoing waves through the interface $z=0$ and at the same time to prescribe the given incoming signal. 
This means that the solution to equation~(\ref{eq:1DNLH}) for $z\leq0$ is to be composed of a given incoming wave and the outgoing wave, which is not known ahead of time.
Since for $z\leq0$ the material is a homogeneous linear dielectric with $\lri\equiv1$ and $\epsilon\equiv0$, we have:
\begin{equation}
	E(z)= \Einc e^{ik_0z} + R e^{-ik_0z}, \quad z \leq 0,
	\label{eq:field_outside}
\end{equation}
and the boundary condition is derived from the continuity of $E$ and $E'$ at $z=0$.

Our approach to constructing the discrete boundary condition for the scheme is to approximate~(\ref{eq:field_outside}) using closed form solutions of the corresponding difference equation.
This will provide the value of the solution at the ghost node $E_0$ in terms of that at the boundary node $E_1$ and the incoming beam $\Einc$.
Then, $E_0$ can be eliminated from the equation $F_1[\bvec{E}]=0$. A survey of methods for setting the boundary conditions at external artificial boundaries can be found in~\cite{Tsynkov-98}.
In the context of the one-dimensional NLH, the continuous two-way boundary conditions are discussed in~\cite{chen-mills-PRB-87}. For the multidimensional NLH, the continuous and discrete two-way boundary conditions are constructed and implemented in~\cite{FT:01,FT:04,BFT:05}.

Since $\lri_m\equiv1$ and $\epsilon_m\equiv0$ for $m=0,-1,\dots\,$ (i.e., for $z\leq0$), both the second order approximation and the fourth order approximation of Section~\ref{sec:discrete-approx} reduce to a symmetric constant-coefficient  three-point discretization of the form:
\begin{equation}
    0 = \DSm = L_1E_{m-1} -2 L_0E_m+L_1E_{m+1},
	\quad m = 0,-1,\dots,
    \label{eq:BC_stencil}
\end{equation}
where the coefficients $L_0$ and $L_1$ are different for each specific approximation.
For the second order discretizations~(\ref{eq:SO_stencil}) and
(\ref{eq:alt_SO_stencil}) we have:
\[
    L_0 = \hk^{-2} - \frac{3}{8}, \qquad 
    L_1 = \hk^{-2} + \frac{1}{8},
\]
while for the fourth order discretization~(\ref{eq:FO_stencil}) we have:
\[
    L_0 = \hk^{-2} -\frac{1}{3} - \frac{3}{128}\hk^2, \qquad
    L_1 = \hk^{-2} +\frac{1}{6} + \frac{7}{384}\hk^2.
\]

The general solution of the difference equation~(\ref{eq:BC_stencil}) is 
$\left.C_1 q^m +C_2 q^{-m} \right.$, where 
\begin{equation}
  q = L_0/L_1+ i \sqrt{1-\left(L_0/L_1\right)^2}\ \ \text{and}\ \
  q^{-1} = L_0/L_1- i \sqrt{1-\left(L_0/L_1\right)^2}
  \label{eq:qs}
\end{equation} 
are roots of the characteristic equation 
$L_1q^2-2L_0q+L_1=0$.
As can be easily seen from~(\ref{eq:qs}), 
$|q|=1$ and $q^{-1}=q^*$.
Moreover, one can show that the solution $q^m$ approximates the right-traveling wave $e^{ik_0z}\equiv e^{ik_0h(m-1)}$, and the solution $q^{-m}$ approximates the left-traveling wave $e^{-ik_0z}\equiv e^{-ik_0h(m-1)}$, with respective orders of accuracy (second or fourth), see~\cite{FT:01,FT:04,BFT:05} for more detail.

Consequently, the discrete counterpart of formula~(\ref{eq:field_outside}) for $m\leq 1$ can be written as  
\begin{equation}
	E_m=\Einc q^{m-1}+Rq^{1-m}, 	\quad m = 1,0,-1,\dots \, .
	\label{eq:E_m-outside}
\end{equation}
			From equation~(\ref{eq:E_m-outside}) considered for $m=1$ and for $m=0$ we can express the value of the solution at the ghost node $E_0$ as
\begin{subequations}
	\begin{equation}
		E_0 = (q^{-1}-q)\Einc+qE_1.
		\label{eq:SBCs_ghost-l}
	\end{equation}
	The discrete version of the two-way boundary condition~(\ref{eq:CM_TWBCs})
	at $z=0$ is then obtained by substituting $E_0$ 
	from~(\ref{eq:SBCs_ghost-l}) into the discrete equation $
	F_1[\bvec{E}]=0
	$, i.e., into equation~(\ref{eq:SO_stencil}),~(\ref{eq:alt_SO_stencil}) or 
	(\ref{eq:FO_stencil}) with $m=1$. 
	
	Similarly, the discrete version of the Sommerfeld boundary condition~(\ref{eq:CM_TWBCs}) at~$z=Z_{\max}$ is  
	\begin{equation}
		E_{M+1} = qE_M.
		\label{eq:SBCs_ghost-r}
	\end{equation} 
\end{subequations}
This relation is substituted into the discrete equation $
	F_M[\bvec{E}]=0
$.

\section{Newton's iterations \label{sec:Newtons} }

The discrete approximations~(\ref{eq:SO_stencil}),~(\ref{eq:alt_SO_stencil}) and~(\ref{eq:FO_stencil}) are coupled systems of nonlinear algebraic equations.
In our previous work~\cite{FT:01,FT:04,BFT:05}, we solved similar systems by an iteration scheme based on freezing the nonlinearity~$|E|^2$ in equation~(\ref{eq:NLH}). 
In doing so, we have observed that the convergence of iterations was limited to relatively low-power incoming beams, i.e., weak nonlinearities. 

In this section, we describe a different iteration scheme for solving the NLH~(\ref{eq:1DNLH}) based on Newton's method. 
Newton's method cannot be applied to equation~(\ref{eq:1DNLH}) directly, because $|E|$ is not differentiable 
in the Cauchy-Riemann sense and hence the entire operator is not differentiable in the sense of Frech\'et. 
This difficulty and the way to overcome it are first discussed in Section~\ref{sub:newtons-1D} 
through the consideration of Newton's method for a single-variable complex function.
The method is then extended to multivariable functions in 
Section~\ref{sub:newtons-MD}, and its application to the discretizations of 
Section~\ref{sec:discrete-approx} is considered in Section~\ref{sub:stencil-variation}.
Note that the particular implementation of Newton's method presented hereafter leads
to a convenient block tridiagonal structure of the Jacobians that enables efficient numerical 
inversion (${\cal O}(M)$ time).

\subsection{\label{sub:newtons-1D}A single complex variable}

The basic idea is, in fact, quite simple --- while the function $|E|^2$ is not differentiable with respect to $E$, it is differentiable with respect to $\RE(E)$ and $\IM(E)$ as a function of two real variables. Hence, Newton's linearization can be obtained if one complex equation is recast as a system of two real equations. 
Let us first recall Newton's method for solving the scalar equation\[
	0=F(E),
\]
where $F$ is differentiable with respect to $E$.
We denote the exact solution by $\tilde{E}$, the  
j-th iterate by $E^{(j)}$, and their difference by $
	\delta E = \tilde{E} - E^{(j)} 
$. Using the
Taylor expansion around $E^{(j)}$ we have: \[
	0 = F(\tilde{E}) 
	= F(E^{(j)} + \delta E)
	= F(E^{(j)}) + \left.
		\frac{dF}{dE}	
	\right|_{E=E^{(j)}} \delta E + \odesq.
\] 
Introducing  the differential of $F$ at $E^{(j)}$:
\begin{equation}
	\delta F = J ( E^{(j)} ) \delta E,\quad\text{where}\quad
	J(E) = \frac{dF}{dE},
	\label{eq:delta_F}
\end{equation}
we can then write:
\begin{equation*}
	\delta F =
	F(E^{(j)} + \delta E) - F(E^{(j)}) + \odesq
	= - F(E^{(j)}) + \odesq,
\end{equation*}
and consequently, 
\begin{equation}
	J ( E^{(j)} ) \delta E = \delta F 
	= - F(E^{(j)}) + \odesq.
	\label{eq:delta_F-eqs-mF}
\end{equation}
Neglecting the $\odesq$ term in~(\ref{eq:delta_F-eqs-mF}) and solving the equation with respect to $\delta E$ we obtain the next iterate $E^{(j+1)}=E^{(j)}+\delta E$:
\begin{equation}
	E^{(j+1)} =  E^{(j)}  
		- \left[ J ( E^{(j)} ) \right]^{-1} 
		F ( E^{(j)} ).
	\label{eq:Newtons-1D}
\end{equation}
If the initial guess $E^{(0)}$ is chosen sufficiently close to $\tilde E$, then the sequence of Newton's iterations~(\ref{eq:Newtons-1D}) is known to converge to the exact solution $\tilde E$ as $j\to\infty$.

Next, we consider the scalar equation
\begin{equation}
	0=F(E) = |E|^2E - 1 = E^*E^2 - 1.
	\label{eq:Newt-example}
\end{equation}
The modulus $|E|$  in~(\ref{eq:Newt-example}) is not differentiable with respect to $E$ in
the Cauchy-Riemann sense. 
However, $F$ is differentiable as a function of two variables
	$\RE(E)$ and $\IM(E)$ or, alternatively, $E$ and $E^*$. Hence,
\begin{equation*}
	0 =  F(\tilde E)=F(E^{(j)}+\delta E) = 
		F(E^{(j)}) 
		+ (E^{(j)})^2\delta E^* 
		+ 2 |E^{(j)}|^2 \delta E 
		+ \odesq.
\end{equation*}
Therefore, since $
	\frac{\partial F} {\partial E} = 2|E|^2
$ and $
	\frac{\partial F} {\partial E^*} = E^2
$, the analogue of~(\ref{eq:delta_F}) is 
\begin{equation}
	\delta F =
	J_1 \delta E + J_2 \delta E^* ,\ \ \text{where}\ \ 
	J_1(E)  =  \frac{\partial F} {\partial E} \ \ ,\ \ 
	J_2(E)  =  \frac{\partial F} {\partial E^*}.
	\label{eq:Newt-example-variation}
\end{equation}
Consequently, the equivalent of~(\ref{eq:delta_F-eqs-mF}) is
\begin{equation}
	J_1(E^{(j)}) \delta E + J_2(E^{(j)}) \delta E^*
		= \delta F 
		= -F(E^{(j)}) 	+ \odesq.
	\label{eq:delta_F-eqs-mF-2}
\end{equation}

To solve equation~(\ref{eq:delta_F-eqs-mF-2}) for $\delta E$ and obtain the equivalent of~(\ref{eq:Newtons-1D}), we separate the real and imaginary parts of the function~$F$ and the independent  variable~$E$.
This is convenient to do by representing them as real $2\times1$ column vectors: \[
	E \mapsto \wh{E} =
    \begin{bmatrix}
        \RE(E) \\
		\IM(E)
    \end{bmatrix}, \qquad 
	F\mapsto \wh{F} =
    \begin{bmatrix}
        \RE(F) \\
		\IM(F)
    \end{bmatrix}. 
\]
Then, multiplication by a complex number and conjugation correspond to matrix operations on~$\mathbb{R}^2$, which leads to a real Jacobian in~(\ref{eq:Newt-example-variation}).
Indeed, multiplication by a complex number $c$ can be represented as 
\begin{equation*}
    c \cdot z \mapsto \wh{\,c\cdot z\,} =
    \begin{bmatrix}
		\RE(c\cdot z) \\
        \IM(c\cdot z)
    \end{bmatrix} = 
    \begin{bmatrix}
        \RE\left(c\right) & -\IM\left(c\right) \\
        \IM\left(c\right) & \RE\left(c\right)
    \end{bmatrix}
	\begin{bmatrix}
        \RE(z) \\
        \IM(z)
    \end{bmatrix}.
\end{equation*}
If we associate a $2\times2$ real matrix $\wwhc{c}$ with a given complex number $c$:
\begin{equation*}
    \wwhc{c} =
    \begin{bmatrix}
        \RE\left(c\right) & -\IM\left(c\right) \\
        \IM\left(c\right) & \RE\left(c\right)
    \end{bmatrix} =
	\RE\left(c\right) 
    	\begin{bmatrix}
        	1 & 0\\
        	0 & 1
    	\end{bmatrix} 
	+ \IM\left(c\right) 
	 \begin{bmatrix}
        0 & -1\\
        1 & 0
    \end{bmatrix},
\end{equation*}
then
\begin{equation*}
    c \cdot z \mapsto
		\wwhc{c} \cdot \wh{z}.
\end{equation*}
Similarly, complex conjugation is a left multiplication by the matrix ${\rm diag}[1,-1]$:
\begin{equation*}
    z^* \mapsto \wh{z^*} =
    \begin{bmatrix}
        1 & 0\\
        0 & -1
	\end{bmatrix}
	\begin{bmatrix}
        \RE(z)\\
        \IM(z)
	\end{bmatrix}
	=
    \begin{bmatrix}
        1 & 0\\
        0 & -1
	\end{bmatrix}
	\wh{z}.
\end{equation*}
Thus, equation~(\ref{eq:Newt-example-variation}) transforms into
\begin{align}
	\delta F 
	& =
	\underbrace{
		\begin{bmatrix}
			2|E^{(j)}|^2 & 0 \\
			0 		& 2|E^{(j)}|^2
		\end{bmatrix}
	}_{\wwh{J_1}}
	\begin{bmatrix}
		\RE(\delta E) \\
		\IM(\delta E)
	\end{bmatrix} 
	+ \underbrace{
		\begin{bmatrix}
			\RE(E^{(j)})^2 & -\IM(E^{(j)})^2 \\
			\IM(E^{(j)})^2 & \RE(E^{(j)})^2
		\end{bmatrix}
	}_{\wwh{J_2}}
    \begin{bmatrix}
        1 & 0\\
        0 & -1
	\end{bmatrix} 
	\begin{bmatrix}
		\RE(\delta E) \\
		\IM(\delta E)
	\end{bmatrix} \nonumber \\
	& =  \left( 
		\wwh{J_1} + 
			\wwh{J_2} 
    		\begin{bmatrix}
        		1 & 0\\
        		0 & -1
			\end{bmatrix} 
		\right) \delta \wh{E}
		\stackrel{\rm def}{=} \wwhc{J}\delta \wh{E},
\end{align}
where  \[
	\wwhc{J}  =  \left( 
		\wwh{J_1} + 
		\wwh{J_2}
    	\begin{bmatrix}
        	1 & 0\\
        	0 & -1
		\end{bmatrix} 
	\right).
\]
Having derived the real Jacobian $\wwhc{J}$, we neglect the quadratic terms in~(\ref{eq:delta_F-eqs-mF-2}) to obtain the following Newton's iteration:
\begin{equation*}
	\wh{E}^{(j+1)} - \wh{E}^{(j)}  = 
	-  \left[ \wwhc{J}( E^{(j)} ) \right]^{-1} 
		\wh{F} ( E^{(j)} ).
\end{equation*}

\subsection{\label{sub:newtons-MD}Extension to multiple variables}

We now apply the procedure outlined in Section~\ref{sub:newtons-1D} to a system of the form $\DS=0$, where 
		$\bvec{E}=\left[E_1,\dots,E_M\right]^T\in \mathbb{C}^M$ and 
		$\bvec{F} = \left[F_1, \dots, F_M \right]^T\in \mathbb{C}^M$.
 We would like to solve the equations using Newton's iterations of the type:  
\begin{equation*}
	\bvec{E}^{(j+1)} - \bvec{E}^{(j)}  = -  \left[ 
			J ( \bvec{E}^{(j)} ) 
		\right]^{-1} 
		\bvec{F} ( \bvec{E}^{(j)} ),
\end{equation*}
where $J(\bvec{E})$ is the appropriate Jacobian of $\DS$. 
As, however, the individual components of the vector $\bvec{F}$ are not differentiable in the Cauchy-Riemann sense with respect to the components of $\bvec{E}$, the Frech\'et differential of $\DS$ and the corresponding Jacobian can only be introduced as  in Section~\ref{sub:newtons-1D}, by recasting the equation using the real and imaginary parts of all variables.
 
As in Section~\ref{sub:newtons-1D}, the variation of $\DS$ 
in terms of the field $\bvec{E}$ and its conjugate $\bvec{E}^*$ is given by
\begin{equation*}
	\delta\DS =
	J_1 \delta \bvec{E} + J_2 \delta \bvec{E}^* ,\ \ \text{where}\ \ 
	J_1  =  \frac{\partial\bvec{F}} {\partial \bvec{E}} \ \ \text{and}\ \  
	J_2  =  \frac{\partial\bvec{F}} {\partial \bvec{E}^*}.
\end{equation*}
Let us represent  $\bvec{E}$ and  
$\bvec{F}$ as $2M\times1$ column vectors with real components:
\begin{eqnarray*}
    \hbvec{E} &=& \begin{bmatrix}
        \RE(E_1), & \IM(E_1), & \dots & \RE(E_m), & \IM(E_m), & \dots & \RE(E_M), & \IM(E_M)
    \end{bmatrix}^T, \\
    \hbvec{F} &=& \begin{bmatrix}
        \RE(F_1),& \IM(F_1), & \dots & \RE(F_m), & \IM(F_m), & \dots & \RE(F_M), & \IM(F_M)
    \end{bmatrix}^T.
\end{eqnarray*}
To obtain the real Jacobian $J$, we will represent the 
complex matrices $J_1$ and $J_2$ as real matrices of dimension $2M\times2M$. 
Let $A$ be a complex $M\times M$ matrix. 
For each entry  $A_{lm}$, we substitute the $2\times2$ real block
 $\wwh{A_{lm}}$:
\begin{equation*}
    A \mapsto \wwhc{A} \eqdef 
     \begin{bmatrix}
        \wwh{A_{11}} &  \dots &  \wwh{A_{1M}} \\
        \vdots &  			\ddots & \vdots\\
        \wwh{A_{M1}} &  \dots &  \wwh{A_{MM}}
     \end{bmatrix}      
	 = \begin{bmatrix}
        \RE(A_{11}) & -\IM(A_{11})    & \dots &   \RE(A_{1M}) & -\IM(A_{1M})\\
        \IM(A_{11}) & \RE(A_{11})     & \dots &   \IM(A_{1M}) & \RE(A_{1M})\\
        \vdots & \vdots             & \ddots    & \vdots & \vdots\\
        \RE(A_{M1}) & -\IM(A_{M1})    & \dots &   \RE(A_{MM}) & -\IM(A_{MM})\\
        \IM(A_{M1}) & \RE(A_{M1})     & \dots &   \IM(A_{MM}) & \RE(A_{MM})
     \end{bmatrix}.
\end{equation*}
Introducing the matrix direct product $A\otimes B$ as the matrix obtained by replacing each entry $A_{lm}$ of $A$ by the block $A_{lm}\cdot B$, we can write:
\begin{equation*}
    \wwhc{A} = 
		\RE\left(A\right)\otimes
    	\begin{bmatrix}
        	1 & 0\\
        	0 & 1
		\end{bmatrix} 
    	+ \IM(A)\otimes
    	\begin{bmatrix}
        	0 & -1\\
        	1 & 0
		\end{bmatrix}.
\end{equation*}
Similarly, the conjugation of a column vector $\bvec{E}$ can be represented as
\begin{equation*}
    \bvec{E}^* \mapsto \wh{\,\bvec{E}^*} = \left(
        \mathbb{I}_M \otimes 
    	\begin{bmatrix}
        	1 & 0\\
        	0 & -1
		\end{bmatrix} 
    \right) \hbvec{E} ,
    \qquad 
        \mathbb{I}_M \otimes 
    	\begin{bmatrix}
        	1 & 0\\
        	0 & -1
		\end{bmatrix} 
        = \begin{bmatrix}
            1 & 0\\
            0 & -1\\
            &  & \ddots\\
            &  &  & 1 & 0\\
            &  &  & 0 & -1 \end{bmatrix}_{2M\times 2M},
\end{equation*}
where $\mathbb{I}_M$ is the $M\times M$ identity matrix.

Then, the differential of the real function $\hbvec{F}$ is:
	\begin{equation*}
    	\delta\hbvec{F}(\hbvec{E})
			= \wwh{J_1} \delta \hbvec{E} 
			+ \wwh{J_2} \delta \wh{\, \bvec{E}^* }
		= \wwhc{J}\delta \hbvec{E},
	\end{equation*}
	where the real Jacobian is given by the $2M\times2M$ matrix
	\begin{equation*}
		\wwhc{J} \eqdef  
			\wwh{J_1} + 
			\wwh{J_2} \cdot \left(
				\mathbb{I}_M \otimes 
    			\begin{bmatrix}
		        	1 & 0\\
        			0 & 1
				\end{bmatrix} 
			\right).
	\end{equation*}

\subsection{ \label{sub:stencil-variation} 
	Differentiation of $\DS$ with respect to $\bvec{E}$ and $\bvec{E}^*$ }

In this section, we discuss the actual differentiation of  $\DS$, i.e., the evaluation of
$J_1  =  \frac{\partial\bvec{F}} {\partial \bvec{E}} \ \text{and}\  
	J_2  =  \frac{\partial\bvec{F}} {\partial \bvec{E}^*}$. 
As we shall see, the tensor notation of Sections~\ref{sub:alt_SO_approx} and \ref{sub:FO_approx} prove extremely useful in this context.

Using the identities
\begin{equation*}
	\frac{\partial E_i^* }{\partial E_k} = 
	\frac{\partial E_i }{\partial E_k^*} = 0
	, \qquad
	\frac{\partial E_i }{\partial E_k} = 
	\frac{\partial E_i^* }{\partial E_k^*} = \delta_{ik} 
	= 	\begin{cases}
			0, & i\neq k\\
			1, & i=k
		\end{cases},
\end{equation*}
we first differentiate the linear terms in $\DS$ and obtain for $q=-1,0,1$:
\begin{align*}
	\frac{\partial}{\partial E_{m+q}^*} 
		(L_1E_{m-1}-2L_0E_m+L_1E_{m+1}) 
	 = & \> 0,\\
	\frac{\partial}{\partial E_{m+q}} 
		(L_1E_{m-1}-2L_0E_m+L_1E_{m+1}) 
	 = & \> L_1\delta_{-1,q}-2L_0\delta_{0,q}+L_1\delta_{1,q},
\end{align*}
where  the notation $L_0,\>L_1$ for the coefficients of the scheme was introduced in Section~\ref{sub:BCs}.
The nonlinear terms of the second order scheme~(\ref{eq:SO_stencil}) are differentiated as
\begin{equation*}
    \frac{ \partial } { \partial E_m } \left|E_m \right|^2 E_m  
		= 2 \left|E_m\right|^2 ,\qquad 
    \frac{ \partial } { \partial E_m^* } \left|E_m \right|^2 E_m = E_m^2,   
\end{equation*}
and similarly for $\left|E_{m\pm1}\right|^2E_{m\pm1}$.

For the alternative second order scheme~(\ref{eq:alt_SO_stencil}) we have:
\begin{equation*}
    \frac{\partial}{\partial E_{m+q}^*}
		\sum_{i,j,k=0}^1 g_{ijk} E_{m+i}^*E_{m+j}E_{m+k}
	= \sum_{i,j,k=0}^1 g_{ijk} \delta_{iq} E_{m+j}E_{m+k}
	 = \sum_{j,k=0}^1 g_{qjk} E_{m+j} E_{m+k} \, , 
\end{equation*}
and, using the symmetry of $g_{ijk}$:
\begin{align*}
    \frac{\partial}{\partial E_{m+q}}
		\sum_{i,j,k=0}^1 g_{ijk} E_{m+i}^*E_{m+j}E_{m+k}	
	& = \sum_{i,k=0}^1 g_{iqk} E_{m+i}^* E_{m+k} 
		+ \sum_{i,j=0}^1 g_{ijq} E_{m+i}^* E_{m+j} \\ 
	& = 2 \sum_{i,j=0}^1 g_{ijq} E_{m+i}^* E_{m+j} \,.
\end{align*}

Similarly, the nonlinear terms of the fourth order scheme~(\ref{eq:FO_stencil}) are differentiated as
\begin{equation*}
	 \frac{\partial}{\partial E_{m+q}}
	 \sum_{i=0}^3 f_i v_i^+ =
	 \sum_{i=0}^3 f_i \frac{\partial v_i^+}{\partial E_{m+q}}, \qquad
	 \frac{\partial}{\partial E_{m+q}^*}
	 \sum_{i=0}^3 f_i v_i^+ =
	 \sum_{i=0}^3 f_i \frac{\partial v_i^+}{\partial E_{m+q}^*}, 
\end{equation*}
and
\begin{align*}
    \frac{\partial}{\partial E_{m+q}}
		\sum_{i,j,k=0}^3 g_{ijk} \cdot (v_i^+)^* v_j^+ v_k^+ 	
	& = \sum_{i,j,k=0}^3 g_{ijk} \left( 
			v_i^+ v_j^+ \frac{\partial (v_k^+)^*}{\partial E_{m+q}} 
			+ 2 (v_i^+)^* v_j^+ \frac{\partial v_k^+}{\partial E_{m+q}} 
		\right), \\ 
    \frac{\partial}{\partial E_{m+q}^*}
		\sum_{i,j,k=0}^3 g_{ijk} \cdot (v_i^+)^* v_j^+ v_k^+ 	
	& = \sum_{i,j,k=0}^3 g_{ijk} \left( 
			v_i^+ v_j^+ \frac{\partial (v_k^+)^*}{\partial E_{m+q}^*} 
			+ 2 (v_i^+)^* v_j^+ \frac{\partial v_k^+}{\partial E_{m+q}^*} 
		\right).
\end{align*}

\subsection{Choice of the initial guess \label{sub:choice} }

Convergence of Newton's iterations is known to be sensitive to how close the initial guess ${\bi E}^{(0)}$ happens to be to the solution ${\bi E}$.
Hereafter, we use two different strategies for choosing the initial guesses.

When we test the convergence of Newton's iterations in the vicinity of the exact solution (of the discrete system of equations), we use for the initial guess the closed form solution of the continuous problem~(\ref{eqs:formulation}) of Chen and Mills~\cite{chen-mills-PRB-87,chen-mills-PRB-87-ML}. 
Indeed, we can expect this solution to be either $\oh{2}$ or $\oh{4}$ close to the exact solution of the discrete system of equations $\DS=0$. 
This expectation, which is based on the accuracy analysis of Sections~\ref{sub:SO_approx}, \ref{sub:alt_SO_approx}, and \ref{sub:FO_approx} and does not involve a stability proof, is later corroborated experimentally (see Section~\ref{sec:results}). 

A more interesting case, of course, is when the solution is not known ahead of time.\footnote{This 
would be the case, in particular, for the multidimensional NLH.
}
In order to  choose the initial guess in this case, we adopt a continuation approach in the nonlinearity parameter $\epsilon$.
Namely, we increase $\epsilon$ in a series of increments:
\begin{equation*}
	\epsilon_1 < \epsilon_2 < \ldots < \epsilon_n,
\end{equation*}	
where at the j-th stage we apply Newton's method to the NLH~(\ref{eq:1DNLH}) with 
$\epsilon=\epsilon_j$ using the solution from the j-1 stage with $\epsilon_{j-1}$ as the initial guess. 
In doing so, the value of $\epsilon_n$ is the target nonlinearity parameter for a given computation, and it can be large.
At the beginning stage $j=0$, the initial guess may be chosen either as the solution of the linear problem with $\epsilon=0$, or as the solution obtained by iteration schemes based on freezing the nonlinearity as in~\cite{FT:01,FT:04,BFT:05,Suryanto:2002,Suryanto:2003}, which converge for weak nonlinearities.

\section{\label{sec:method-summary} Summary of the numerical method}

An integral formulation of the NLH~(\ref{eq:1DNLH}) is  discretized on the grid~(\ref{eq:grid}) and written in the form $\DS=0$. The operator $\DS$ is given by 
(\ref{eq:SO_stencil}),~(\ref{eq:alt_SO_stencil}) or~(\ref{eq:FO_stencil}) at the interior nodes $m=1,\dots,M$, while the ghost nodes $E_0$ and $E_{M+1}$ are specified by~(\ref{eq:SBCs_ghost-l}) and~(\ref{eq:SBCs_ghost-r}), respectively.
The resulting system of nonlinear equations is linearized:
\begin{equation*}
	\delta\DS =
	J_1 \delta \bvec{E} + J_2 \delta \bvec{E}^* ,\qquad 
	J_1 \eqdef \frac{\partial\bvec{F}} {\partial \bvec{E}} , \quad 
	J_2 \eqdef \frac{\partial\bvec{F}} {\partial \bvec{E}^*}, 
\end{equation*}
 where $J_1$ and $J_2$ are calculated in Section~\ref{sub:stencil-variation}. 
An equivalent linearized form is obtained using the $\mathbb{R}^{2M}$ representation of Section~\ref{sub:newtons-MD}:
\begin{equation*}
    \delta \hbvec{F}  
	= \wwhc{J}\delta \hbvec{E}
		,\qquad \wwhc{J} \eqdef  
			\wwh{J_1} + 
			\wwh{J_2} \cdot \left(
				\mathbb{I}_M \otimes 
    			\begin{bmatrix}
		        	1 & 0\\
        			0 & -1
				\end{bmatrix} 
			\right).
\end{equation*}
Subsequently, the real Jacobian $\wwhc{J}$ is used to build the sequence of  Newton's iterations:
\begin{equation*}
	\hbvec{E}^{(j+1)} - \hbvec{E}^{(j)}  = -  \left[ 
			\wwhc{J} ( \bvec{E}^{(j)} ) 
		\right]^{-1} 
		\hbvec{F} ( \bvec{E}^{(j)} ).
\end{equation*}

As we shall see, it is sometimes useful to use a relaxation scheme:
\begin{equation}
	\hbvec{E}^{(j+1)} - \hbvec{E}^{(j)}  = - \omega  \left[ 
			\wwhc{J} ( \bvec{E}^{(j)} ) 
		\right]^{-1} 
		\hbvec{F} ( \bvec{E}^{(j)} ).
	\label{eq:newtons-SOR}
\end{equation}
where  the relaxation parameter is typically chosen in the range $0.1\leq\omega\leq0.3$. 
The value $\omega=1$ reduces~(\ref{eq:newtons-SOR}) back to the original Newton's method. 

The initial guess~${\bi E}^{(0)}$ is taken as the closed form continuous solution when the convergence of Newton's iterations is first studied. 
In general, continuation by the nonlinearity parameter $\epsilon$ is used to compute the solutions for strong nonlinearities.

The last important component of the overall numerical method is the inversion of the Jacobian. 
As the problem we are currently solving is one-dimensional, we are using a direct sparse solver to evaluate $\left[	\wwhc{J} ( \hbvec{E}^{(j)} ) \right]^{-1} $ for every $j=0,1,2,\ldots$.
For all our schemes, both second order and fourth order accurate, the Jacobian has three non-zero
diagonals composed of $2\times2$ blocks, which enables an efficient solution by a sparse method.

\section{\label{sec:results}Numerical results}

In this section, we experimentally assess the computational error and the convergence of iterations for the new method, by comparing it with other methods that we have used previously to solve the NLH~\cite{FT:01,FT:04,BFT:05}. 
Since in the one-dimensional case the closed form solutions are available explicitly, the numerical error is evaluated directly.  
All computations in this section are conducted with double
precision. 

\subsection{\label{sub:ref-methods}Reference methods}

In Section~\ref{subsub:loc_conv}, we compare {\em convergence} of Newton's iterations with that of the iterations based on freezing the nonlinearity:
\begin{equation}
		\frac{d^2E^{(j+1)}}{dz^2} + \lri(z)E^{(j+1)} 
		+ \epsilon(z) |E^{(j)}|^2E^{(j+1)}=0.
	\label{eq:NL-iterations}
\end{equation}
For every $j=0,1,2,\ldots$, the next iterate $E^{(j+1)}$ is obtained by solving the linear, variable coefficients, differential equation~(\ref{eq:NL-iterations}).
This ''freezing''  approach was used in our earlier work~\cite{FT:01,FT:04,BFT:05} and also by Suryanto et al.\ ~\cite{Suryanto:2002,Suryanto:2003}. 
In addition, we use a relaxation method based on~(\ref{eq:NL-iterations}):
\begin{equation}
	E^{(j+1)} = (1-\omega)E^{(j)} + \omega  E^{(j+1/2)},
	\label{eq:NL-iterations-SOR}
\end{equation}
where $ E^{(j+1/2)}$ in~(\ref{eq:NL-iterations-SOR}) is the solution of~(\ref{eq:NL-iterations}).
Note that~(\ref{eq:NL-iterations-SOR}) is an analogue of~(\ref{eq:newtons-SOR}).

In  Section~\ref{subsub:res:accuracy} we evaluate the {\em error} of the compact schemes built in Section~\ref{sec:discrete-approx}.
We compare it with the error of the standard central-difference discretizations of order two:
\begin{equation}
        \frac{
            E_{m-1}-2E_m+E_{m+1}
        }{
            h^2
        }
        + k_0^2(\lri_m 
        + \epsilon_m |E_m|^2)E_m =0
    \label{eq:old_SO_stencil}
\end{equation}
and of order four:
\begin{equation}
        \frac{
            -E_{m-2} + 16E_{m-1} - 30E_m + 16E_{m+1} - E_{m+2}
        }{
            12h^2
        } 
	  	+ k_0^2(\lri_m 
        + \epsilon_m |E_m|^2)E_m = 0.
    \label{eq:old_FO_stencil}
\end{equation}
%
In all the simulations, we made sure that the iterations' convergence was sufficient to enable a robust evaluation of the discretization error, i.e., to distinguish between the error of the difference scheme and the error due to the possible ``underconvergence'' of our iterations. 
Furthermore, we verified that the new iterative method (Newton's) and the freezing iterative method~(\ref{eq:NL-iterations}), when they both converge, provide the same error (Section~\ref{subsub:res:accuracy}).

\subsection{\label{sub:Newtons-convergence}Convergence}

In this section, we discuss convergence of Newton's method and compare it with that of the nonlinear iterations (\ref{eq:NL-iterations}) and (\ref{eq:NL-iterations-SOR}).
The parameters used are $\lri\equiv 1$, $k_0=8$, $Z_{\max}=10$, and a uniform nonlinearity profile $\epsilon(z)\equiv const$.
Note that $\lri\equiv 1$ corresponds to the case of the linear index of refraction being the same inside and outside the Kerr medium.
For these parameters, the first nonuniqueness region occurs around $
	\epsilon=\epscrit\approx0.72
$ (see Figure~\ref{fig:transmittance}).

\subsubsection{\label{subsub:loc_conv}Local convergence}

\begin{wrapfigure}{r}{0.45\textwidth}
        \begin{center}
                \includegraphics[scale=0.25]{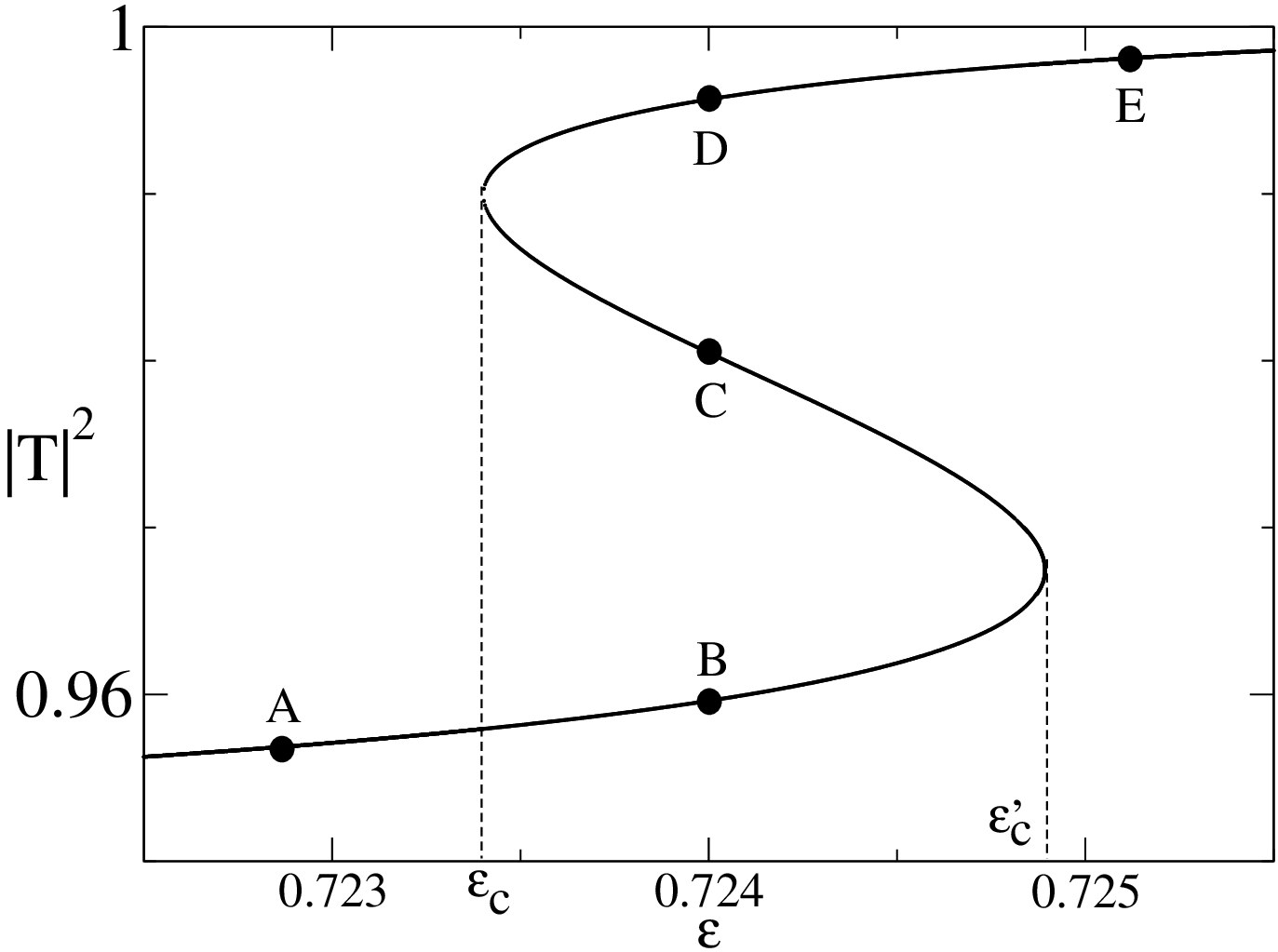}
                \caption{
                        Local convergence experiments for Newton's iterations performed in the region of the first switchback of the one-dimensional NLH~(\ref{eq:1DNLH}) (see Figure~\ref{fig:transmittance}).
                        Each of the initial guesses near~A--E converged to the corresponding discrete  solution.
                        In addition, the continuation approach with~A as the initial guess and~$\epsilon=0.724$ converged to~B, while continuation with~E as the initial guess and~$\epsilon=0.724$ converged to~D.
                }
                \label{fig:Newton-on-switchback}
        \end{center}
\end{wrapfigure}%
The goal of the first series of computations is to determine how the magnitude of nonlinearity (i.e., the value of~$\epsilon$) affects the convergence of Newton's iterations relative to that of nonlinear iterations~(\ref{eq:NL-iterations}).
To achieve this goal, we choose the initial guess to be the pointwise values of the continuous exact solution on the grid, which, as noted, is close to the actual discrete solution for fine grids.
To distinguish between the issues related to iterations' convergence and those pertinent to a specific discretization, we choose one particular scheme, the simplest second order scheme~(\ref{eq:old_SO_stencil}), for all the convergence experiments in this subsection.

The nonlinear iteration scheme~(\ref{eq:NL-iterations}) converges for $\epsilon<0.167\approx0.23\epscrit$. 
Its relaxation analogue (\ref{eq:NL-iterations-SOR}) with $\omega=0.1$ allows us to increase the convergence range up to about $0.3\approx0.4\epscrit$.
Decreasing the value of $\omega$ does not seem to have a significant effect.
For $\epsilon$ above these thresholds the iterations diverge, and the divergence occurs also for other discretizations that we have used (Section~\ref{sec:discrete-approx}) rather than  only for (\ref{eq:old_SO_stencil}). 
Therefore, it shall be interpreted as a limitation of the iteration procedure itself.
This divergence is not related to the onset of nonuniqueness in the NLH, because the convergence breaks down far below the nonlinearity threshold for uniqueness~$\epscrit$. 

In contradistinction to that, Newton's method convergence for $\left.\epsilon\in[0,0.9]\right.$, except near the switchback points $\left.\epsilon=\epscrit\right.$ and $\left.\epsilon=\epscrit'\right.$, where $
		\left.	\left( \frac{dT}{d\epsilon} \right)^{-1} =0	\right.
$ 
(see Figure~\ref{fig:Newton-on-switchback}). 
As expected, the convergence of Newton's iterations considerably slows down as $|\epsilon-\epscrit|$ or $\left.|\epsilon-\epscrit'|\right.$ becomes small, and eventually, close enough to a given switchback point, the iterations diverge 
(the Jacobians degenerate at the switchback points). 
Other than that, the method shows robust convergence.
Specifically, the method converges when the solutions are non-unique:
When the initial guess was close to one of the points B, C, or D in Figure~\ref{fig:Newton-on-switchback}, which correspond to $\epsilon=0.724$ inside the first region of nonuniqueness, the method converged to the respective discrete solution.
This indicates that if the grid is sufficiently fine to tell between two close solutions inside the switchback (see Section~\ref{subsub:res:accuracy}),
Newton's method has an adequate domain of convergence. 
Similar results were obtained for $\epsilon=0.834$, which is in the middle
of the second switchback region $0.828\lesssim\epsilon\lesssim0.839$.

Even at the highest nonlinearity we have tried, $\epsilon=3\approx4\epscrit$, Newton's iterations still converge.
The value $\epsilon=3$ is well beyond the first switchback.
Moreover, it is an extremely high nonlinearity in two respects:
First, for $\epsilon=3$ there are $7$ distinct, i.e., nonunique solutions (we tested the convergence to the highest power solution).
Second, at this value the nonlinear response is so large that it would cause a breakdown and ionization of the  actual physical material in an experimental setting, which renders the original Kerr model inapplicable.
We note that we did not observe in our simulations any convergence deterioration of Newton's method around $\epsilon=3$ compared with smaller $\epsilon$'s, it only requires 3 to 6 iterations to drive the residual down by 10 orders of magnitude.
We thus assume that most likely Newton's method would have  converged for much higher values of $\epsilon$ as well, although it has not been tried because of the physical irrelevance.

As normally expected from Newton's method, the iterations converge rapidly, at a quadratic rate.
In most of the cases that we studied in this section, it took 4--6 iterations to reduce the original residual by 9 to 11 orders of magnitude. 
As has been mentioned, the only situation when Newton's convergence may noticeably slow down is for $\epsilon$ near the switchback points $\epscrit$ and $\epscrit'$, where the tangent to the curve $T(\epsilon)$ becomes vertical, see  Figure~\ref{fig:Newton-on-switchback}. 
Convergence of the iteration scheme~(\ref{eq:NL-iterations}) which is based on freezing the nonlinearity, is much slower.
It is estimated as linear based on experimental evidence, and on the fact that it can 
be interpreted as a fixed-point iteration. 

\subsubsection{\label{subsub:glob_conv} Continuation approach}

Having seen that Newton's method is locally convergent, we would like to test its performance for initial guesses that are not necessarily close to the solution.
Our first observation is that when the solution to the linear problem is used as the initial guess, i.e.,~$E^{(0)}=e^{ik_{0}z}$, then the iterations converge for $0\leq \epsilon\leq0.08$, and diverge for $\epsilon>0.08$.
Therefore, for larger values of $\epsilon$, we employ a continuation heuristics, increasing the nonlinearity in a series of increments (see Section~\ref{sub:choice}).
{\em We emphasize that this version of the method uses no  a~priori knowledge of
the solutions sought for.} For the experiments in this section, we use the 
compact fourth order scheme~(\ref{eq:FO_stencil}).

\begin{figure}[h]
	\begin{centering}
		\includegraphics[clip,scale=0.27]{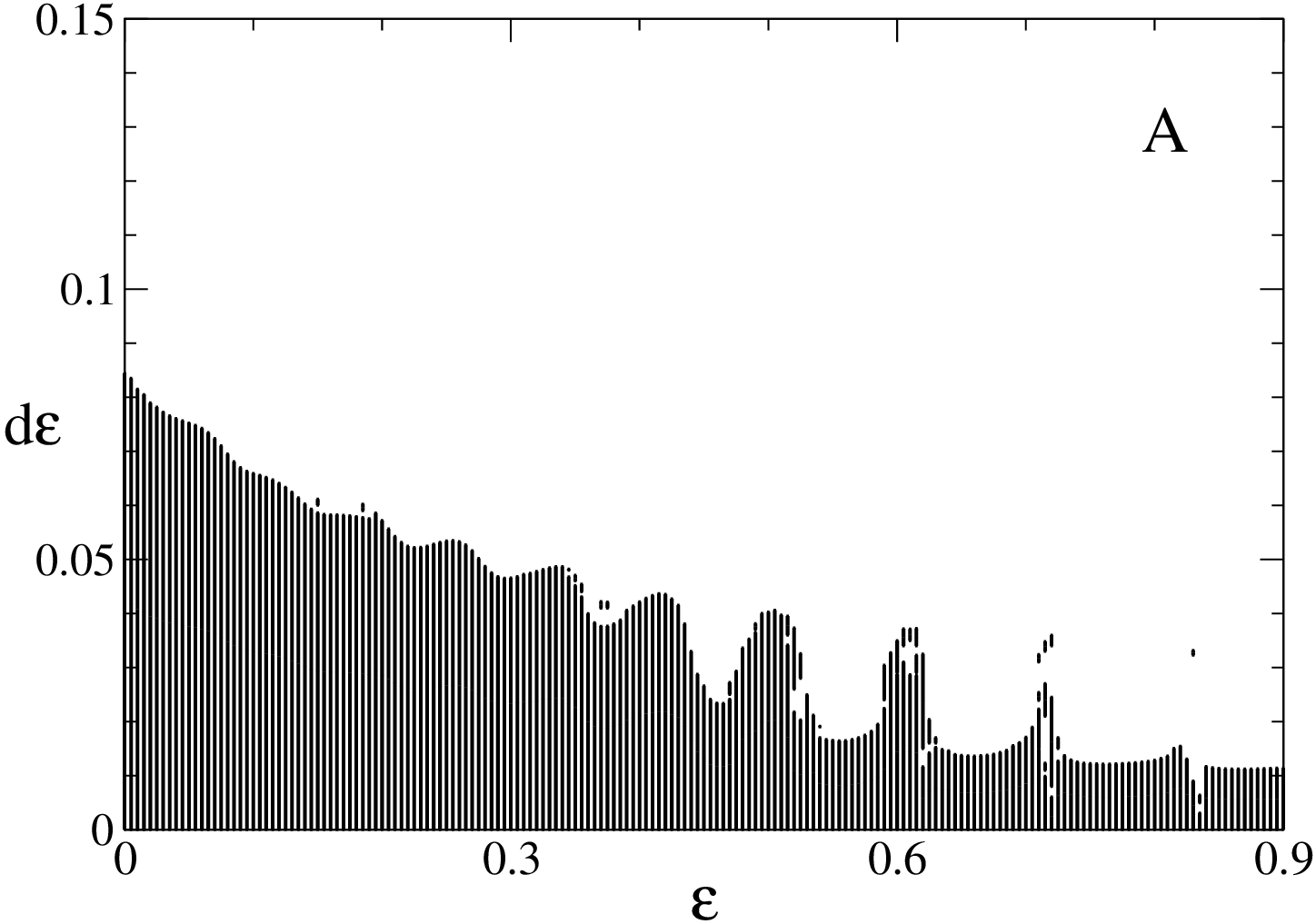}
		\includegraphics[clip,scale=0.27]{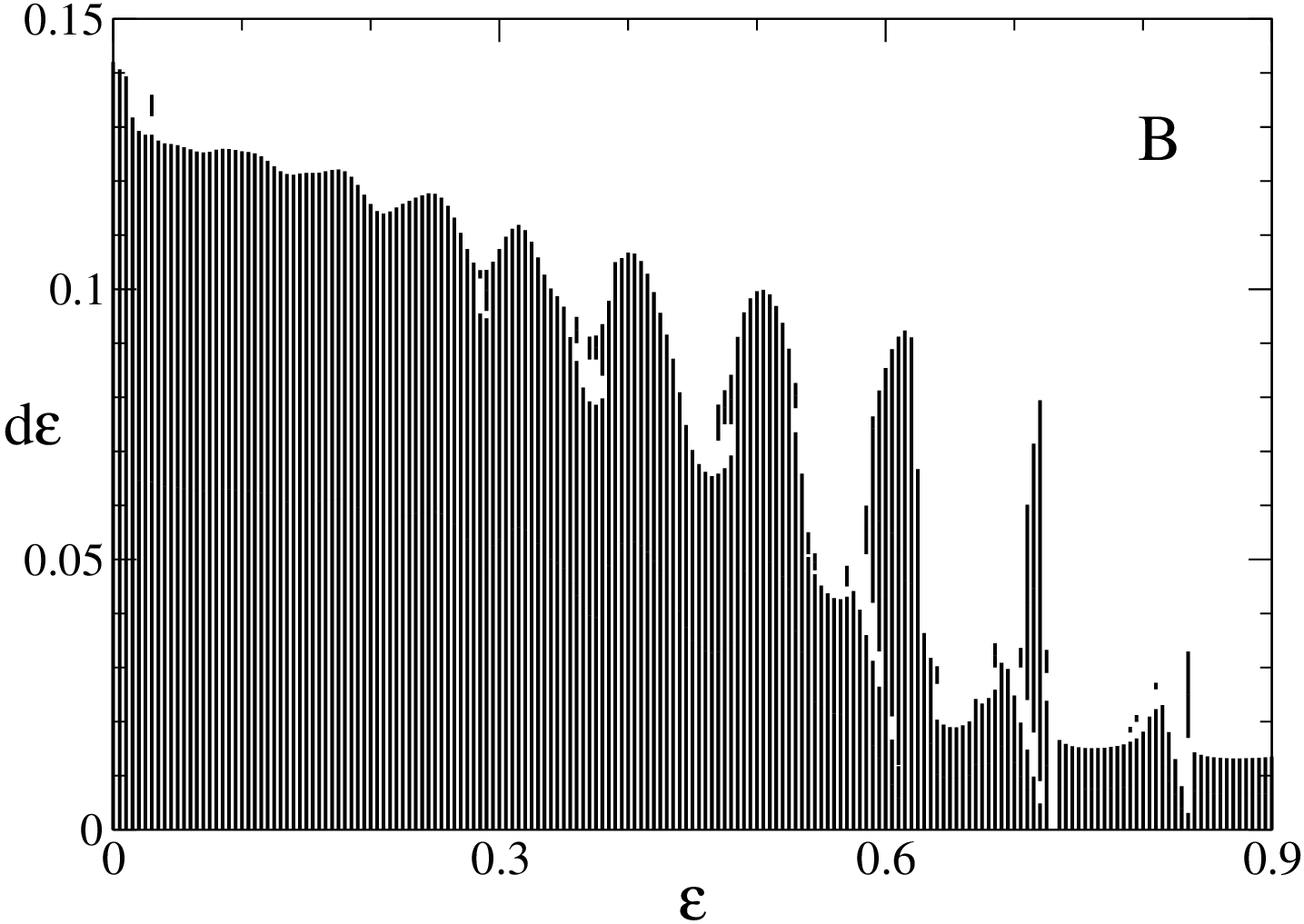}
	\caption{
		\label{fig:continuation-1}
		A) The allowable positive increments $d\epsilon = \epsilon_{\rm f}-\epsilon_{\rm i}$
		for which the continuation method works, i.e., for which Newton's method converges;
        $\lri\equiv1$, $k_0=8$ and $Z_{\max}=10$. 
		The iterations were defined as converged if the residual decreased 
		by a factor of $10^{6}$  in 20 iterations.
		B) Same as (A), for Newton's method with relaxation~(\ref{eq:newtons-SOR}), with $\omega=0.3$.
		The iterations were defined as converged if the residual decreased 
		by a factor of $10^{6}$ in 60 iterations.
	}
	\end{centering}
\end{figure}

\begin{figure}[h]
	\begin{centering}
		\includegraphics[clip,scale=0.27]{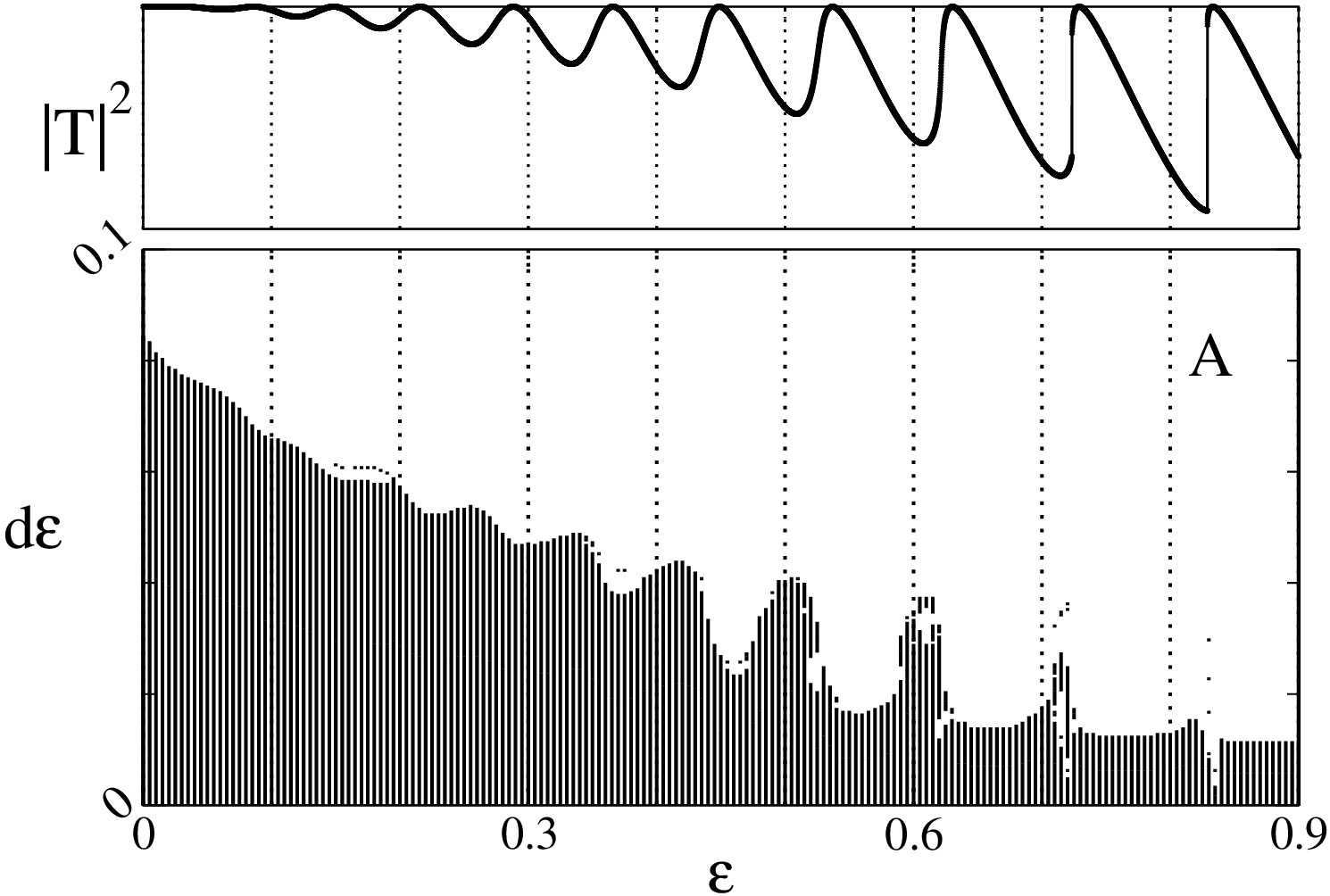}
		\includegraphics[clip,scale=0.27]{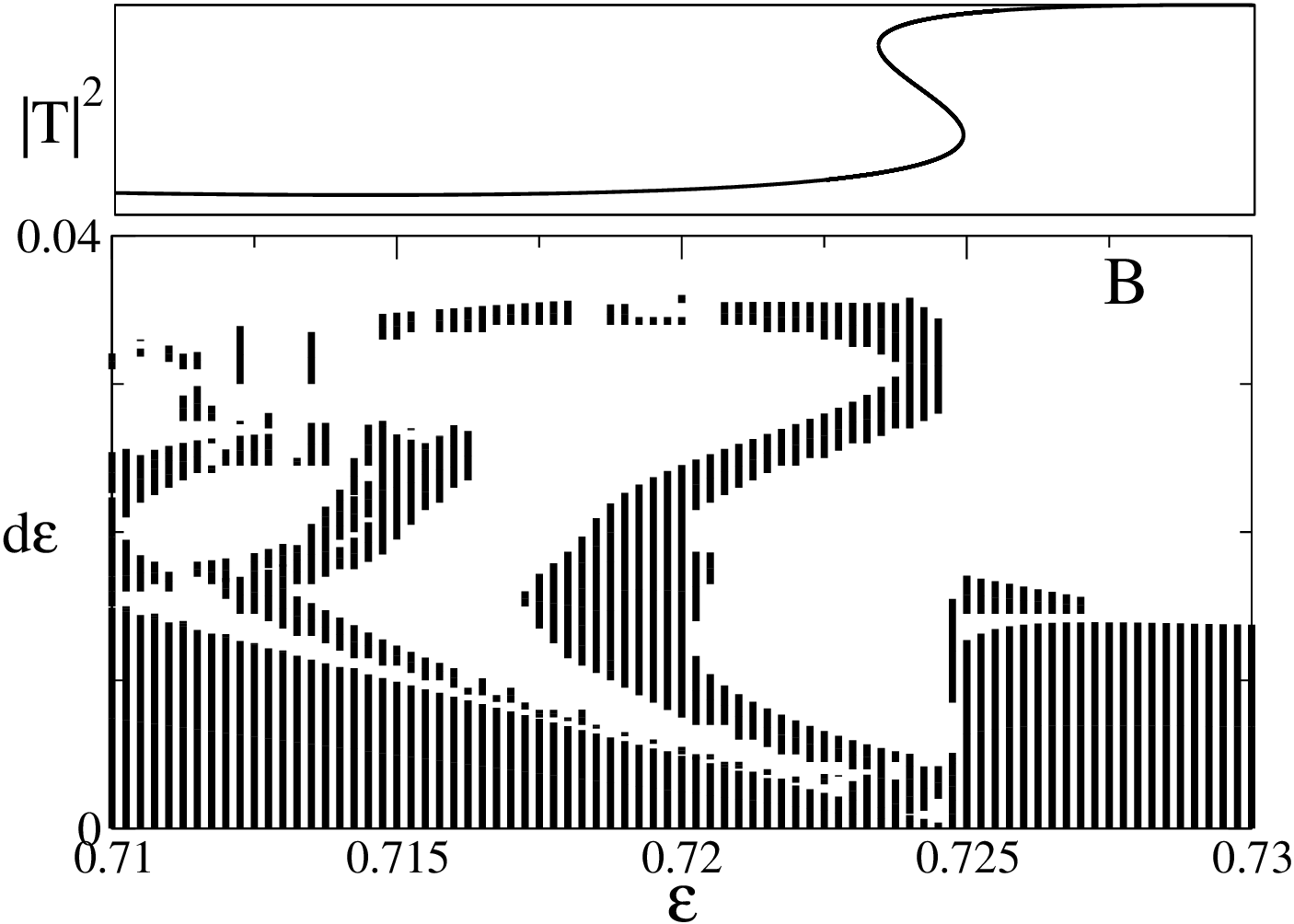}
	\caption{
		\label{fig:continuation-2}
		A) Same as (\ref{fig:continuation-1}A), plotted together with the transmittance $T(\epsilon)$ (see Figure~\ref{fig:transmittance}).
		B) Same as (A), zooming on the first switchback region
	}
	\end{centering}
\end{figure}

In order to quantify the performance of the continuation heuristic, we tested, 
for initial values of $\epsilon_{\rm i}$ in $[0,0.9]$, the ranges of allowable 
positive increments $d\epsilon=\epsilon_{\rm f}-\epsilon_{\rm i}$ 
for which Newton's method would still converge.\footnote{
We are primarily interested in
the positive increments because our key objective for employing the continuation strategy 
is to obtain suitable initial guesses for Newton's method when it is applied to high energy cases.
}
In other words, for each pair $(\epsilon_{\rm i},\,d\epsilon>0)$ 
we applied Newton's method for the NLH with nonlinearity $
	\epsilon_{\rm f}=\epsilon_{\rm i} + d\epsilon,
$ and with the initial guess $E^{(0)}$ given by the solution with $\epsilon=\epsilon_{\rm i}$.
The results are displayed in Figures~\ref{fig:continuation-1} and~\ref{fig:continuation-2}.
Several observations can be made.
First, at higher nonlinearities the allowable increments are generally smaller, see Figure~\ref{fig:continuation-1}(A).
Second, as can be seen by observing the adjacent transmittance graph in Figure~\ref{fig:continuation-2}(A), the allowable increments are highly correlated with the transmittance, which can be viewed as an indicator for the distance between the NLH solutions.
Specifically, at the first and second switchback regions, see Figure~\ref{fig:continuation-2}(B), the allowable values of $d\epsilon$ demonstrate a rather irregular behavior.
It is most important however, that the allowable values of $d\epsilon$ {\em do not decrease to zero}, so that the continuation strategy for Newton's method can traverse through the first and second switchback regions.

The previous tests were rerun with the relaxation version (\ref{eq:newtons-SOR}) of Newton's method.
The results are presented in Figure~\ref{fig:continuation-1}(B).
One can see that the relaxation considerably increases the allowable increments, especially at the switchback regions.

We next study the performance of Newton's method inside the nonuniqueness region.
When the initial guess was the solution slightly before the first switchback on the $T(\epsilon)$ curve, at point~A in Figure~\ref{fig:Newton-on-switchback}, and the value of $\epsilon$ was chosen within the nonuniqueness region: $\epsilon=0.724$, the method converged to the lower branch of the switchback, i.e., to the solution denoted by B in Figure~\ref{fig:Newton-on-switchback}.
Similarly, selecting the initial guess slightly past the switchback, 
at point~E in Figure~\ref{fig:Newton-on-switchback}, and again taking 
$\epsilon=0.724$ 
(negative increments are not shown in Figures~\ref{fig:continuation-1} and
\ref{fig:continuation-2}), 
facilitated convergence to the higher branch of the switchback curve, i.e., to the solution~D in Figure~\ref{fig:Newton-on-switchback}.
This behavior agrees with the standard notion of a hysteresis loop for a bistable device.


It is also important to note that the continuation strategy for Newton's 
method can ``hop over'' (at least) the first and second switchback regions,
which is an efficient way of reaching the regions of high nonlinearity. 
For example, a transition across the first switchback, from point A to point E in Figure~\ref{fig:Newton-on-switchback}, is possible by choosing the solution at point A as the initial guess for computing the solution with the value of $\epsilon$ that corresponds to E.
In this context we should mention that the first two nonuniqueness regions are rather narrow. 
For wider nonuniqueness regions that correspond to larger values of $\epsilon$, 
using a combination of the continuation in $\epsilon$ and relaxation can be beneficial.
Indeed, even though more iterations will be required for convergence with relaxation, larger allowable increments $d\epsilon=\epsilon_{\rm f}-\epsilon_{\rm i}$ will help traverse those wider regions of nonuniqueness, see  Figure~\ref{fig:continuation-1}(B).

\subsection{\label{sub:accuracy} Computational error}

\subsubsection{\label{subsub:res:accuracy}Homogeneous medium (discontinuities at the boundaries)}

In this section, we consider the case of a homogeneous Kerr medium (see formula (\ref{eq:homogeneous-material})). 
Hence, the discontinuities are only at $z=0$ and $z=Z_{\max}$.
The error of the solutions computed with the new schemes~(\ref{eq:SO_stencil}), (\ref{eq:alt_SO_stencil}), and~(\ref{eq:FO_stencil}), as well as the reference schemes~(\ref{eq:old_SO_stencil}) and~(\ref{eq:old_FO_stencil}), is reported in Table~\ref{tab:schemes-accuracy}.
All computations are done using Newton's solver.
The error is defined as the difference between the computed solution and the closed form Chen and Mills solution~\cite{chen-mills-PRB-87}, and is evaluated in the maximum ($l_\infty$) norm.

\begin{table}
\noindent \begin{centering}\begin{tabular}{|c|c||c|c|c|c|c||c|}
\hline 
$\lri$&
$\epsilon$&
\multicolumn{6}{c|}{$\hk\equiv hk_{0}$}\tabularnewline
\hline
\hline 
&
&
$8\cdot10^{-1}$&
$8\cdot10^{-1.5}$&
$8\cdot10^{-2}$&
$8\cdot10^{-2.5}$&
$8\cdot10^{-3}$&
error$\left(\hk\right)$\tabularnewline
\hline
\hline 
&
&
\multicolumn{6}{c|}{Standard centered-difference $\mathcal{O}\left(h^{2}\right)$ discretization
(\ref{eq:old_SO_stencil})}\tabularnewline
\hline 
$1.01^{2}$&
$0.01$&
-&
$0.230$&
$0.0228$&
$2.28\cdot10^{-3}$&
$2.28\cdot10^{-4}$&
$3.56\cdot\hk^{2}$\tabularnewline
\hline 
$1.3^{2}$&
$0.845$&
-&
-&
$0.16$&
$8.15\cdot10^{-3}$&
$8.26\cdot10^{-4}$&
$12.7\cdot\hk^{2}$\tabularnewline
\hline 
\multicolumn{1}{|c|}{}&
&
\multicolumn{6}{c|}{Standard centered-difference $\mathcal{O}\left(h^{4}\right)$ discretization
(\ref{eq:old_FO_stencil})}\tabularnewline
\hline 
$1.01^{2}$&
$0.01$&
$0.187$&
$2.01\cdot10^{-3}$&
$2.73\cdot10^{-5}$&
$9.97\cdot10^{-7}$&
$9.15\cdot10^{-8}$&
${\scriptstyle 0.45\cdot\hk^{4}\,+\,0.0024\cdot\hk^{2}}$\tabularnewline
\hline 
$1.3^{2}$&
$0.845$&
-&
$0.15$&
$0.093$&
$5.40\cdot10^{-4}$&
$5.38\cdot10^{-5}$&
${\scriptstyle 24\cdot\hk^{4}\,+\,0.84\cdot\hk^{2}}$\tabularnewline
\hline 
\multicolumn{1}{|c|}{}&
&
\multicolumn{6}{c|}{Finite-volume $\mathcal{O}\left(h^{2}\right)$ discretization (\ref{eq:SO_stencil}) }\tabularnewline
\hline 
$1.01^{2}$&
$0.01$&
-&
$0.107$&
$1.07\cdot10^{-2}$&
$1.07\cdot10^{-3}$&
$1.07\cdot10^{-4}$&
$1.68\cdot\hk^{2}$\tabularnewline
\hline 
$1.3^{2}$&
$0.845$&
-&
$6.82\cdot10^{-2}$&
$8.07\cdot10^{-3}$&
$8.03\cdot10^{-4}$&
$8.01\cdot10^{-5}$&
$1.26\cdot\hk^{2}$\tabularnewline
\hline 
\multicolumn{1}{|c|}{}&
&
\multicolumn{6}{c|}{Alternative finite-volume $\mathcal{O}\left(h^{2}\right)$ discretization
(\ref{eq:alt_SO_stencil}) }\tabularnewline
\hline 
$1.01^{2}$&
$0.01$&
-&
$0.109$&
$1.09\cdot10^{-2}$&
$1.09\cdot10^{-3}$&
$1.09\cdot10^{-4}$&
$1.71\cdot\hk^{2}$\tabularnewline
\hline 
$1.3^{2}$&
$0.845$&
-&
-&
$2.36\cdot10^{-2}$&
$2.01\cdot10^{-3}$&
$1.98\cdot10^{-4}$&
$3.72\cdot\hk^{2}$\tabularnewline
\hline 
\multicolumn{1}{|c|}{}&
&
\multicolumn{6}{c|}{Finite-volume $\mathcal{O}\left(h^{4}\right)$ discretization (\ref{eq:FO_stencil})}\tabularnewline
\hline 
$1.01^{2}$&
$0.01$&
$0.121$&
$1.29\cdot10^{-3}$&
$1.28\cdot10^{-5}$&
$1.28\cdot10^{-7}$&
$1.33\cdot10^{-9}$&
$0.314\cdot\hk^{4}$\tabularnewline
\hline 
$1.3^{2}$&
$0.845$&
-&
$8.16\cdot10^{-2}$&
$9.12\cdot10^{-5}$&
$9.13\cdot10^{-7}$&
$9.16\cdot10^{-9}$&
$2.23\cdot\hk^{4}$\tabularnewline
\hline
\end{tabular}\par\end{centering}

\caption{\label{tab:schemes-accuracy}Error for the 3 schemes of Section~\ref{sec:discrete-approx}
and 2 reference schemes of Section~\ref{sub:ref-methods}; $\left.Z_{\max}=10\right.,\left.k_{0}=8\right.$.
The entries are empty for cases wherein Newton's iteration diverged. }
\end{table}

The discrete approximations of Section~\ref{sec:discrete-approx} are designed to retain their order of accuracy in the presence of material discontinuities.
In order to test this, for each scheme we consider two cases, see Table~\ref{tab:schemes-accuracy}.
The first case, $\lri=1.01^2,\ \epsilon=0.01$, corresponds to a small discontinuity at the boundary and a weak nonlinearity.
Note that the quantities $\sqrt{\lri}-1=0.01$ and $\epsilon$ characterize the difference between the linear and nonlinear indices of refraction inside and outside the medium, see Section~\ref{sec:backgr}.
The second case, $\lri=1.3^2,\ \epsilon=0.845$, corresponds to a large discontinuity at the boundary and an ${\cal O}(1)$ nonlinearity ($\epsilon/\lri=0.5$).

In the first case, $\lri=1.01^2,\ \epsilon=0.01$, the computations can also
be repeated using the original iterative solver~(\ref{eq:NL-iterations}), because for this choice of parameters it still converges.
Having done that, we determined that the accuracy of the corresponding solution was the same as the accuracy of the solution obtained using Newton's method.
This indicates that the errors  presented in Table~\ref{tab:schemes-accuracy} are indeed the approximation errors of the discrete schemes and should not be attributed to the solver.
For the case with  higher nonlinearity, $\epsilon=0.845$, only Newton's iterations converged.

The functional dependence of the error on the dimensionless grid resolution 
$
	\left. \hk	=	k_0h	=	\frac{k_0 Z_{\max}} M	\right.
$ is shown in the rightmost column of Table~\ref{tab:schemes-accuracy}.
It is obtained by a weighted least squares fit.
Considering the reference methods of Section~\ref{sub:ref-methods}, the three-point central-difference approximation~(\ref{eq:old_SO_stencil}) displays a second order convergence.
The five-point central-difference approximation~(\ref{eq:old_FO_stencil}), however, displays a fourth order convergence for (relatively) low grid resolutions and small discontinuities.
For high grid resolutions and large discontinuities, however, its accuracy deteriorates and shows a second order convergence.
The limited ability of the reference methods to handle discontinuities is also reflected by the fact that the actual errors, and hence the coefficients in front of $\hk^2$ and $\hk^4$ increase substantially for larger discontinuities. 
As mentioned in Section~\ref{sec:backgr}, the five-node discretization~(\ref{eq:old_FO_stencil}) is particularly sensitive to the presence of discontinuities.

On the other hand, the errors of the new second order discretizations~(\ref{eq:SO_stencil}) and~(\ref{eq:alt_SO_stencil}), as well as that of the fourth order scheme 
(\ref{eq:FO_stencil}), are hardly affected by the increase of the discontinuity. 
Indeed, for larger discontinuities at the boundary, the improvement over~(\ref{eq:old_SO_stencil}) ranges from a factor of 4 for~(\ref{eq:alt_SO_stencil}) to a factor of 10 for (\ref{eq:SO_stencil}). The improvement of (\ref{eq:FO_stencil}) over (\ref{eq:old_FO_stencil})
is even more substantial.

We also see that~(\ref{eq:SO_stencil}) yields better accuracy (smaller errors) than (\ref{eq:alt_SO_stencil}). 
Indeed, intuitively one can expect that the integration of the interpolation of $|E|^2E$ will approximate $\int |E|^2Edz$ better than the integration of the interpolation of $E$ cubed.
However, as we do not have $\frac{d(|E|^2E)}{dz}$ or $\frac{d^2(|E|^2E)}{dz^2}$, the discretization~(\ref{eq:SO_stencil}) cannot be extended to fourth order accuracy.
There, perhaps, could be other approaches, such as the interpolation of the
amplitude and phase of $E$. They, however, do not provide an obvious venue to the fourth
order either.

Regarding the new fourth order discretization~(\ref{eq:FO_stencil}), we can see from Table~\ref{tab:schemes-accuracy} that it is indeed fourth order accurate for both small and large material discontinuities.
A minor increase of the error for larger $\epsilon$ can be observed, which is natural to expect for solutions with sharper variations.
Altogether, for the cases reported in Table~\ref{tab:schemes-accuracy}
scheme~(\ref{eq:FO_stencil}) has proven 
up to 6000 times more accurate than the standard five-node
central-difference scheme~(\ref{eq:old_FO_stencil}).

\begin{figure}[H]
\begin{center}
\includegraphics[clip=,width=4in]{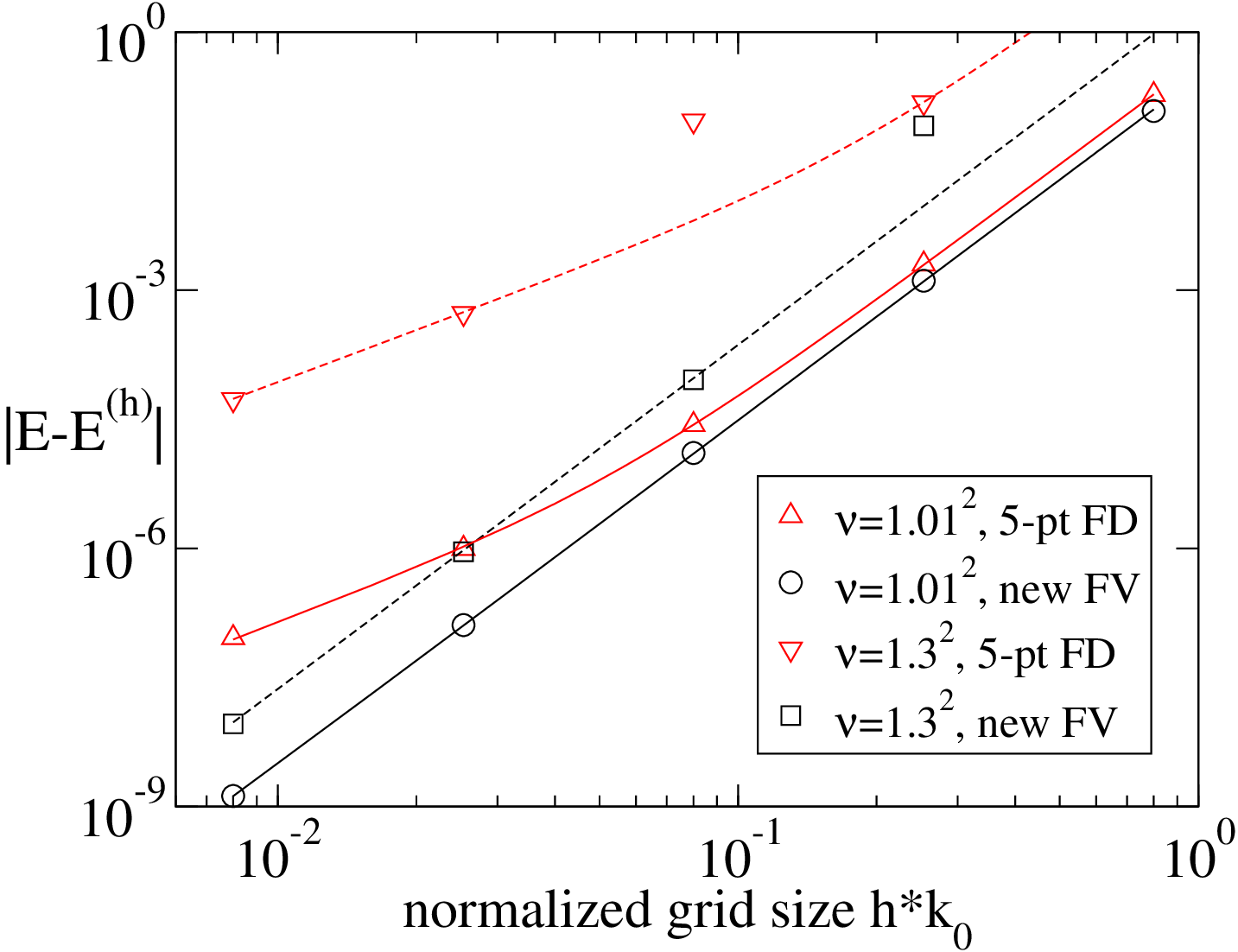}
\caption{Computational error as a function of the grid size for schemes
(\ref{eq:FO_stencil}) [labeled
``new FV''] and (\ref{eq:old_FO_stencil}) [labeled ``5-pt FD''].}
\label{log-log}
\end{center}
\end{figure}

To provide a more descriptive and more intuitive account of our grid convergence results, we
present a log--log plot of the error as it depends on the grid size for our fourth order schemes,
see Figure~\ref{log-log}. 
The data used for Figure~\ref{log-log} are the same as those in Table~\ref{tab:schemes-accuracy}. Once can
clearly see that the accuracy of the original fourth order scheme of \cite{FT:01,FT:04} deteriorates on
fine meshes, whereas the new scheme (\ref{eq:FO_stencil}) maintains its fourth order.

Finally, we compare the minimum grid resolutions required by different schemes (second and fourth order) to distinguish between the solutions inside the region of nonuniqueness (points B, C, and D in Figure~\ref{fig:Newton-on-switchback}) and thus enable convergence of Newton's iterations. 
As could be expected, the fourth order scheme used for computations of Section~\ref{subsub:glob_conv} took roughly ten times fewer points per wavelength than the second order scheme used for computations of Section~\ref{subsub:loc_conv}.

\subsubsection{Layered medium}

Here, we apply Newton's method along with the fourth order scheme~(\ref{eq:FO_stencil}) 
to solve the NLH for a piecewise-constant material.
The configuration is that of a two-layer Kerr slab:
\begin{equation}
	\label{eq:two-layered-config}
	\lri(z) = 
		\begin{cases}
			1.21 & z\in[0,5) \\
			1.69, & z\in(5,10]
		\end{cases},
	\qquad
	\epsilon(z) = 
		\begin{cases}
			0.1210 & z\in[0,5) \\
			0.5070, & z\in(5,10]
		\end{cases}.
\end{equation}
The material coefficients are therefore discontinuous at $z=5$, as well as at the boundaries $z=0$ and $z=10$.
The value of the linear wavenumber is $k_0=8$. 

The computed solution is compared with the closed form solutions obtained by Chen and Mills in \cite{chen-mills-PRB-87-ML}.
The results are given in Table~\ref{tab:schemes-accuracy-BL}; they corroborate the designed fourth order accuracy of the method.

\begin{table}[h]
\noindent \begin{centering}\begin{tabular}{|c|c|c|c|c||c|}
\hline 
\multicolumn{6}{|c|}{$\hk\equiv hk_{0}$}\tabularnewline
\hline
\hline 
$4\cdot10^{-1}$&
$4\cdot10^{-1.5}$&
$4\cdot10^{-2}$&
$4\cdot10^{-2.5}$&
$4\cdot10^{-3}$&
error$\left(\hk\right)$\tabularnewline
\hline
\hline 
$3.70\cdot10^{-2}$&
$3.72\cdot10^{-4}$&
$3.69\cdot10^{-6}$&
$3.69\cdot10^{-8}$&
$3.93\cdot10^{-10}$&
$1.42\cdot\hk^{4}$\tabularnewline
\hline
\end{tabular}\par\end{centering}

\caption{\label{tab:schemes-accuracy-BL}Error for the two-layered configuration
(\ref{eq:two-layered-config}). The finite-volume $\mathcal{O}\left(h^{4}\right)$
discretization (\ref{eq:FO_stencil}) was used in combination with
Newton's method for $\left.k_{0}=8\right.$. }
\end{table}

\subsection{Computational efficiency}
\label{sec:eff}

Having addressed the issues of convergence and accuracy, we would also like to comment on the
numerical efficiency of our method.

\begin{table}[H]
\noindent \begin{centering}
\begin{tabular}{|c||c|c|c|c|c|}
\hline 
Grid dimension $M$ & $10^{2}$ & $10^{2.5}$ & $10^{3}$ & $10^{3.5}$ & $10^{4}$\tabularnewline
\hline
\hline 
CPU time (sec) & $0.105$ & $0.333$ & $1.06$ & $3.39$ & $11.1$\tabularnewline
\hline
\end{tabular}
\par\end{centering}
\caption{\label{tab:times}Mean CPU times for a single Newton's iteration
of the finite volume scheme (\ref{eq:FO_stencil}) on
AMD Athlon64 at 2200MHz in Matlab 7.3.0 under
Linux.}
\end{table}

The CPU times for one Newton's iteration summarized in Table~\ref{tab:times} clearly indicate
that the complexity scales linearly as a function of the grid dimension. Moreover, as the
number of Newton's iterations required to obtain the solution typically does not depend
on the grid dimension (see Section~\ref{sub:Newtons-convergence}), we can say that
the overall complexity of the proposed method also depends linearly on the grid.
Of course, it is natural to expect that the methods based on shooting
\cite{chen-mills-PRB-87,chen-mills-PRB-87-ML,%
baghdasaryan-99,midrio-01,kwan-lu-04,petracek-06} will perform faster for a 
one-dimensional problem than our method that involves a full fledged
approximation of the boundary value problem for the NLH. The shooting-based methods, however,
will not generalize to multiple space dimensions.

\section{\label{sec:discussion} Discussion and future work}

\subsection{Discussion}
\label{sub:disc}

In this study, we approximated  the one-dimensional NLH using new compact finite volume 
schemes of orders four and two (the latter predominantly for reference purposes). 
The fourth order approximation of nonlinear terms was the most challenging part, 
and it required the use of Birkhoff-Hermite interpolation (see Lemma~\ref{lem:cubic-interp} 
and Appendix~\ref{app:interp_accuracy}).  
For actual implementation, the automation of the  computation of the coefficients of the scheme was crucial (see Appendix~\ref{app:SE}). 
Note that as we have interpolated the field using third degree polynomials, the cubic nonlinearity
inside every cell was represented by the polynomials of a relatively high degree --- degree 9. 
We, however, have not seen any adverse implications of that in our simulations.

Let us also mention that the piecewise interpolating polynomial of
the Birkhoff-Hermite type\footnote{This is
a function on $[0,Z_{\max}]$ that on every
 grid cell coincides with the corresponding cubic 
polynomial obtained by Lemma~\ref{lem:cubic-interp}.}
is not, generally speaking, 
equivalent to the standard Schoenberg cubic spline
(see, e.g., \cite[Section 2.3.2]{mybook}).
Indeed, the Schoenberg spline takes only the nodal values of the interpolated function
as input, and is built so that its first and second derivatives are
continuous at the nodes. In doing so, the equations for the coefficients of the
spline become coupled across the entire grid. In contradistinction to that, all
individual polynomials of the Birkhoff-Hermite interpolation are built
independently of one another inside their respective cells. Moreover, the
nodal values of the second derivative are required as input
in addition to the nodal values of the function. In doing so, even if the first
and second derivatives of the interpolated function are continuous everywhere,
the first derivative of the interpolating polynomial may be discontinuous at the nodes.
However, the mismatch may not exceed ${\cal O}(h^3)$,
because inside every cell the first derivative
is approximated with third order accuracy.

Note also that as an alternative to the integral formulation (Section~\ref{sub:FV_formulation})
and the finite volume scheme built uniformly across the entire domain,
one could have used compact finite differences on the regions of smoothness coupled with the condition
of continuity of the first derivative at the interfaces. The latter can be built into the scheme,
say, via one-sided
differences. This approach, however, is not equivalent to ours and may, in our opinion, come
short at least along the following two lines. First, the uniformity of the approach will be lost
--- another discontinuity introduced in the domain will require special treatment and hence the scheme
will have to be rebuilt. Second, compactness of the approximation will be compromised because of
the long one-sided stencils near the discontinuities and as such, the resulting matrix will have
a higher bandwidth.

The non-reflecting two-way artificial boundary conditions (Section~\ref{sub:BCs}) were designed similarly to our previous work \cite{FT:01,FT:04,BFT:05}; they are based on the analysis of the waves governed by the discrete equation.
The boundary conditions prescribe the impinging wave that drives the problem and at the same time enable the propagation of all the outgoing waves. 
The important difference compared to \cite{FT:01,FT:04} is, however, that for a compact scheme, even fourth order accurate, it is sufficient to consider only one ghost node outside the computational domain, whereas for the five-point central-difference approximation (\ref{eq:old_FO_stencil}) we had to introduce two ghost nodes.
Indeed, for the linear homogeneous five-point discretization, an additional evanescent
mode always exists, which needs to be handled with care, see \cite{FT:01,FT:04}.
For the compact three point discretization, however, no such mode exists, and the construction of the boundary-condition is greatly simplified.
In both cases, the additional assumption that we used when calculating the value of
the solution at the ghost nodes is that outside the domain of interest the field is governed by the linear constant coefficient Helmholtz equation.

The analysis in the paper establishes 
the formal accuracy of our schemes (i.e., it is the analysis of consistency).
We do not, however, derive any rigorous error estimates because the problem is nonlinear. 
Instead, we study the computational error experimentally (see Section~\ref{sub:accuracy}). 
By comparing our numerical solutions with the closed form solutions of \cite{chen-mills-PRB-87,chen-mills-PRB-87-ML}, we have been able to demonstrate that in all the cases our schemes possess the design rate of grid convergence.
Besides, we provide a convergence proof for a linear layered medium (see Appendix~\ref{app:linear_interface}).

Our nonlinear solver for the discretized NLH exploits Newton's iterations.
However, the nonlinearity in equation (\ref{eq:1DNLH}) is nondifferentiable in the sense of Frech\'et for complex-valued solutions $E$.
Therefore, we present a convenient mechanism for transforming  the nonlinear systems of equations to the representation in real variables, which enables Newton's linearization.
The results fully justify this additional effort. 
Indeed, Newton's iterations allow us to solve the NLH for very high nonlinearities, addressing the full range of nonlinearities interesting from the standpoint of physics, and beyond, to the level of the actual material breakdown.
Moreover, even though Newton's method has been applied to problems with Kerr 
nonlinearity previously \cite{gomez-04}, our current implementation is particularly well suited
for the one-dimensional NLH as it yields block tridiagonal Jacobians.

We now compare our current work with other studies available in the literature on the numerical
solution of boundary value problems for the NLH:
our previous work~\cite{FT:01,FT:04,BFT:05}, and Suryanto \etal~\cite{Suryanto:2002,Suryanto:2003}.
In terms of discrete approximations, these previous studies did not guarantee a fourth order approximation for materials with discontinuities. We have shown this directly for the discretization of our work~\cite{FT:01,FT:04,BFT:05} (Section~\ref{subsub:res:accuracy}), 
while for Suryanto's finite element discretization which accounts for discontinuities, the nonlinearity is approximated only to the second order.
We again emphasize that to the best of our knowledge, the current method is the {\em first ever high-order approximation of the NLH with material discontinuities}.
The additional improvement is due to the Newton's solver.
All iterative schemes used previously were based on freezing the nonlinearity
\cite{FT:01,FT:04,BFT:05,Suryanto:2002,Suryanto:2003}.
As we have seen, this freezing approach cannot be used beyond a certain $\epsilon$ threshold, unrelated to the uniqueness of the solutions.
We note that the freezing approach was used by Suryanto \etal\ to solve the NLH for the cases when the solution is not unique.
They report, however, that their setup was that of a highly-grated material with a defect, and that often for such setups the threshold for non-uniqueness is much lower (in fact,  lowering of the threshold was one of the goals in \cite{Suryanto:2002,Suryanto:2003}).
This is in agreement with our own observations: The freezing approach fails at a certain nonlinearity threshold unrelated to the solution uniqueness.
Therefore, the results show that Newton's method, compared with the commonly used freezing approach, allows for the much high levels of nonlinearity.
Apparently, {\em this is the first numerical method for the NLH that works at such high nonlinearities.} 
To summarize, compared to \cite{FT:01,FT:04,BFT:05,Suryanto:2002,Suryanto:2003}, the approach of the current paper enables efficient discrete approximation for a problem with material discontinuities and allows solution for high levels of nonlinearity. 
We expect that it will provide a basis for the future extension to the case of multiple space dimensions (see Section~\ref{sub:future}).

Let us also mention a few additional studies that have something in common but are not as close to the current work.
In \cite{mckenna-93}, Choi and McKenna analyzed a  somewhat different equation:
$\Delta u + u^3 = 0$. They could employ the mountain pass ideas 
because the boundary condition was homogeneous Dirichlet and hence $u\equiv0$ was a solution. 
This approach, however, will not
apply to the NLH, which is normally to be driven by a given incoming wave at the boundary.
 
In yet another series of papers, Kriegsmann and Morawetz solve a linear Helmholtz equation with variable coefficients \cite{KM-80} and a focusing NLH \cite{KM-81} in two space dimensions, and then Bayliss, Kriegsmann and Morawetz consider a defocusing NLH \cite{BKM-83}. 
They employ second order approximations, and the solver is based on integration in real time (i.e., using the wave equation) and applying the principle of limiting amplitude.
The problem they solved is very different though, so at the moment we cannot compare their method with ours.

\subsection{Possible future extensions}
\label{sub:future}

The method can be extended to the case of a quintic nonlinearity, $\sigma=2$.
This will involve evaluation of the fifth order tensor coefficients [cf.\ formula (\ref{eq:FO_tensor_elements})]: 
\begin{equation*}
	g_{ijklm} = \int F_iF_j F_k F_l F_m d\zeta.
\end{equation*}
This is a straightforward, though tedious extension, 
for which the automatic generation of tensor elements will be a necessity.

The method can also be extended to the case of piecewise smooth material coefficients $\lri(z)$ and $\epsilon(z)$, as opposed to only piecewise constant coefficients that we have analyzed in the paper.
Approximating the quantities $\nu(z)$ and $\epsilon(z)$ by cubic polynomials within each grid cell: \[
	\lri(z_m \pm \zeta h)= \sum_{j=0}^3c_j^\pm \zeta^j
	, \qquad
	\epsilon(z_m \pm \zeta h)= \sum_{k=0}^3d_k^\pm \zeta^k,
\]
one can then substitute these approximations into formulae~(\ref{eq:1S_2nd_derivatives_1DNLH}) for one-sided second derivatives, and into the definitions~(\ref{eq:FO_tensor_elements}) for the coefficients $f_i$ and $g_{ijk}$. 

Likewise, linear and nonlinear absorption can be modeled by allowing the material coefficients to become 
complex. 
Note, however, that in this case the tensor elements $g_{ijk}$ will also become complex,
\[ 	g_{ijk} = \int F_i^*F_jF_k d\zeta, \]
and will lose their symmetry with respect to the indices $i,j,k$ .

Furthermore, for the {\em linear} Helmholtz equation, which corresponds to the case $\epsilon\equiv 0$ in this paper, the scheme we have used to approximate (\ref{eq:VI-integrals}) to fourth order accuracy can be extended to arbitrarily high orders at virtually no computational cost.
For example, using the first three even (one-sided) derivatives at $z_m$: $
	\lbrace
		E_m,
		E_{m+}'',
		E_{m+}^{(4)}
	\rbrace
$ and at $z_{(m+1)-}$: $
	\lbrace
		E_{m+1},
		E_{(m+1)-}'',
		E_{(m+1)-}^{(4)}
	\rbrace
$, one can construct the Birkhoff-Hermite quintic polynomial: \[
	P_5\left(
	\zeta; \;
		E_m,
		E_{m+}'',
		E_{m+}^{(4)},
		E_{m+1},
		E_{(m+1)-}'',
		E_{(m+1)-}^{(4)}
	\right),
\] such that \[
	\left.E(z_m+\zeta h) = P_5(\zeta) + \oh{6} \right.,
\] and then use it to approximate the integrals in (\ref{eq:VI-integrals}).
The values of the one-sided derivatives are again obtained from the equation: $
	\left. E^{(4)}=-k_0^2\lri E''=k_0^4\lri^2E \right.
$, etc. 
Note that this extension cannot be used for the NLH.

			From the standpoint of physics, a very useful extension could be that of considering the vectorial NLH, when no assumption of the linear polarization of the field is made.
Building boundary conditions for this case may require special care.

On the numerical side, to improve the quality of approximation a nonuniform 
grid can, in principle, be used that would be  better suited for resolving sharp variations of the solution.
In general, however, the structure of the solution is not known ahead of time, and therefore, a methodology of this type can only be adaptive.

Improvements can also be introduced aimed at reducing the CPU time for the method proposed in this paper.
For example, the summation $\sum g_{ijk}$ was performed in Sections~\ref{sub:FO_approx} and \ref{sub:stencil-variation} without using the symmetry of the tensor $g_{ijk}$, see Table~\ref{tab:compact-FO-coefs}. 
Taking it into account could decrease the cost of constructing the Jacobians.
We believe, however, that it can only benefit the one-dimensional problem because in 2D the overall cost will most likely be dominated by the inversion of the Jacobian.

The extension of utmost interest to us, from the standpoint of both theory and applications, is to the multidimensional case, see equation~(\ref{eq:NLH}).
It is well known that under the paraxial approximation, the NLH reduces to the nonlinear Schr\"odinger equation, which possesses singular solutions.
Therefore, the question of global existence for the solutions of the NLH 
in similar configurations is of a substantial 
mathematical and physical interest (see \cite{FT:01,FT:04,BFT:05} and the bibliography 
there for more detail).
Currently, the only analytical result in this regard is due to Sever~\cite{sever-06}, 
who proved existence for the NLH with real Robin boundary conditions.
However, the radiation boundary conditions which model the physical problem do not lead to
linearized self-adjoint formulations. Hence, the question of global existence in this case
remains outstanding.

We re-emphasize that none of the shooting-type methods \cite{chen-mills-PRB-87,chen-mills-PRB-87-ML,%
baghdasaryan-99,midrio-01,kwan-lu-04,petracek-06} that are apparently faster than ours in 1D can
be generalized to multiple space dimensions.
Hence, the only viable option in multi-D is to approximate on the grid and solve numerically
the boundary value problem for the NLH. 
Construction of a compact finite volume discretization in multi-D is possible, although it will not be an automatic generalization of what has been done in the 1D case.
The use of Newton's method will be of foremost importance, because  as we have seen, a simpler 
iteration scheme has severe convergence limitations. Hence, the key contribution to the overall computational
cost in multi-D will be from the inversion of the Jacobians  --- large, sparse, non-Hermitian matrices.
The use of direct solvers does not seem feasible for those dimensions that will provide
a sufficiently fine grid resolution. The only viable alternative is the preconditioned Krylov subspace
iterations, and as such, finding a good preconditioner will be in the focus of the study.
Besides, convergence of Newton's iterations slows down if two solutions in the region of
nonuniqueness are close to one another, such as near the switchback points in 
Figure~\ref{fig:Newton-on-switchback}. In the multidimensional case, however, we do not know the structure of
the NLH solutions ahead of time. Thus, numerical experiments will play a key role for
fine-tuning the method.

\appendix

\section{\label{app:E_z-continuity} Continuity conditions at material interfaces}

For a linearly polarized plane wave that impinges normally on the interface
$z={\rm const}$, we may assume without loss of generality that the electromagnetic
field has the form: 
\[
	\bvec{E} = \left[E_1, 0, 0\right], 
	\qquad 
	\bvec{H} = \left[0, H_2, 0\right].
\]
The tangential component of the electric field must be continuous across the
interface, see, e.g., \cite{land8,Jackson-98}. In our case, this implies the 
continuity
of $E_1\equiv E_1(z)$. The same is also true for the tangential component of the
magnetic field $H_2\equiv H_2(z)$, 
which we do not consider explicitly in the current framework.
The continuity of $H_2$, however, allows us to establish another important condition
for $E_1$. The time-harmonic form of the Faraday's law (a part of the 
Maxwell system of equations) reads:
\[
	-\,\frac{i\omega\mu}{c}\bvec{H}={\rm curl}\bvec{E},
\] and taking into account that $E_3\equiv0$ we have:
\begin{equation*}
	-\,\frac{i\omega\mu}{c}H_2
		= \frac{ \partial E_1 }{ \partial z } 
		- \frac{ \partial E_3 }{ \partial x	}
		= \frac{ \partial E_1 }{ \partial z }.
\end{equation*}
Then, disregarding all possible magnetization effects, 
i.e., assuming that the magnetic permeability is equal to 1 
(which is certainly legitimate for optical frequencies), 
we obtain that the first derivative of the electric field $
	\frac{\partial E_1}{\partial z}\equiv \frac{d E(z)}{d z}
$ is also continuous across any interface $z={\rm const}$, and hence everywhere.

\section{\label{app:linear_interface} Error estimate in the linear case}

For a linear medium ($\epsilon\equiv0$) and piecewise constant refraction index $\lri(z)$, we will show that the fourth order scheme (\ref{eq:FO_stencil}) indeed converges with the design rate of ${\cal O}(h^4)$ as $h\longrightarrow0$. 
Let $\lri\left(z\right)$ be a step function: 
\begin{equation*}
	\lri\left(z\right)= 
	\begin{cases}
		\lri_{\lf}=1, & z<0\\
		\lri_{\rt}\not=1, & z>0
	\end{cases}	,\qquad 
	z\in\left[-Z_{\max},Z_{\max}\right].
\end{equation*}
Let the solution be driven by the impinging wave $\Einc=e^{i\sqrt{\lri_\lf}k_0z}
\equiv e^{ik_0z}$,
and let it satisfy the boundary conditions [cf.\ formulae (\ref{eq:CM_TWBCs})]:
\begin{equation*}
	\left.\left(ik_0+\frac{d}{dz}\right)E\right|_{z=-Z_{\max}}
		= 2ik_0 , \qquad 
	\left.\left(ik_1-\frac{d}{dz}\right)E\right|_{z=Z_{\max}}
		=0,
\end{equation*}
where $k_{1}=\sqrt{\lri_{\rt}}k_{0}$. 

Then, the continuous solution of the problem is 
\begin{equation}
	E(z) =  
	\begin{cases}
		e^{ik_0z}+Re^{-ik_0z}, & z<0,\\
		Te^{ik_1z}, & z>0,
	\end{cases} 
	\label{eq:cont_sol}
\end{equation}
where the transmission and reflection coefficients are given by
\begin{equation*}
	T=\frac{2}{ 1+\sqrt{ \lri_\rt/\lri_\lf} }
	\quad\text{and}\quad 
	R=\frac{
		1-\sqrt{ \lri_\rt/\lri_\lf }
	} {
		1+\sqrt{ \lri_\rt/\lri_{\lf} }
	}.
\end{equation*}

On the uniform grid $z_m=mh$, $m=0,\pm1,\pm2,\ldots$, we define:
\begin{equation*}
	\lri_m=
		\begin{cases} 
			\lri_\lf, & m<0,\\
			\lri_\rt, & m\geq 0. 
		\end{cases}
\end{equation*}
The fourth order discretization~(\ref{eq:FO_stencil}) then reduces to 
\begin{equation}
		L_1(\lri_{m-1})E_{m-1}
		-(L_0(\lri_{m-1})+L_0(\lri_m))E_m
		+L_1(\lri_m)E_{m+1}=0,\label{eq:disc}
\end{equation}
where  
\begin{equation*}
	L_0(\lri) = 
		\hk^{-2}-\frac{1}{3}\lri-\frac{3}{128}\lri^2\hk^2
	\quad \text{and}\quad
	L_1(\lri) = 
		\hk^{-2}+\frac{1}{6}\lri+\frac{7}{384}\lri^2\hk^2,
\end{equation*}
and $\hk=hk_0$.
For $m<0$ and, independently, for $m>0$, the fundamental set of solutions of the difference equation~(\ref{eq:disc}) is $\lbrace q_\lri^m, q_\lri^{-m}\rbrace$, where $q_\lri$ and $q_\lri^{-1}$ are roots of the characteristic equation $L_{1}(\lri)q-2L_{0}(\lri)+L_{1}(\lri)q^{-1}=0$, and $\lri=\lri_\lf$ or $\lri_\rt$, respectively. 
The roots are given by the following expressions:
\begin{equation*}
	q_\lri=L_0/L_1+i\sqrt{1-(L_0/L_1)^{2}}, \qquad 
	q_\lri^{-1}=q_\lri^*.
\end{equation*}
The Taylor expansion yields: $
	\left.
		q_\lri = e^{i\sqrt{\lri}\hk} + \mathcal{O}(\hk^5),
	\right.
$ which means that $q_\lri^m$ approximates the right-traveling 
wave~$e^{i\sqrt{\lri}k_0z_m}\equiv e^{i\sqrt{\lri}k_0hm}$, 
while its conjugate $q_\lri^{-m}$ 
approximates the left-traveling 
wave~$e^{-i\sqrt{\lri}k_0z_m}\equiv e^{-i\sqrt{\lri}k_0hm}$, with fourth order
accuracy on any finite interval of the independent variable $z$:
\begin{equation}
\label{eq:wave_appr}
\begin{aligned}
\max_m|q_\lri^m-e^{i\sqrt{\nu}k_0z_m}|\leq &\> {\rm const}\cdot h^4,\\
\max_m|q_\lri^{-m}-e^{-i\sqrt{\nu}k_0z_m}|\leq &\> {\rm const}\cdot h^4.
\end{aligned}
\end{equation}

Similarly to (\ref{eq:cont_sol}), the discrete solution of the problem is constructed in the form:
\begin{equation}
	E_m = 
		\begin{cases}
			q_\lf^m+R^{(h)}q_\lf^{-m}, & m\leq0,\\
			T^{(h)}q_\rt^{m}, & m\geq0,
		\end{cases}
	\label{eq:disc_sol}
\end{equation}
where $q_\lf\stackrel{\rm def}{=}q_\lri$ for $\lri=\lri_\lf$, 
$q_\rt\stackrel{\rm def}{=}q_\lri$ for $\lri=\lri_\rt$, and
the reflection and transmission coefficients are obtained from the condition of continuity 
at $m=0$:
\begin{subequations}
	\label{eq:m=0}
	\begin{equation}
		\label{eq:cont_m=0}
		1+R^{(h)}=T^{(h)},
	\end{equation}
	and from the difference equation (\ref{eq:disc}) at $m=0$, which reads:
	\begin{equation}
		\label{eq:de_m=0}
		\begin{aligned}
		L_1(\lri_\lf)(q_\lf^{-1}+ &\> R^{(h)}q_\lf)\\
			- &\> (L_0(\lri_\lf)+ L_0(\lri_\rt))T^{(h)}\\
				& \qquad\qquad+ L_1(\lri_\rt)T^{(h)}q_\rt=0.
		\end{aligned}
	\end{equation}
\end{subequations}
Solving the system of equations (\ref{eq:m=0}) with respect to $R^{(h)}$ and $T^{(h)}$ and using the Taylor expansion of the resulting solution, one can show that $R^{(h)}=R+\oh{4}$ and $T^{(h)}=T+\oh{4}$.
These relations, along with estimates (\ref{eq:wave_appr}), imply that the discrete solution $E_m$ of (\ref{eq:disc_sol}) converges to the continuous solution $E(z)$ of (\ref{eq:cont_sol}) with the rate $\oh{4}$ as $h\longrightarrow0$.

\section{\label{app:interp_accuracy}Birkhoff-Hermite interpolation 
	(proof of Lemma~\ref{lem:cubic-interp}) }

A large body of work has been done by different authors on Birkhoff-Hermite interpolation.
Nonetheless, 
for the completeness of our analysis we present an
elementary convergence proof in the case of cubic polynomials.
It is self-contained and does not require any additional facts from the literature.

It will be convenient to make the change of variables: $x=z-z_\nph$, so that $ x\in\left[-\frac{h}{2},\frac{h}{2}\right]$.  
With respect to the new coordinate $x$, the material discontinuities are allowed at $x\pm\frac{h}{2}$, whereas on the interval $(-\frac{h}{2},\frac{h}{2})$ and, in particular, at the cell center $x=0$, the solution is $C^\infty$.

A cubic polynomial $P_3(x)$ that satisfies $
    P_3\left(\pm\frac{h}{2}\right) = E_{\pm\frac{h}{2}}
$ and $
    P_3''\left(\pm\frac{h}{2}\right) = E_{\pm\frac{h}{2}}''
$ is
\begin{equation}
\label{eq:P_x}
\begin{aligned}
    P_3(x) & = 
        \left(
            \frac{1}{2}-\frac{x}{h}
        \right) \left(
            E_{-\frac{h}{2}}
            - \frac{h^2}{6} E_{-\frac{h}{2}}''
        \right)
        + \frac{h^2}{6} E_{-\frac{h}{2}}''
            \left( \frac{1}{2}-\frac{x}{h} \right)^3  \\
        & 
        +\left(
            \frac{1}{2}+\frac{x}{h}
        \right) \left(
            E_{\frac{h}{2}}
            - \frac{h^2}{6} E_{\frac{h}{2}}''
        \right)
        + \frac{h^2}{6} E_{\frac{h}{2}}''
            \left( \frac{1}{2}+\frac{x}{h} \right)^3.
\end{aligned}
\end{equation}
It is unique since the four parameters $E_{\pm\frac{h}{2}}, \ E_{\pm\frac{h}{2}}''$ uniquely determine the four coefficients $c_j$ of $P_3(x)=\sum_{j=0}^3c_jx^j$ via the solution of the corresponding $4\times4$ linear system.

Next, we prove that the polynomial $P_3(x)$ is indeed a fourth order 
approximation of the field $E(x)$.
Differentiating $P_3(x)$ of (\ref{eq:P_x}) three times at $x=0$, 
we have: 
\begin{alignat*}{2}
	P_3(0) = &\>
		\frac{ E_{\frac{h}{2}} + E_{-\frac{h}{2}} } 2
		-\frac{ h^2 } 8 
			\frac{ E_{\frac{h}{2}}'' + E_{-\frac{h}{2}}''} 2, \quad &
        P_3''(0) = &\>
                \frac{ E_{\frac{h}{2}}'' + E_{-\frac{h}{2}}'' } 2,\\
        P_3'(0) = &\>
                \frac{ E_{\frac{h}{2}} - E_{-\frac{h}{2}} }{ h }
                -\frac{ h^2 }{ 24 }
                        \frac{ E_{\frac{h}{2}}'' - E_{-\frac{h}{2}}'' }{ h }, &
        P_3^{(3)}(0) = &\>
                \frac{ E_{\frac{h}{2}}'' - E_{-\frac{h}{2}}'' }{ h }.
\end{alignat*}
Then, expressing $E_{\pm\frac{h}{2}}$ and $E_{\pm\frac{h}{2}}''$ 
with the help of the Taylor formulae for $E(x)$ and $E''(x)$ at $x=0$, we obtain:
\begin{equation}
\label{eq:P_deriv}
\begin{aligned}
	P_3(0)  = &\> \left(
			E(0) + \frac{h^2}{8}E''(0) + \oh{4}
		\right) - \frac{h^2}{8} \left(
			E''(0) + \oh{2}
		\right)\\ = &\> E(0)+ \ \oh{4},\\
        P_3'(0)  = &\> \left(
                        E'(0) + \frac{h^2}{24}E^{(3)}(0) + \oh{4}
                \right)
                - \frac{h^2}{24} \left(
                        E^{(3)}(0) + \oh{2}
                \right)\\ = &\> E'(0)+ \oh{4}, \\
	P_3''(0)  = &\> E''(0)+\oh{2}, \\
	P_3^{(3)}(0)  = &\> E^{(3)}(0)+\oh{2}.
\end{aligned}
\end{equation}
Since $P_3(x)$ is a cubic polynomial, we can write:
\begin{equation*}
	P_3(x)=\sum_{k=0}^3\frac{P_3^{(k)}(0)}{k!}x^k. 
\end{equation*}
Moreover, as $E(x)$ is smooth on $\left(-\frac{h}{2},\frac{h}{2}\right)$, the
Taylor formula yields:
\begin{equation*}
	E(x)=\sum_{k=0}^3\frac{E^{(k)}(0)}{k!}x^k+{\cal O}(h^4),\quad x\in
\left(-\frac{h}{2},\frac{h}{2}\right). 
\end{equation*}
Hence, using equalities (\ref{eq:P_deriv}), we obtain:
\begin{equation*}
P_3(x)-E(x)=
	\sum_{k=0}^3 \frac{P_3^{(k)}(0)-E^{(k)}(0)}{k!}x^k + \oh{4} = \oh{4}, \quad x\in
\left(-\frac{h}{2},\frac{h}{2}\right).
\end{equation*}

\section{\label{app:SE}Software engineering}

Although calculating the $64$ coefficients $g_{ijk}$ in (\ref{eq:FO_tensor_elements}) is 
straightforward, it is a tedious and error prone task. 
As such, it is a natural choice for automation.
Note that automation will become an absolute necessity should we wish 
to extend the method of this paper, say, to a quintic nonlinearity
, or a multidimensional setting.

In the current paper, we developed simple scripts which
automate the calculation of the constants $g_{ijk}$. 
The general approach is to use a template file to generate a different 
Maple~script for each coefficient.
For $\left.i,j,k=0,\dots,3\right.$, 
Maple's symbolic utilities calculate the function $\left.g_{ijk}(\lri,\hk)\right.$, 
while its code generation utilities then generate the required Matlab 
function to evaluate the expression.

The scripts are available under the GPL license at  the following URL:
{\tt http://www.tau.ac.il/$\sim$guybar/1DNLH.}

\bibliographystyle{elsart-num}
\bibliography{NLH,NLS}

\end{document}